\documentclass[nofootinbib,aps,11pt]{revtex4}
\usepackage{graphicx}
\usepackage{cancel}
\usepackage{amssymb}
\usepackage{textcomp}
\usepackage{amsmath}
\usepackage{bm}
\usepackage{times}
\usepackage{epsfig}
\usepackage{color}
\def\br{{\rm BR}}
\def\vd{v^{}_\Delta}
\def\lsim{\mathrel{\raise.3ex\hbox{$<$\kern-.75em\lower1ex\hbox{$\sim$}}}}
\def\gsim{\mathrel{\raise.3ex\hbox{$>$\kern-.75em\lower1ex\hbox{$\sim$}}}}

\def\non{\nonumber}

\def\gev{\,{\rm GeV}}

\def\ev{\,{\rm eV}}
\def\to{\rightarrow}

\def\beq{\begin{equation}}
\def\eeq{\end{equation}}
\def\be{\begin{equation}}
\def\ee{\end{equation}}
\def\bea{\begin{eqnarray}}
\def\eea{\end{eqnarray}}
\begin{document}
\begin{titlepage}
\begin{flushright}
MADPH-08-1510 \\
NSF-KITP-08-65
\end{flushright}
\vspace{-2cm}
%
\title{NEUTRINO MASSES AND THE LHC: TESTING TYPE II SEESAW}
\author{Pavel Fileviez P{\'e}rez$^{\bf1}$}
\email{fileviez@physics.wisc.edu}
\author{Tao Han~$^{\bf1,2}$}
\email{than@hep.wisc.edu}
\author{Guiyu Huang~$^{\bf1}$}
\email{ghuang@hep.wisc.edu}
\author{Tong Li~$^{\bf1,3}$}
\email{tli8@wisc.edu}
\author{Kai Wang~$^{\bf1}$}
\email{wangkai@hep.wisc.edu}
\vspace{3.0cm}
\affiliation{
{\small $^{\bf1}$ \textit{Department of Physics, University of Wisconsin, Madison, WI 53706, USA }}\\
{\small $^{\bf2}$ \textit{KITP, University of California, Santa Barbara, CA 93107, USA }}\\
{\small $^{\bf3}$ \textit{Department of Physics, Nankai University, Tianjin 300071, P.R.~China }}}
\date{\today}
\pagestyle{plain}
\begin{abstract}
We demonstrate how to systematically test a well-motivated mechanism
for neutrino mass generation (Type-II seesaw)  at the LHC, in which
a Higgs triplet is introduced.
In the optimistic scenarios with a small Higgs triplet vacuum expectation
value $\vd < 10^{-4}$ GeV, one can look for clean signals of lepton number
violation in the decays of doubly charged ($H^{\pm\pm}$) and singly
charged ($H^{\pm}$) Higgs bosons to distinguish the Normal
Hierarchy (NH), the Inverted Hierarchy (IH) and the
Quasi-Degenerate (QD) spectrum for the light neutrino masses.
The observation of either $H^+ \to \tau^+\bar \nu$ or $H^+ \to e^+\bar \nu$
will be particularly robust for the spectrum test since
they are independent of the unknown Majorana phases.
The $H^{++}$ decays moderately depend on a Majorana phase
$\Phi_2$ in the NH, but sensitively depend on $\Phi_1$ in the IH.
In a less favorable scenario $\vd > 2 \times 10^{-4}$ GeV,
when the leptonic channels are suppressed, one needs to observe
the decays $H^+ \to W^+ H_1$ and $H^+ \to t \bar b$ to confirm the
triplet-doublet mixing which in turn implies the existence of the
same gauge-invariant interaction between the lepton doublet
and the Higgs triplet responsible for the neutrino mass generation.
In the most optimistic situation, $v_{\Delta} \sim10^{-4}$ GeV, both
channels of the lepton pairs and gauge boson pairs may be available
simultaneously. The determination of their relative branching
fractions would give a measurement for the value of $v_{\Delta}$.
\end{abstract}
\maketitle

\end{titlepage}

\section{INTRODUCTION}

The existence of massive neutrinos~\cite{review} is a strong motivation
for physics beyond the Standard Model (SM). As pointed out a long time
ago by Weinberg~\cite{Weinberg}, there is just one dimension-five operator
relevant for neutrino masses in the context of the Standard Model:
$\left( \kappa/\Lambda \right) l_L^{} l_L^{} HH$, where $l_L^{}$ and $H$
are the leptonic and Higgs $SU(2)_L$ doublets. After the electroweak symmetry
breaking (EWSB), the Majorana mass of the neutrinos reads
as $m_\nu \sim \kappa v^2_0/\Lambda$, where $v_0 \approx 246$ GeV is the
SM Higgs vacuum expectation value (vev). The smallness of $m_\nu \lsim 1$ eV
is thus understood by the ``seesaw" spirit if $\Lambda \gg v_0$. Assuming
that the coupling $\kappa$ of the dimension-five operator is the order
of unity, the observed neutrino masses imply that $\Lambda \lsim 10^{14-15}$ GeV.
The crucial issue is to understand the origin of this operator in a given
extension of the SM in order to identify the dimensionless coupling $\kappa$
and the mass scale $\Lambda$ at which the new physics enters.
This dimension five operator thus guides us to look for extensions of Standard
Model in which the neutrino masses are generated in a UV complete formalism.

There are four simple renormalizable extensions of the Standard Model with
minimal addition to generate neutrino Majorana masses conceivable to agree with
the experimental observations:
\begin{itemize}
\item  \textit{Type I seesaw mechanism}~\cite{TypeI}:
One can add at least two fermionic singlets $N_i$ and the
neutrino masses are $m_\nu \sim y^2_D v^2_0/ M_N$, where
$y_D^{}$ is the Yukawa coupling and $M_N$ is the
right-handed neutrino mass, which sets the new physics scale $\Lambda$.
If $y_D \simeq 1$ and $M_{N} \approx 10^{14-15}$ GeV,
one obtains the natural value for the neutrino masses $m_\nu \approx 1$ eV.
\item \textit{Type II seesaw mechanism}~\cite{TypeII}:
The Higgs sector of the Standard Model is extended by adding an
$SU(2)_L$ Higgs triplet $\Delta$. The neutrino masses are
$m_\nu \approx Y_\nu  v_{\Delta}$, where  $v_{\Delta}$ is the
vacuum expectation value (vev) of the neutral component of the
triplet and $Y_\nu$ is the Yukawa coupling. With a doublet and
triplet mixing via a dimensional parameter $\mu$, the EWSB leads
to a relation $v_{\Delta} \sim \mu v_0^2/M_{\Delta}^2$,
where $M_{\Delta}$ is the mass of the triplet. In this case the
scale $\Lambda$ is replaced by $M_{\Delta}^2/\mu$, and a natural
setting would be for $Y_\nu \approx 1$ and
$\mu \sim M_{\Delta} \approx 10^{14-15}$ GeV.
\item \textit{Type III seesaw mechanism}~\cite{TypeIII}:
Adding at least two extra matter fields in the adjoint
representation of $SU(2)_L$ and with zero hypercharge, one can
generate neutrino masses, $m_\nu \sim y^2 v^2_0/M$. Therefore,
the high scale $\Lambda$ is replaced by the mass of the extra
fermions in the adjoint representation.
\item \textit{Hybrid seesaw mechanism}~\cite{TypeI-III}:
One SM fermionic singlet $N$ and one fermion in the adjoint representation
of $SU(2)_L$ are added. This is a combination of Type I and Type III but with
the same minimal fermionic content. This mechanism has a very simple and
unique realization in the context of grand unified theories~\cite{TypeI-III}.
\end{itemize}
In the case of Left-Right symmetric models~\cite{LR} both Type I and Type II seesaw are present.
Alternatively, neutrino masses can be generated by radiative corrections~\cite{Babu-Zee}.

To test the above seesaw mechanisms one needs to search for the effects
of lepton number violation in their unique way. In particular, direct observations
of the new heavy states responsible for the seesaw mechanisms would be
more conclusive. While the seesaw spirit resides in the existence of a much
higher scale $\Lambda \gg v_0$, rendering the new states experimentally inaccessible
in the foreseeable future, this may not be necessary the case.
For recent studies where the seesaw mechanism could happen at a
very low scale see~\cite{deGouvea:2006gz}. A light $SU(2)_L$ triplet
field responsible for Type II seesaw can be present in the context
of a minimal grand unified theory~\cite{MGUT}. Low scale
Type III seesaw was also studied in~\cite{Bajc:2007zf}.

The Large Hadron Collider (LHC) at CERN will soon take us to a
new frontier with unprecedented high energy and luminosity. Major discoveries
of exciting new physics at the Terascale are highly anticipated.
It is thus pressing to investigate the physics potential of the LHC in
connection with the new physics for the neutrino mass generation.
Searching for heavy Majorana neutrinos at hadron colliders have been
considered by many authors \cite{goran}. The interests for the LHC
have been lately renewed \cite{Han:2006ip,more,Franceschini:2008pz}.
However, it is believed
that any signal of $N$ would indicate a more subtle mechanism beyond
the simple Type I seesaw due to the otherwise naturally small mixing
$V_{N\ell}^2 \sim m_\nu/M_N$ between $N$ and the SM leptons.

In this paper,
we investigate the possibility to test the Type II seesaw mechanism at the LHC.
Several earlier studies for certain aspects of the Type II seesaw model
at the LHC exist \cite{Chun,taurec,last,Thomas,Raidal,Akeroyd,Chao:2007mz}.
We systematically explore the parameter space in the model.
Guided by the neutrino oscillation experiments,
we first establish the preferred parameter regions by reproducing the light neutrino
mass and mixing patterns. We then go on to predict the corresponding signatures
at the LHC. We find that in the optimistic scenarios, by
identifying the flavor structure of the lepton number violating
decays of the charged Higgs bosons, one can establish the neutrino mass
pattern of the Normal Hierarchy, Inverted Hierarchy or Quasi-Degenerate.
We emphasize the crucial role of the singly charged
Higgs boson decays. The associated pair production
of $H^{\pm\pm} H^{\mp}$ is essential to test the triplet
nature of the Higgs field. The observation of either
$H^+ \to \tau^+ \bar \nu$ or $H^+ \to e^+\bar \nu$ will be
particularly robust for the test since they are independent
of the unknown Majorana phases. Combining with the doubly charged
Higgs decay, for instance $H^{++} \to e^+ \mu^+, e^+\tau^+,\mu^+\tau^+$,
one will even be able to probe the Majorana phases.
We investigate in great detail all the issues mentioned
above, showing all the possibilities to test this
appealing mechanism for the neutrino masses at the
Large Hadron Collider. A summary of our main results
appeared in an early publication~\cite{Letter}.

The outline of the paper is as follows: In Section II we present
the Type II seesaw mechanism  and discuss its main
predictions. In Section III the constraints on the physical
Higgs couplings coming from neutrino oscillation experiments
are investigated.  The general features of the Higgs decays are discussed in
Section IV.
In Section V we study the predictions for the Higgs decays
in this theory. Taking into account the effect of neutrino
masses and mixing we show the different predictions
for the branching fractions of all lepton number violating
decays $H^{++} \to e_i^+ e_j^+$ and $H^+ \to e^+_i \bar{\nu}$,
where $e_i=e,\mu,\tau$. We discuss the possibility to identify
the spectrum for neutrino masses if all the lepton violating decays
are measured at the LHC or at future colliders. The possibility
to get the information about the Majorana phases
from Higgs decays is discussed. The most important production mechanisms
at the LHC are discussed in Section VI.  In Section VII,
we discuss the necessary steps for testing the Type II seesaw at the LHC, and
we draw our conclusions.

\section{THE TYPE II SEESAW MECHANISM FOR NEUTRINO MASSES}

The Type II seesaw mechanism~\cite{TypeII} is one of the most appealing scenarios
for the generation of neutrino masses. In this section we discuss in detail this
mechanism and its main predictions. In order to realize the so-called Type II
seesaw mechanism for neutrino masses one has to extend the Higgs sector of
the Standard Model. In this case the Higgs sector of the theory
is composed of the SM Higgs $H \sim (1,2,1/2)$ and
an $SU(2)_L$ scalar triplet $\Delta \sim (1,3,1)$. The matrix
representation of the triplet reads as
\begin{equation}
\Delta = \left( \begin{array} {cc}
 \delta^+/\sqrt{2}  &  \delta^{++} \\
 \delta^0 & - \delta^+/\sqrt{2}
\end{array} \right).
\end{equation}
The kinetic terms and the relevant interactions in this theory are given by
\begin{eqnarray}
{\cal L}_{\text{TypeII}} &=& (D_\mu H)^\dagger (D^\mu H) \ + \  \text{Tr} (D_\mu \Delta)^\dagger (D^\mu \Delta)
\ + \ {\cal L}_{Y} \ - \ V (H,\Delta),
\label{Lagrangian}
\end{eqnarray}
where the needed interaction to generate neutrino masses reads as
\begin{eqnarray}
{\cal L}_{Y} & = & - Y_\nu \ l_L^T \ C \ i \sigma_2 \ \Delta \ l_L \ + \ \text{h.c.},
\label{Yukawa}
\end{eqnarray}
and the scalar interactions are given by
\begin{eqnarray}
V(H,\Delta) & = & - m_H^2 \ H^\dagger H \ + \ \frac{\lambda}{4} (H^\dagger H)^2 \ + \ M_{\Delta}^2 \ Tr \Delta^\dagger \Delta
\ + \ \left( \mu \ H^T \ i \sigma_2 \ \Delta^\dagger H \ + \ h.c.\right) \ + \nonumber \\
&+& \ \lambda_1 \ (H^\dagger H) Tr \Delta^\dagger \Delta \ + \ \lambda_2 \ \left( Tr \Delta^\dagger \Delta \right)^2 \ +
\ \lambda_3 \ Tr \left( \Delta^\dagger \Delta \right)^2 \ + \ \lambda_4 \ H^\dagger \Delta \Delta^\dagger H.
\label{Potential}
\end{eqnarray}
In the above equations the Yukawa coupling $Y_\nu$ is a $3\times 3$
symmetric complex matrix. $l_L^T = (\nu_L^T, \ e_L^T)$, C is the charge
conjugation operator, and $\sigma_2$ is the Pauli matrix.
Since we are mainly interested in a heavy Higgs triplet,
typically $M_{\Delta}^2 > v_0^2/2$, we will neglect the
contributions coming from the terms proportional to
$\lambda_1$, $\lambda_2$, $\lambda_3$ and $\lambda_4$.
The detailed structure and interactions of this Higgs
sector will be presented in Appendix A.

Let us discuss some important features of this model for neutrino masses:
\begin{itemize}
\item Imposing the conditions of global minimum one finds that
\begin{eqnarray}
&& - m_H^2 \ + \ \frac{\lambda}{4} v_0^2 \ - \ \sqrt{2} \ {\mu} \ v_{\Delta}= 0,
\qquad \text{and} \qquad v_{\Delta} = \frac{\mu \ v_0^2}{ \sqrt{2} \ M_{\Delta}^2},
\label{vev3}
\end{eqnarray}
where $v_0$ and $v_\Delta$ are the vacuum expectation values
of the Higgs doublet and triplet, respectively, with
$v_0^2 + v_\Delta^2 \approx (246\ {\rm GeV})^2$.
Due to the simultaneous presence of the Yukawa coupling $Y_\nu$ in Eq.~(\ref{Yukawa}),
and the term proportional to the $\mu$ parameter in Eq.~(\ref{Potential}),
the lepton number is explicitly broken in this theory. Therefore,
one expects that the neutrino Majorana mass term has to be
proportional to $Y_\nu \times \mu$.
\item Once the neutral component  in $\Delta$ gets the vev, $v_{\Delta}$
as in Eq.~(\ref{vev3}), the neutrinos acquire a Majorana mass given by
the following expression:
\begin{eqnarray}
M_{\nu}= \sqrt{2} \ Y_\nu \ v_{\Delta} = Y_\nu\  \frac{\mu \ v_0^2}{  M_{\Delta}^2},
\label{type2}
\end{eqnarray}
which is the key relation for the Type II seesaw scenario.
\item After the electroweak symmetry breaking, there are seven physical
massive Higgs bosons left in the spectrum:
\begin{eqnarray}
H_1 &=& \cos \theta_0 \ h^0 \ + \ \sin \theta_0 \ \Delta^0,
\quad
H_2 = - \sin \theta_0 \ h^0 \ + \ \cos \theta_0 \Delta^0,
\qquad
\text{with} ~~ \theta_0\approx {2v_\Delta\over v_0},
\\
A &=& - \sin \alpha \ \xi^0 \ + \ \cos \alpha \ \eta^0,\qquad \text{with} \qquad \alpha \approx {2v_\Delta\over v_0},
\\
H^{\pm} &=& - \sin \theta_{\pm} \ \phi^{\pm} \ + \ \cos \theta_{\pm} \ \delta^{\pm},\qquad
\text{with} \qquad \theta_+ \approx {\sqrt 2v_\Delta\over v_0},
\\
\text{and} && \qquad H^{\pm\pm}=\delta^{\pm\pm}, \qquad \text{with mass} \qquad M_{\delta^{++}} = M_{\Delta},
\label{higgses}
\end{eqnarray}
where $H_1$  is SM-like (doublet) while the rest of the Higgs states
are all $\Delta$-like (triplet), and
$$M_{H_2}\simeq M_{A} \simeq M_{H^{\pm}} \simeq M_{H^{++}}=M_{\Delta}.$$
\item Working in the physical basis for the fermions we find that
the Yukawa interactions can be written as
\begin{eqnarray}
& & \nu_L^T \ C \ \Gamma_+ \ H^+ \ e_L, \qquad \text{and} \qquad e_L^T \ C \ \Gamma_{++} \ H^{++} \ e_L,
\label{Physical-Yukawa}
\end{eqnarray}
where
\begin{eqnarray}
\Gamma_+  =  \cos \theta_+ \ \frac{m_\nu^{diag}}{v_{\Delta}} \ V_{PMNS}^\dagger, & & \qquad \text{and} \qquad
\Gamma_{++} =  V_{PMNS}^* \ \frac{m_{\nu}^{diag}}{\sqrt{2} \ v_{\Delta}} \ V_{PMNS}^{\dagger}=Y_\nu.
\label{gamma}
\end{eqnarray}
The values of the physical couplings $\Gamma_+$ and $\Gamma_{++}$ are
thus governed by the spectrum and mixing angles for the active neutrinos.
Therefore, one can expect that the lepton-number violating decays of the Higgs bosons,
$H^{++} \to e_i^+ e_j^+$ and $H^{+} \to e_i^+ \bar{\nu}\ (e_i=e,\mu,\tau)$ will
be characteristically different in each spectrum for neutrino masses.

\item Higgs-Gauge Interactions: The doubly charged Higgs has
only one coupling to gauge bosons, $H^{\pm \pm} W^{\mp} W^{\mp}$,
which is proportional to the vev of the triplet field $v_{\Delta}$.
In the case of the singly charged Higgs there are two relevant
couplings for the decays into gauge bosons, $H^{\pm} W^{\mp} H_1$
and $H^{\pm} W^{\mp} Z$. As for the heavy neutral
Higgs $H_2$ one finds that its coupling to $W$'s is further suppressed.
The only relevant couplings for the decays are
$H_2 ZZ$ and $H_2 H_1 H_1$, see Appendix A for details.
\end{itemize}

These are the main properties and predictions of this simple
extension of the Standard Model where the neutrino masses
are generated through the Type II seesaw mechanism.

\section{CONSTRAINTS ON THE PHYSICAL PARAMETERS}

In this section we discuss the constraints coming from neutrino experiments,
rare decays and collider experiments on the physical parameters in this
theory for neutrino masses.

\subsection{Constraints From Neutrino Oscillation Experiments}

The relevant physical Yukawa couplings of the singly and doubly charged Higgs bosons for the leptonic
decays are given by Eq.~(\ref{gamma}). In order to understand the constraints coming from
neutrino physics let us discuss the relation between the neutrino masses
and mixing. The leptonic mixing matrix is given by
\beq
V_{PMNS}=
\left(
\begin{array}{lll}
 c_{12} c_{13} & c_{13} s_{12} & e^{-\text{i$\delta $}} s_{13}
   \\
 -c_{12} s_{13} s_{23} e^{\text{i$\delta $}}-c_{23} s_{12} &
   c_{12} c_{23}-e^{\text{i$\delta $}} s_{12} s_{13} s_{23} &
   c_{13} s_{23} \\
 s_{12} s_{23}-e^{\text{i$\delta $}} c_{12} c_{23} s_{13} &
   -c_{23} s_{12} s_{13} e^{\text{i$\delta $}}-c_{12} s_{23} &
   c_{13} c_{23}
\end{array}
\right)\times \text{diag} (e^{i \Phi_1/2}, 1, e^{i \Phi_2/2})
\eeq
where $s_{ij}=\sin{\theta_{ij}}$, $c_{ij}=\cos{\theta_{ij}}$,
$0 \le \theta_{ij} \le \pi/2$ and $0 \le \delta \le 2\pi$.
The phase $\delta$ is the Dirac CP-violating phase, while $\Phi_i$ are the Majorana phases.
The experimental constraints on the neutrino masses and mixing parameters,
at $2\sigma$ level~\cite{Schwetz}, are
\bea
7.3 \times 10^{-5} \ev^2 \  < & \Delta m_{21}^2 & < \  8.1 \times 10^{-5} \ev^2, \\
2.1 \times 10^{-3} \ev^2 \  < & |\Delta m_{31}^2| & < \  2.7 \times 10^{-3} \ev^2, \\
                   0.28 \  < & \sin^2{\theta_{12}} & < \  0.37, \\
                   0.38 \  < & \sin^2{\theta_{23}} & <\  0.63, \\
                          & \sin^2{\theta_{13}} & <\  0.033,
\eea
and from cosmological observations
\beq
\sum_{i=1}^3 m_i  < \ 1.2 \ \ev.
\eeq
For a complete discussion of these constraints see reference~\cite{review}.
In this section we focus mainly on the case of Normal Hierarchy (NH),
$\Delta m_{31}^2 > 0$, and Inverted Hierarchy (IH) spectrum, $\Delta m_{31}^2 < 0$,
neglecting the Majorana phases.

\begin{figure}[tb]
\includegraphics[scale=1,width=8.0cm]{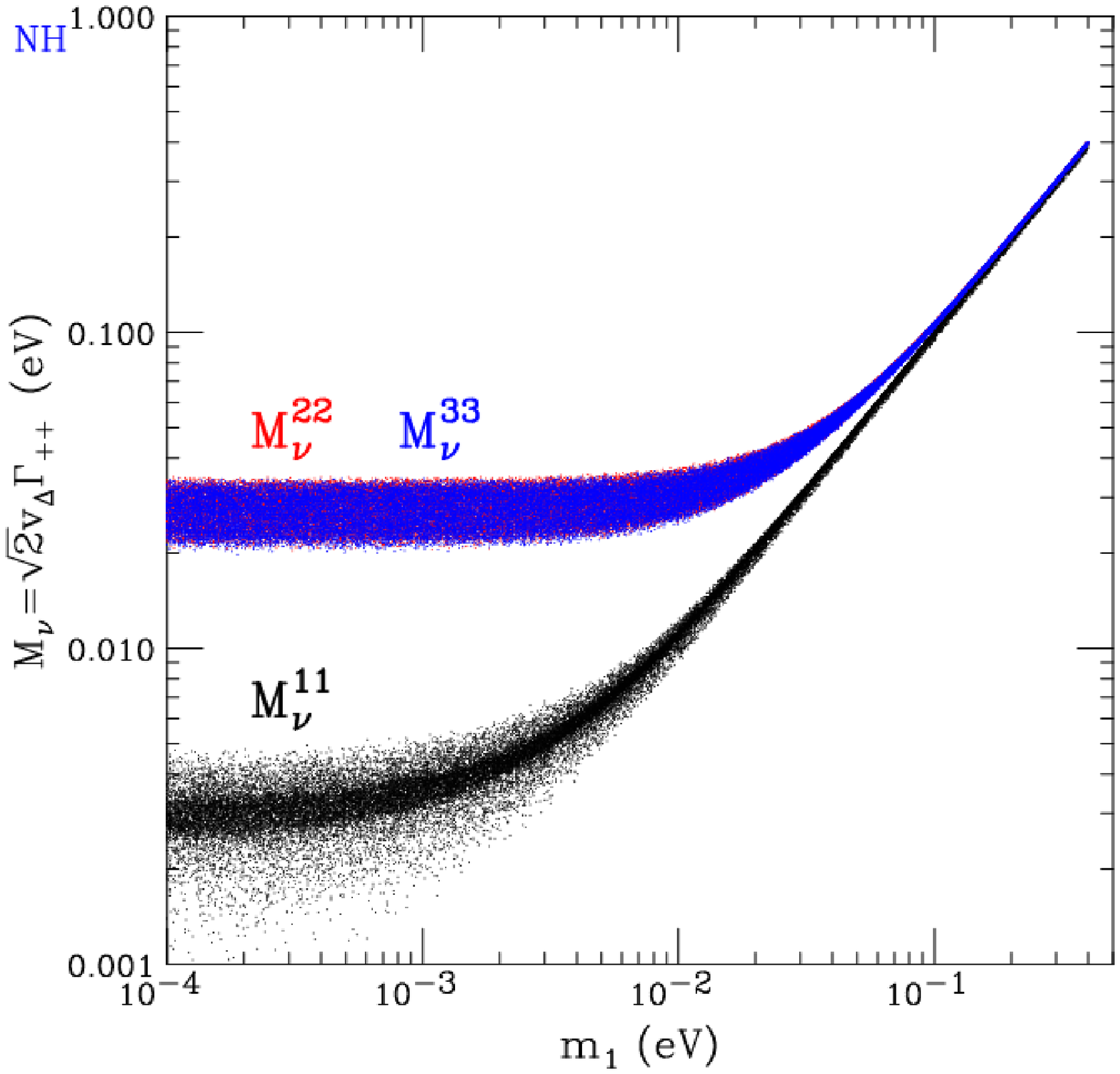}
\includegraphics[scale=1,width=8.0cm]{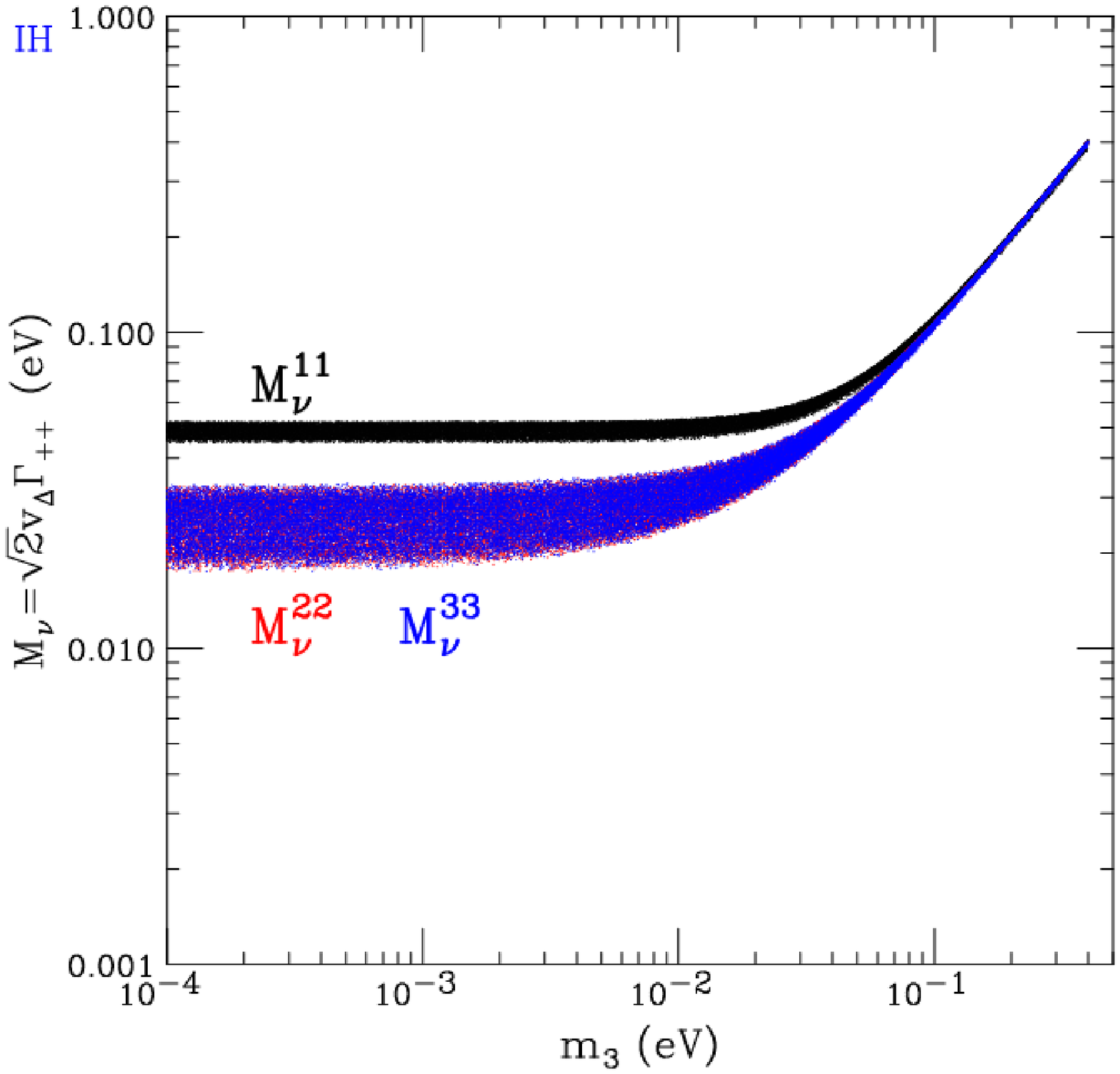}
\caption{Constraints on the diagonal elements of the neutrino mass
matrix $M_\nu$ versus the lowest neutrino mass for (a) NH (left) and (b) IH (right)
when $\Phi_1 =0$ and $\Phi_2 = 0$.}
\label{mii}
\end{figure}

\begin{figure}[tb]
\includegraphics[scale=1,width=8.0cm]{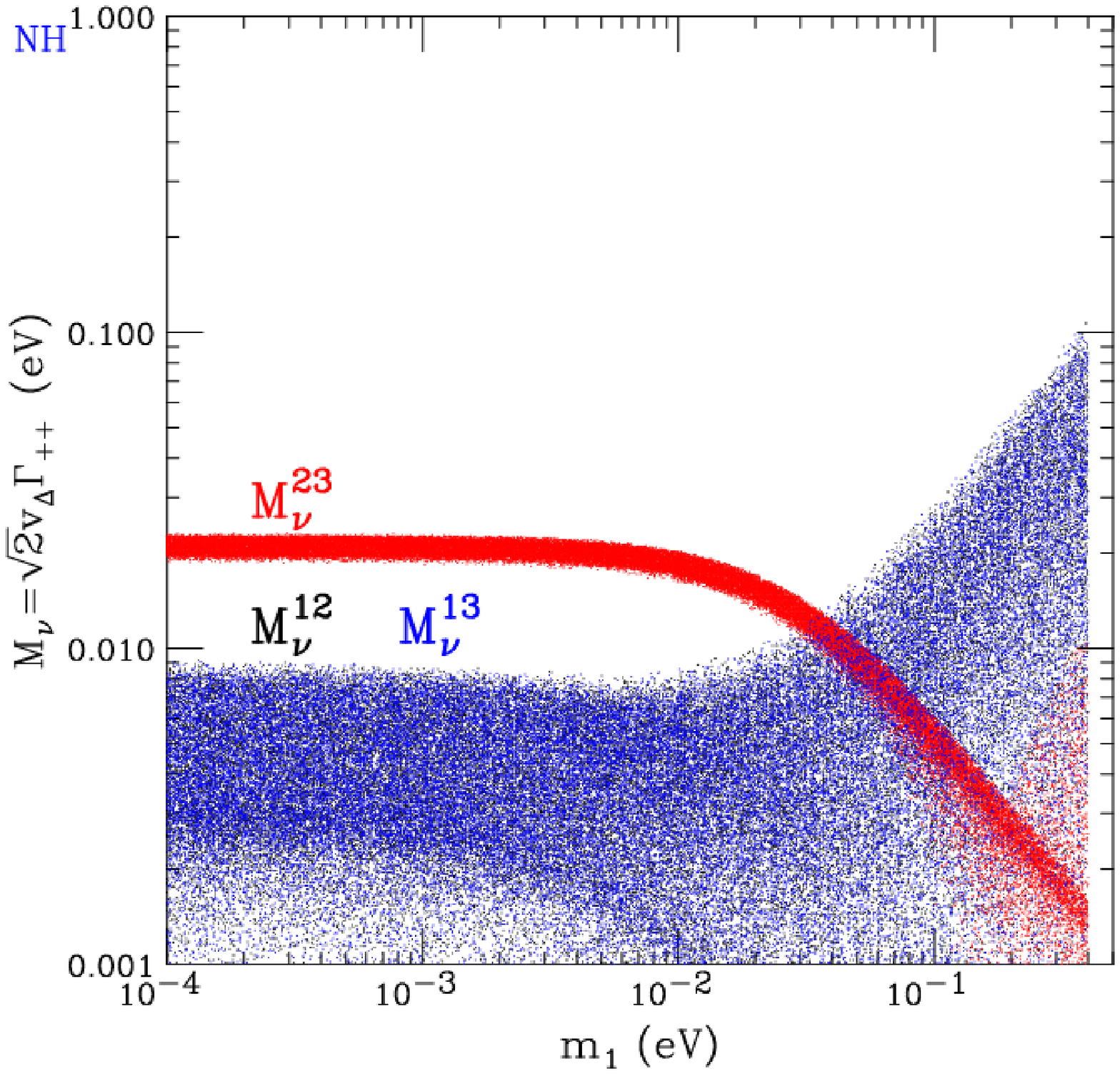}
\includegraphics[scale=1,width=8.0cm]{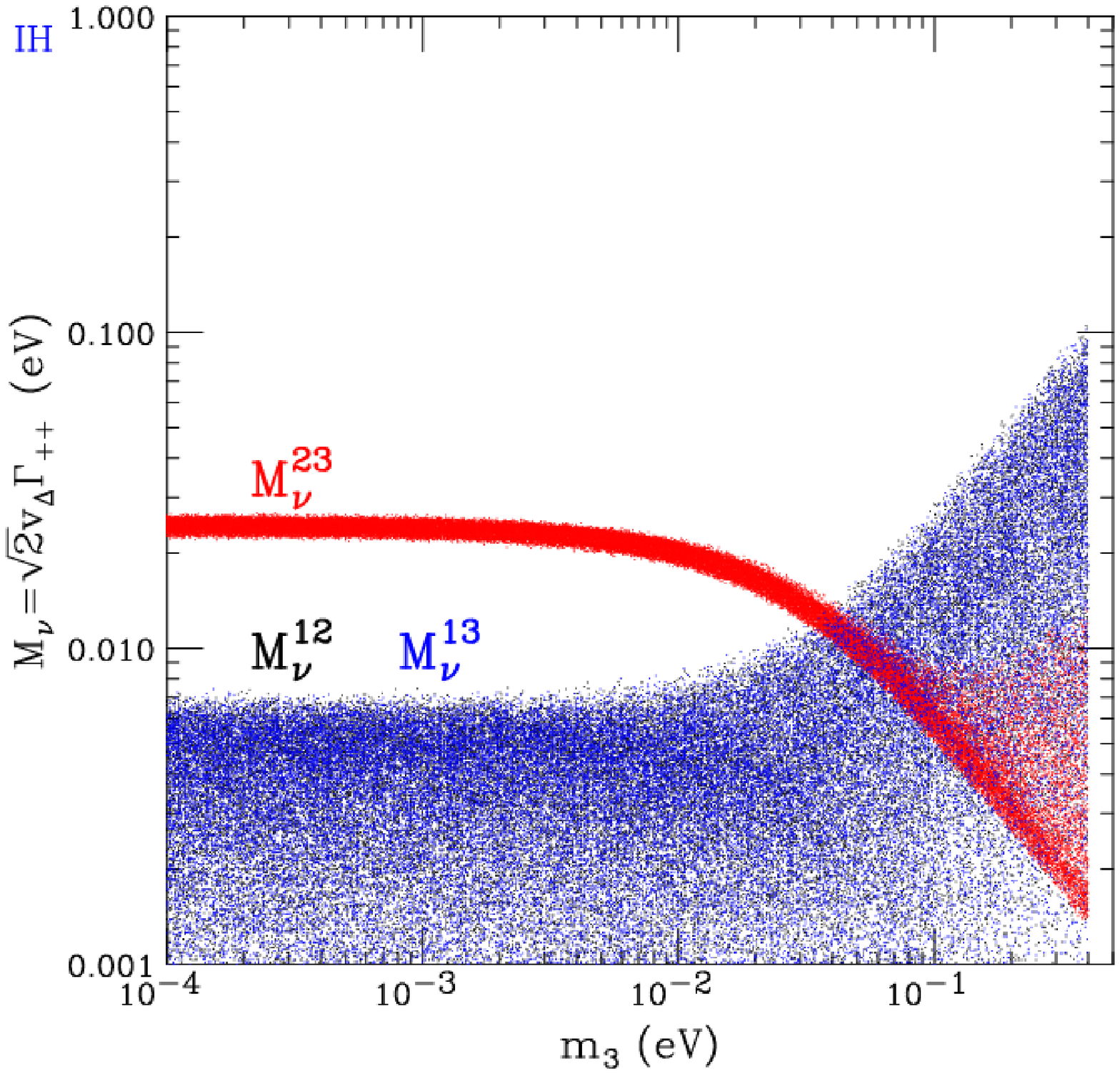}
\caption{Constraints on the off-diagonal elements of the neutrino mass
matrix $M_{\nu}$ versus the lowest neutrino mass
for (a) NH (left) and (b) IH (right) when $\Phi_1 =0$ and $\Phi_2 = 0$.}
\label{mij}
\end{figure}

\begin{figure}[tb]
\includegraphics[scale=1,width=8.0cm]{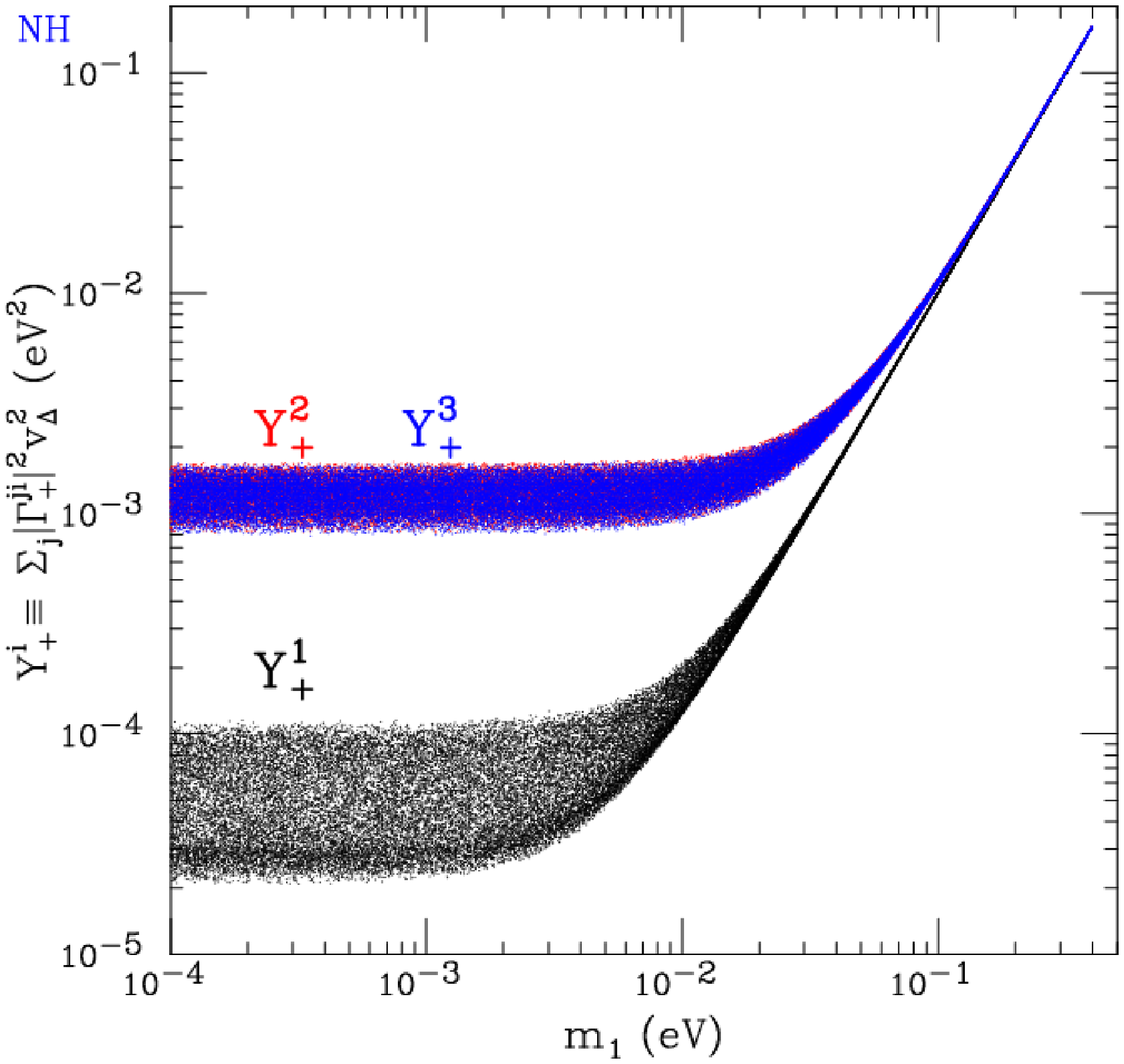}
\includegraphics[scale=1,width=8.0cm]{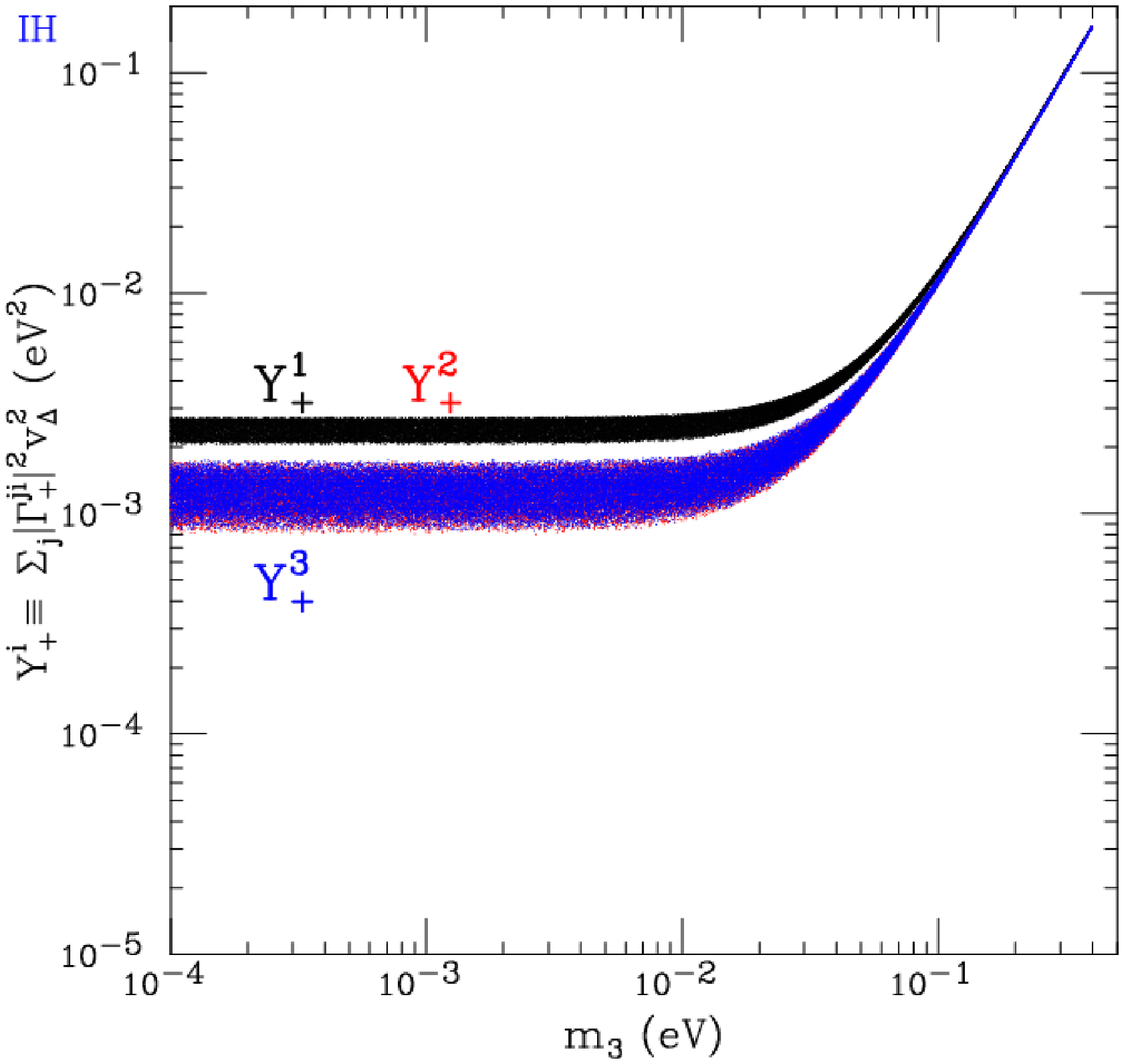}
\caption{Constraints on the coupling squared for $H^{+}$,
$Y_+^i\equiv \sum_j |\Gamma_{+}^{ji}|^2 v^2_{\Delta}$,
 versus the lowest neutrino mass for (a) NH (left) and (b) IH (right).
}
\label{Yi}
\end{figure}

Using the above experimental constraints, we first show the
allowed values for the  neutrino mass matrix  $M_{\nu}$ as seen
in Figs.~\ref{mii} and \ref{mij}, as a function of the lightest neutrino mass.
These results directly reflect the patterns of the neutrino mass and mixing:
$M_{\nu}^{11} \ \ll \ M_{\nu}^{22}, M_{\nu}^{33}$ in the case of NH in Fig.~\ref{mii}(a),
and $M_{\nu}^{11} \ > \ M_{\nu}^{22}, M_{\nu}^{33}$ in the case of IH in Fig.~\ref{mii}(b).
For the off-diagonal elements, $M_{\nu}^{23}$ takes the largest
values in each spectrum due to the large atmospheric mixing angle as seen
in  Fig.~\ref{mij}. Also seen is the ``quasi-degenerate" case for
$m_1\approx m_2 \approx m_3 > |\Delta m_{31}|$, where the flavor-diagonal
elements are about equal.
Since $\Gamma_{++}=M_{\nu}/\sqrt{2} v_{\Delta}$,
the constraints on the neutrino mass matrix elements directly translate into the
physical couplings of $H^{++}$ that govern its decay widths.
As for the coupling of the singly charged Higgs boson, we sum over the final
state neutrinos since they are experimentally unobservable. Thus
the relevant couplings are written as
\beq
Y_+^i\equiv \sum_{j=1}^3 |\Gamma_{+}^{ji}|^2 v^2_{\Delta} \quad (i=1,2,3\ {\rm for\ charged\ leptons}\
e,\mu,\tau).
\eeq
The allowed values are shown in Fig.~\ref{Yi}. Similar to the situations for
$H^{++}$, $Y_+^1 \ \ll \ Y_+^2, Y_+^3$ in the NH
and $Y_+^1 \ > \ Y_+^2, Y_+^3$ in the IH.


\subsection{Rare Decays}

The charged Higgs bosons may mediate tree-level  lepton flavor violation processes,
leading to some stringent constraints on the model parameters,
see reference \cite{Gavela} for a recent comprehensive analysis.
In the model under consideration, the most important constraint comes from
 the process $\mu \to 3 e$ via the doubly charged Higgs.
The branching fraction is given by
\begin{equation}
\text{BR} \left( \mu  \to 3 e \right)\simeq \frac{\Gamma \left( \mu \to 3 e \right)}
{ \Gamma \left( \mu \to \ e \ \nu_\mu \ \bar{\nu}_e \right)}
= \frac{|\Gamma_{++}^{11} \ \Gamma_{++}^{12}|^2}{4 \ M_{\Delta}^4 \ G_F^2}.
\end{equation}
Using the experimental upper bound listed in~\cite{PDG}, BR$(\mu \to 3 e) < 10^{-12}$,
one finds
\beq
|\Gamma_{++}^{11} \ \Gamma_{++}^{12}| \ < \ 2.4 \times 10^{-5} \times \left( M_{\Delta}\over 1~\text{TeV} \right)^2.
\eeq
This in turn, combining with the relation between the Yukawa couplings
and the neutrino mass matrix, gives a lower bound on the vev for
a given value of the triplet mass
\begin{equation}
v_{\Delta}^2 \ > \ 0.2 \times 10^5 \ |M_\nu^{11} M_\nu^{12}| \times \left(\frac{1~\text{TeV} }{ M_\Delta}\right)^2.
\end{equation}
Even in the conservative case, the IH scenario where
$\sqrt{M_\nu^{11} M_\nu^{12}}$ is as large as 0.02 eV,
and for $M_\Delta \sim 1~\text{TeV}$, one obtains $v_\Delta \gtrsim 2~\text{eV}$,
which is not very relevant for our interest.

\subsection{Other Constraints}

There are two dimensionful free parameters $M_\Delta$ and $v_\Delta$
in this theory for neutrino masses. The current constraint
on $M_\Delta$ comes from the direct search for $H^{\pm\pm}$
at the Tevatron  \cite{Tevatron}
\beq
 M_{\Delta} \gsim 110\ {\rm GeV}.
\eeq
The vev of the triplet,
\beq
1 \ \text{eV} \lsim \ v_{\Delta} \ \lsim 1 \ \text{GeV},
\eeq
where the lower bound is based on the naturalness consideration
from neutrino masses,  and upper bound is from the constraint of the
electroweak $\rho$-parameter \cite{GunionDawson}.

\section{GENERAL FEATURES OF HIGGS DECAYS}

In this section we study all decays of the physical Higgs bosons
in the theory neglecting the leptonic mixings. In this theory
one has seven physical Higgs bosons, the CP-even neutral scalars
$H_1$ (SM-like), and $H_2$ ($\Delta$-like), a CP-odd neutral scalar
$A$, two singly charged Higgs bosons $H^{\pm}$, and two doubly charged
Higgs bosons $H^{\pm \pm}$.
Their decay partial widths are given in Appendix B.

\subsection{Doubly Charged Higgs Boson Decays}

\begin{figure}[tb]
\includegraphics[scale=1,width=8cm]{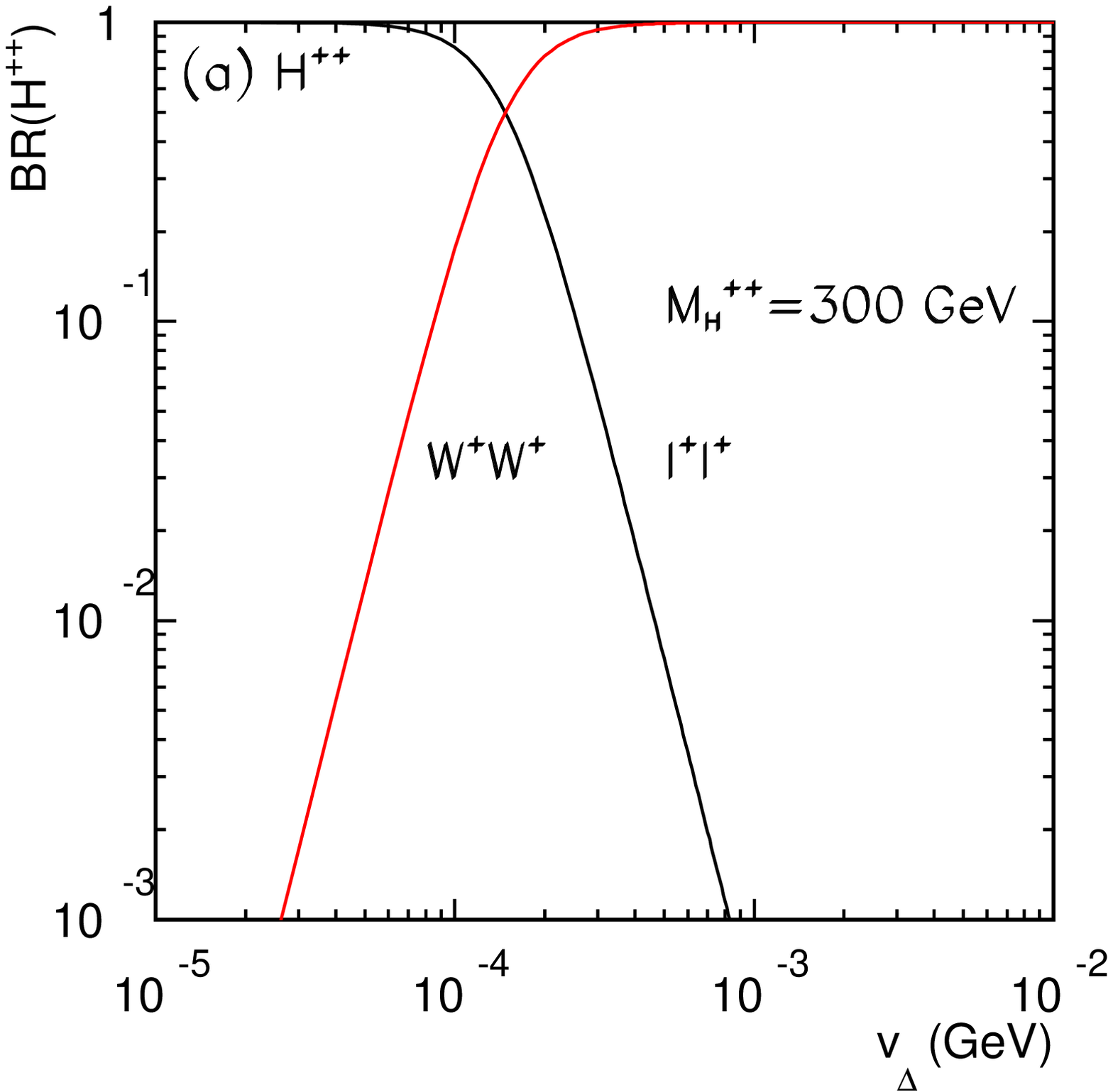}
\includegraphics[scale=1,width=8cm]{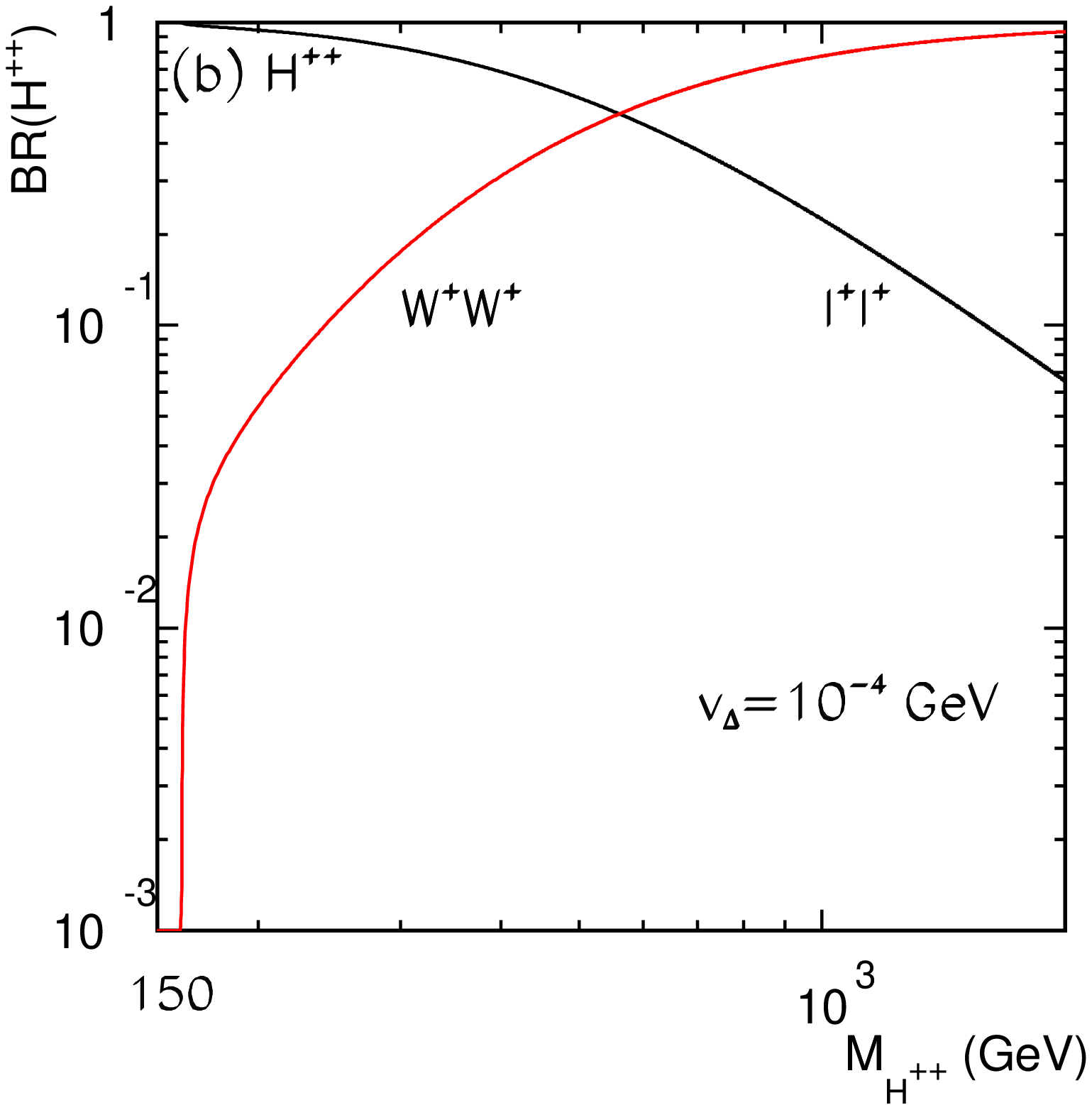}
\caption{Branching fractions of the doubly charged Higgs boson decay versus
(a) $v_{\Delta}$ for $M_{H^{++}}=300$ GeV, and  (b)
 $M_{H^{++}}$ for $v_{\Delta}=10^{-4}$ GeV.}
\label{hpp1}
\end{figure}

The possible decays of the doubly charged Higgs bosons,
$H^{\pm \pm}$, are the lepton number violating decays
$H^{++} \to e_i^+ e^+_j$, where $e_i=e,\mu,\tau$,
and the decays into two $W$'s. The decay rates for the
lepton number violating decays are:
\begin{equation}
\Gamma \left( H^{++} \to e_i^+ e_j^+ \right)=
\frac{|M_\nu^{ij}|^2}{8 \pi ( 1 \ + \ \delta_{ij}) v_{\Delta}^2} M_{H^{++}}
\end{equation}
where $M_\nu^{ij} $ is the neutrino mass matrix and $\delta_{ij}$ is the
Kronecher's delta. In the case of the decays into $W$'s the decay rates are
given by
\begin{equation}
\Gamma \left( H^{++} \to W^+_T W^+_T \right) = \frac{2 M_W^4 v_{\Delta}^2}{\pi v_0^4 M_{H^{++}}}
\left( 1 - \frac{4 M_W^2}{M_{H^{++}}^2}\right)^{1/2}
\end{equation}
and
\begin{equation}
\Gamma \left( H^{++} \to W^+_L W^+_L \right) = \frac{v_{\Delta}^2 M_{H^{++}}^3}{4 \pi v_0^4}
\left( 1 - \frac{4 M_W^2}{M_{H^{++}}^2}\right)^{1/2} \left( 1 - \frac{2 M_W^2}{M_{H^{++}}^2}\right)^{2}
\end{equation}
where $W_L$ and $W_T$ stand for the longitudinal and transverse polarizations of the $W$
gauge boson, respectively. The decays into leptons are proportional
to the Yukawa coupling for neutrinos while the decays into two $W$'s are
proportional to the vev. The relative decay branchings can be estimated by
\begin{eqnarray}
{\Gamma(H^{++}\to e^+_i e^+_j) \over \Gamma(H^{++}\to W^+ W^+) }
\approx { | \Gamma_{++} |^2 M_{H^{++}} \over M_{H^{++}}^3  v_\Delta^2 / v_0^4 }
\approx \left( { m_\nu \over M_{H^{++}}  } \right)^2
 \left( { v_0 \over v_\Delta^{}  } \right)^4 .
 \label{ratio1}
\end{eqnarray}
Taking  $ m_\nu / M_{H^{++}}\sim$1 eV$/$1 TeV,  one finds that these two decay
modes are comparable when $v_\Delta^{} \approx 10^{-4}\ {\rm GeV}$.
The branching fractions for the decays of the doubly charged Higgs,
BR$(H^{++})$, are shown in Fig.~\ref{hpp1},
assuming that the Yukawa matrix $Y_\nu$ (or $\Gamma_{++}$)
is diagonal, for simple illustration.
In Fig.~\ref{hpp1}(a) we plot the branching fractions versus
$v_{\Delta}$ for $M_{H^{++}}=300$ GeV;
while in Fig.~\ref{hpp1}(b) we show BR$(H^{++})$ versus
the doubly charged Higgs mass for $v_{\Delta}=10^{-4}$ GeV.
As seen from Eq.~(\ref{ratio1}) and the figures,  an  important feature is that when
$v_{\Delta}  <  10^{-4}$ GeV the most important
decays are those with a pair of like-sign charged leptons, while for $v_{\Delta} > 10^{-4}$ GeV the
most relevant decays are into two $W$'s.

\subsection{Singly Charged Higgs Boson Decays}

\begin{figure}[tb]
\includegraphics[scale=1,width=8cm]{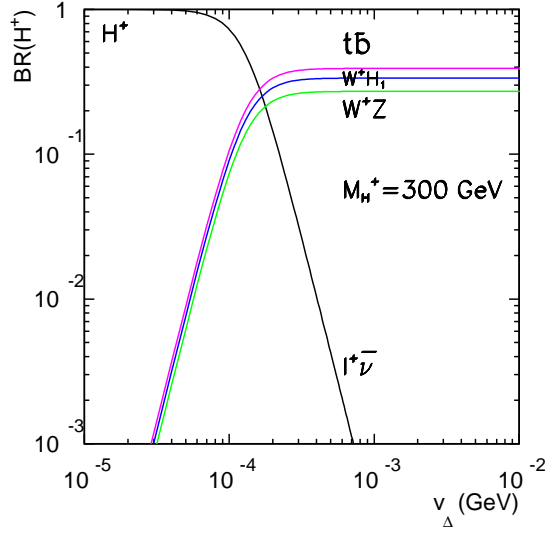}
\caption{Branching fractions of the singly charged Higgs boson decay versus
$v_{\Delta} $ for $M_{H^+}=300$ GeV (In our study we use $M_{H_1}=120$ GeV.).}
\label{hp2}
\end{figure}

In the case of the singly charged Higgs boson,  one has the decays
$H^+ \to e^+_i \bar{\nu}$ proportional to the Yukawa coupling
of neutrinos, $H^+ \to W^+ H_1, \ W^+ Z$, and $H^+ \to t \bar{b}$
proportional to the $v_{\Delta}$. As in the case of the
doubly charged Higgs all decays are connected by the
relation $M_\nu = \sqrt{2} \ Y_\nu \ v_{\Delta}$.
In Fig.~\ref{hp2} one can see the relevant decay channels for $M_{H^{+}}=300$
GeV versus $v_{\Delta}$. The most important channels
for large values of vev are $H^+ \to t \bar{b}$, $H^+ \to W^+ H_1$
and $H^+ \to W^+ Z$, while $H^+ \to e^+_i \bar{\nu}$ is the dominant
channel for small $v_{\Delta}$ when the Higgs mass is below TeV.
Here and henceforth,  we take $M_{H_1}=120$ GeV.
Furthermore,
\beq
{\Gamma(H^{+}\to t \bar b) \over \Gamma(H^{+}\to W^+Z) }
\approx {3 (\vd m_t/v_0^2)^2 M_{\Delta} \over M_\Delta^3   v_\Delta^2 /2 v_0^4 } =
6 \left( { m_t \over M_\Delta  } \right)^2 .
 \nonumber
\eeq
Thus the decays $H^+ \to W^+ Z,\ W^+ H_1$ dominate
over $t\bar b$ for $M_{\Delta} > 400$ GeV.

\begin{figure}[tb]
\includegraphics[scale=1,width=8cm]{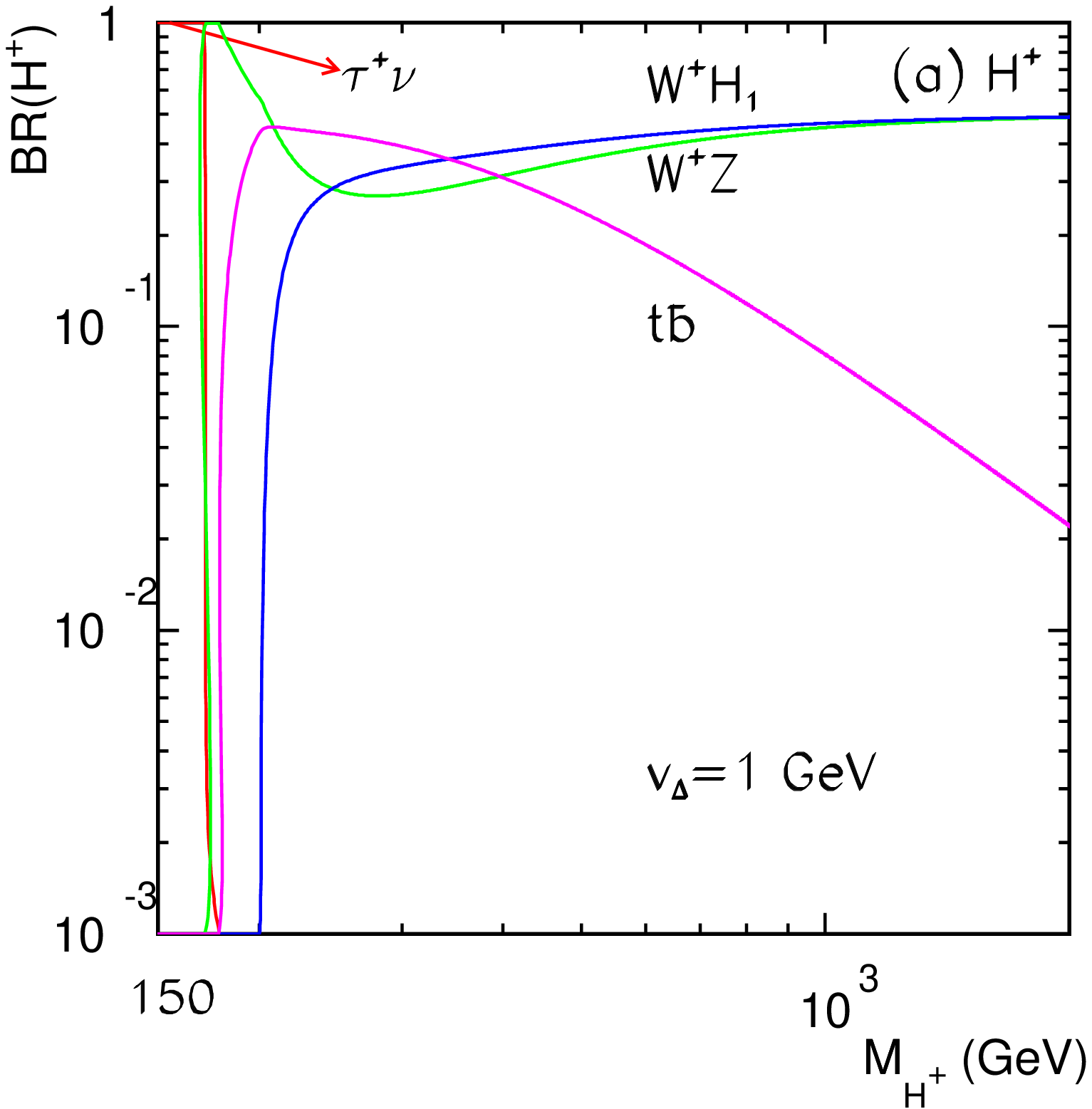}
\includegraphics[scale=1,width=8cm]{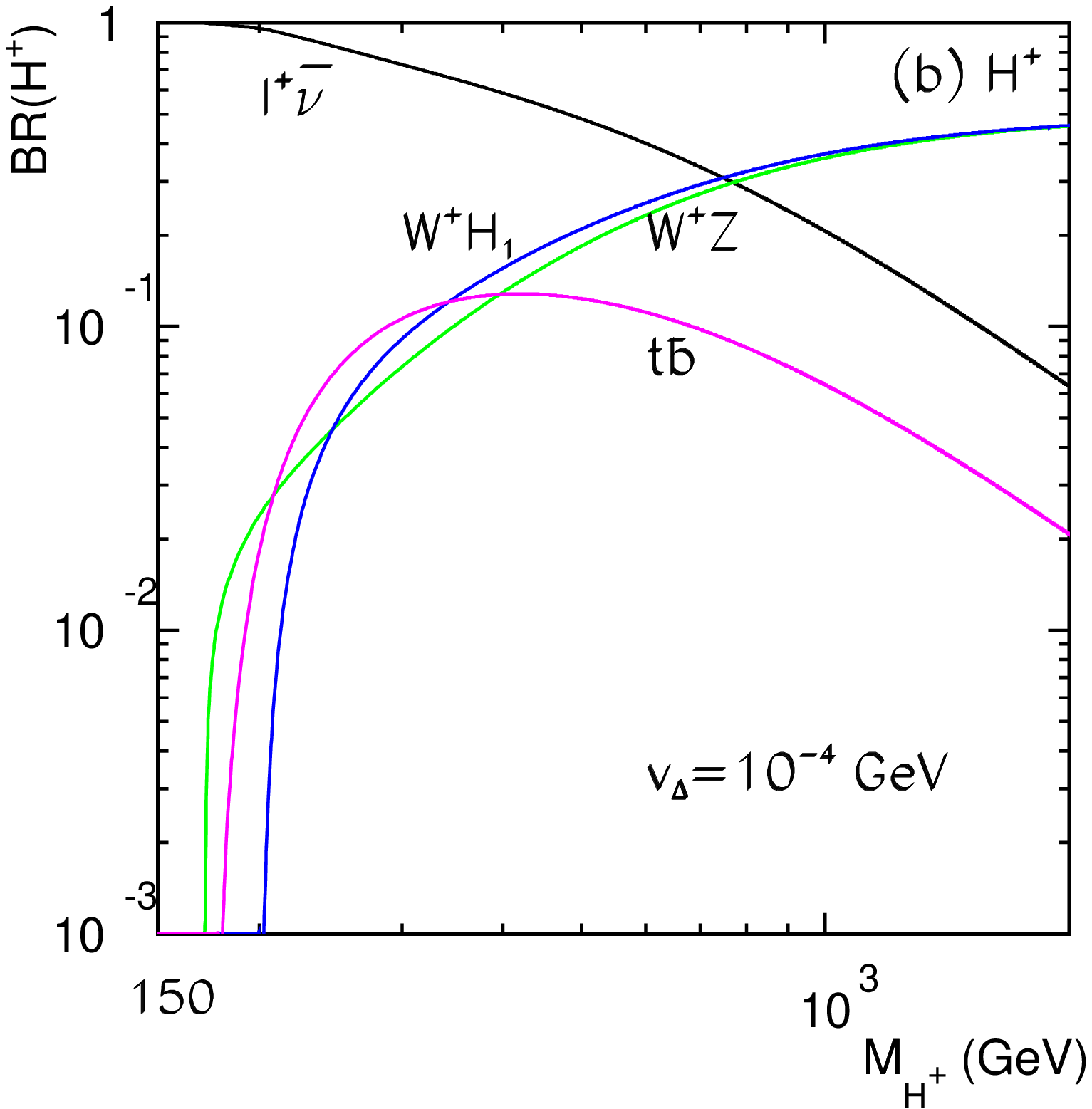}
\caption{Branching fractions of the singly charged Higgs boson decay versus its mass
for (a) $v_{\Delta}= 1$ GeV and (b) $v_{\Delta}= 10^{-4}$ GeV, respectively.}
\label{hp1}
\end{figure}

In Fig.~\ref{hp1}(a) and  Fig.~\ref{hp1}(b) we plot the branching
fractions of the singly charged Higgs boson versus its mass
for $v_{\Delta}= 1$ GeV and $v_{\Delta}= 10^{-4}$ GeV, respectively.
In Fig.~\ref{hp1}(a), below the $WZ$ threshold, it is irrelevant to
our collider search so we neglect the offshell $W^*$/$Z^*$ decay
channels then $H^+\to\tau^+ \nu$ is dominant.

\subsection{CP-Even Heavy Higgs Boson Decays}

\begin{figure}[tb]
\includegraphics[scale=1,width=8cm]{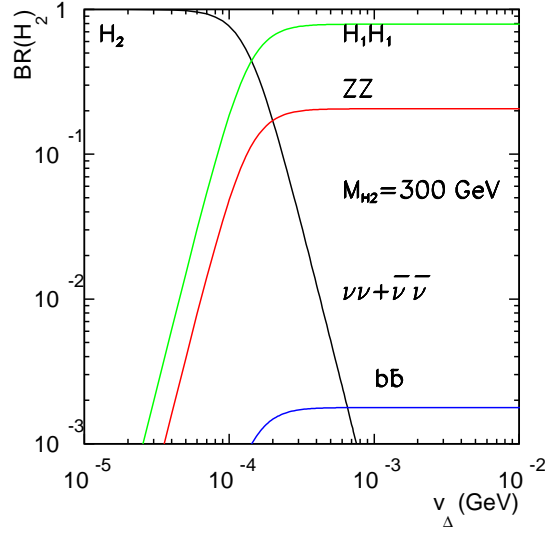}
\caption{Branching fractions of the heavy CP-even  Higgs boson decay
versus $v_{\Delta} $ for $M_{H_2}=300$ GeV.}
\label{h21}
\end{figure}

\begin{figure}[tb]
\includegraphics[scale=1,width=8cm]{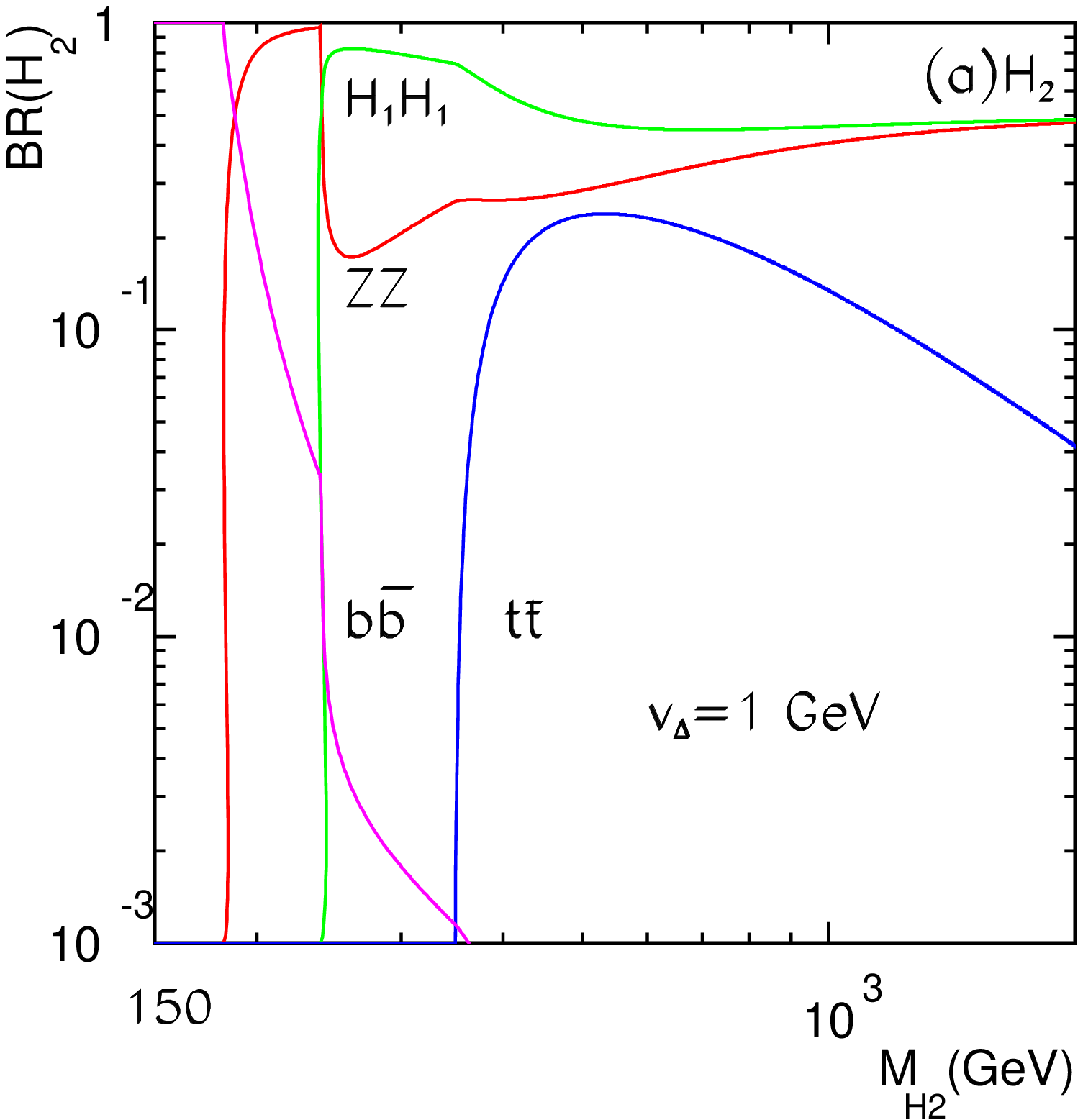}
\includegraphics[scale=1,width=8cm]{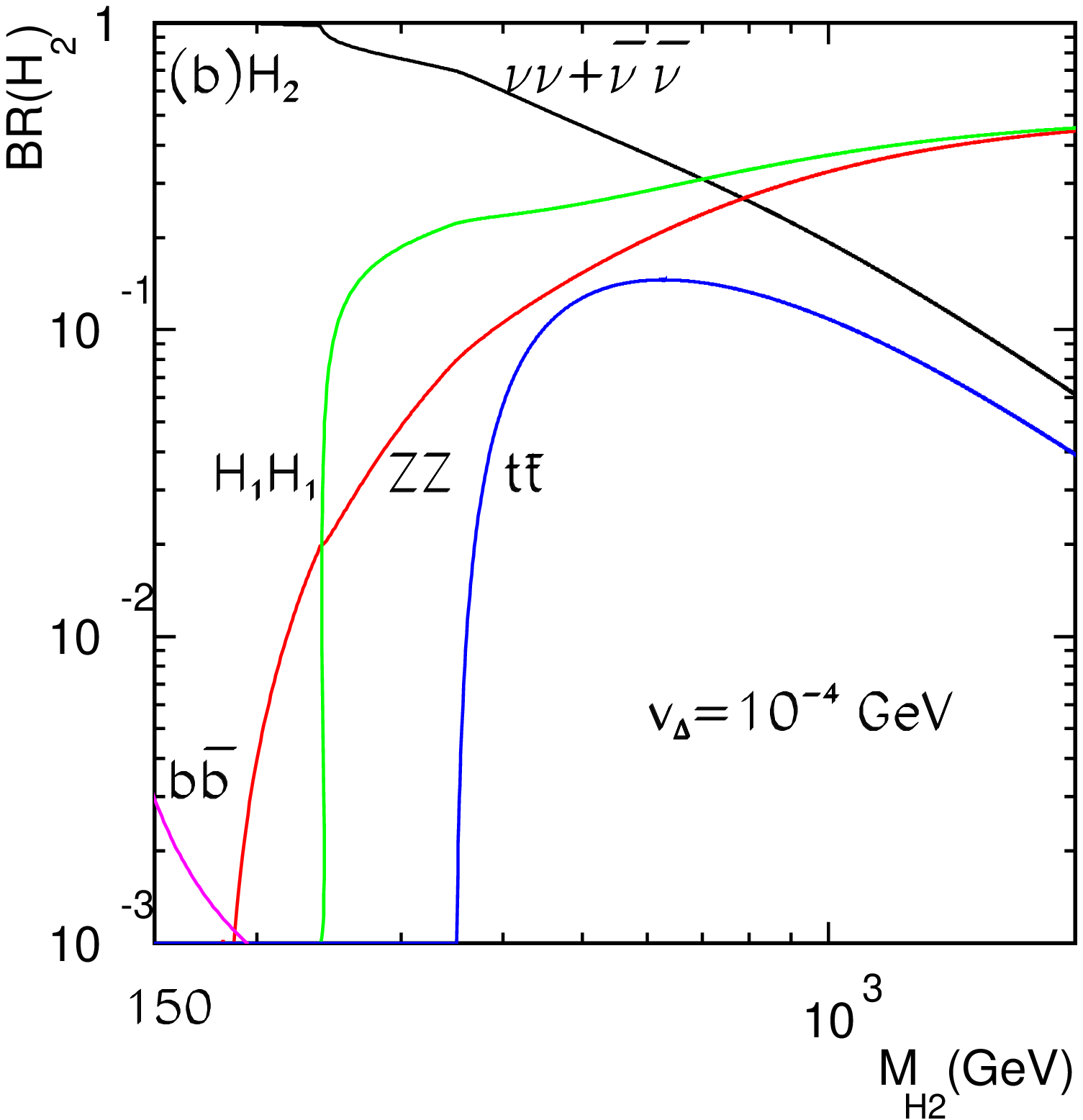}
\caption{Branching fractions of the heavy CP-even  Higgs boson decay  versus its mass
for (a) $v_{\Delta}= 1$ GeV and (b) $v_{\Delta}= 10^{-4}$ GeV, respectively.}
\label{h22}
\end{figure}

The decays of the heavy neutral CP-even neutral scalar
$H_2$ ($\Delta$-like) are shown in Figs.~\ref{h21}
and~\ref{h22}. The most relevant decays  are
$H_2 \to  H_1 H_1, \ Z Z, \ b\bar{b},\ t \bar{t}$  proportional
to $v_{\Delta}$, and the decays into a pair of neutrinos proportional to
the Yukawa couplings. As for all physical Higgs bosons in the theory
all decays are connected by the neutrino mass relation in
Eq.~(\ref{type2}). As we can appreciate from the Figs.~\ref{h21}
and~\ref{h22} when the $v_{\Delta}$ is large $H_2 \to H_1  H_1$
and $H_2 \to Z  Z$ are the most relevant channels. In this model
the channel $H_2 \to W^+ W^-$ is highly suppressed being zero at leading
order (see Appendix B for details). As one expects the decays into neutrinos and
antineutrinos become important below  $M_{\Delta}\sim $ TeV and for small $v_{\Delta}$.

\subsection{CP-Odd Heavy Higgs Boson Decays}

\begin{figure}[tb]
\includegraphics[scale=1,width=8cm]{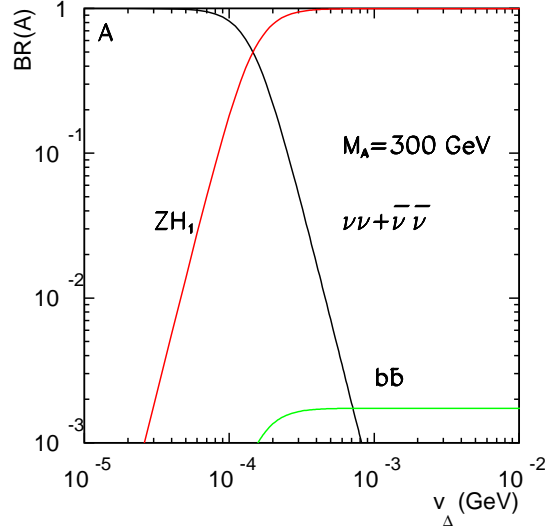}
\caption{Branching fractions of the heavy CP-odd  Higgs boson decay
 versus $v_{\Delta} $ for $M_{A}=300$ GeV.}
\label{hs1}
\end{figure}

\begin{figure}[tb]
\includegraphics[scale=1,width=8cm]{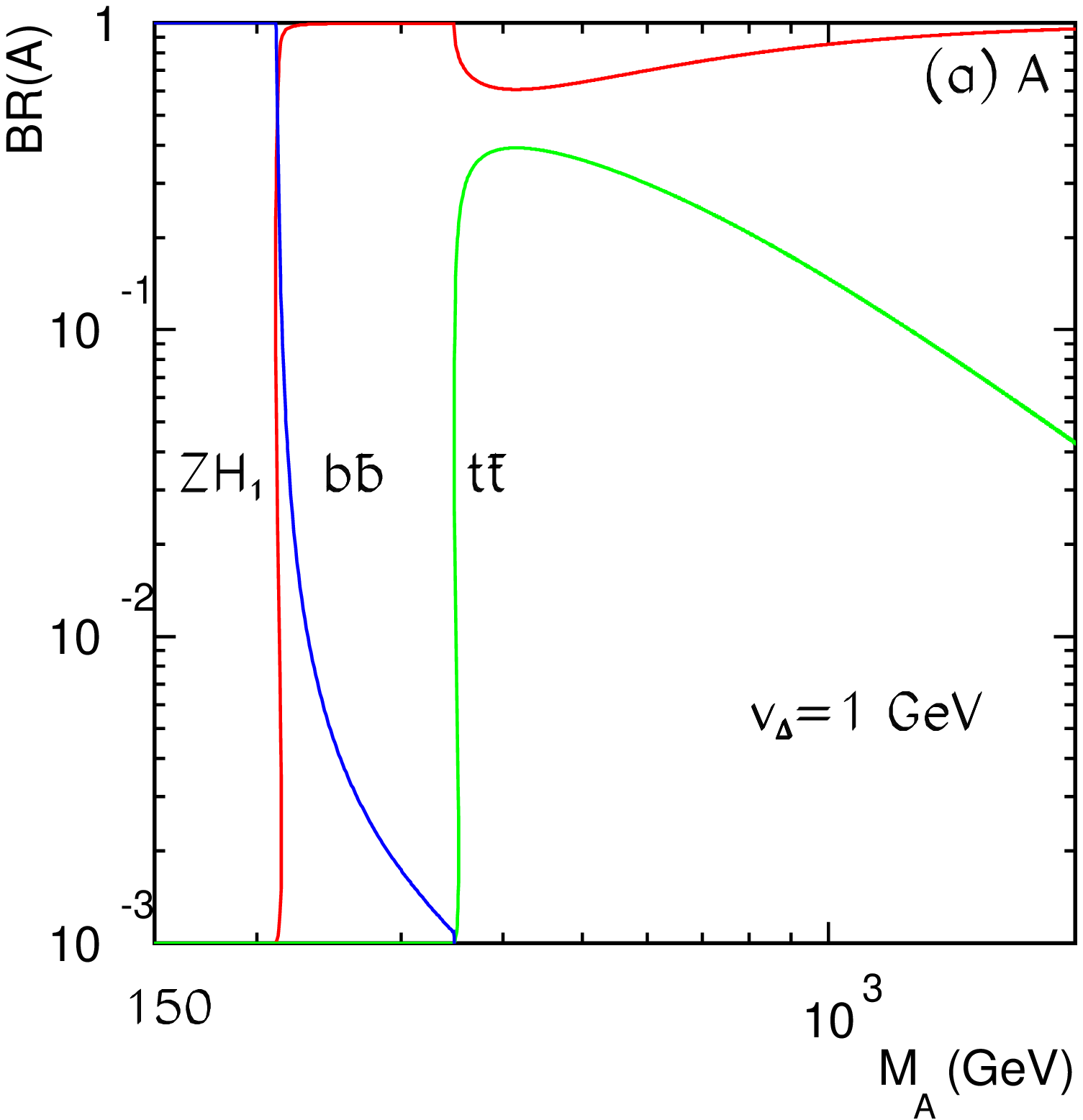}
\includegraphics[scale=1,width=8cm]{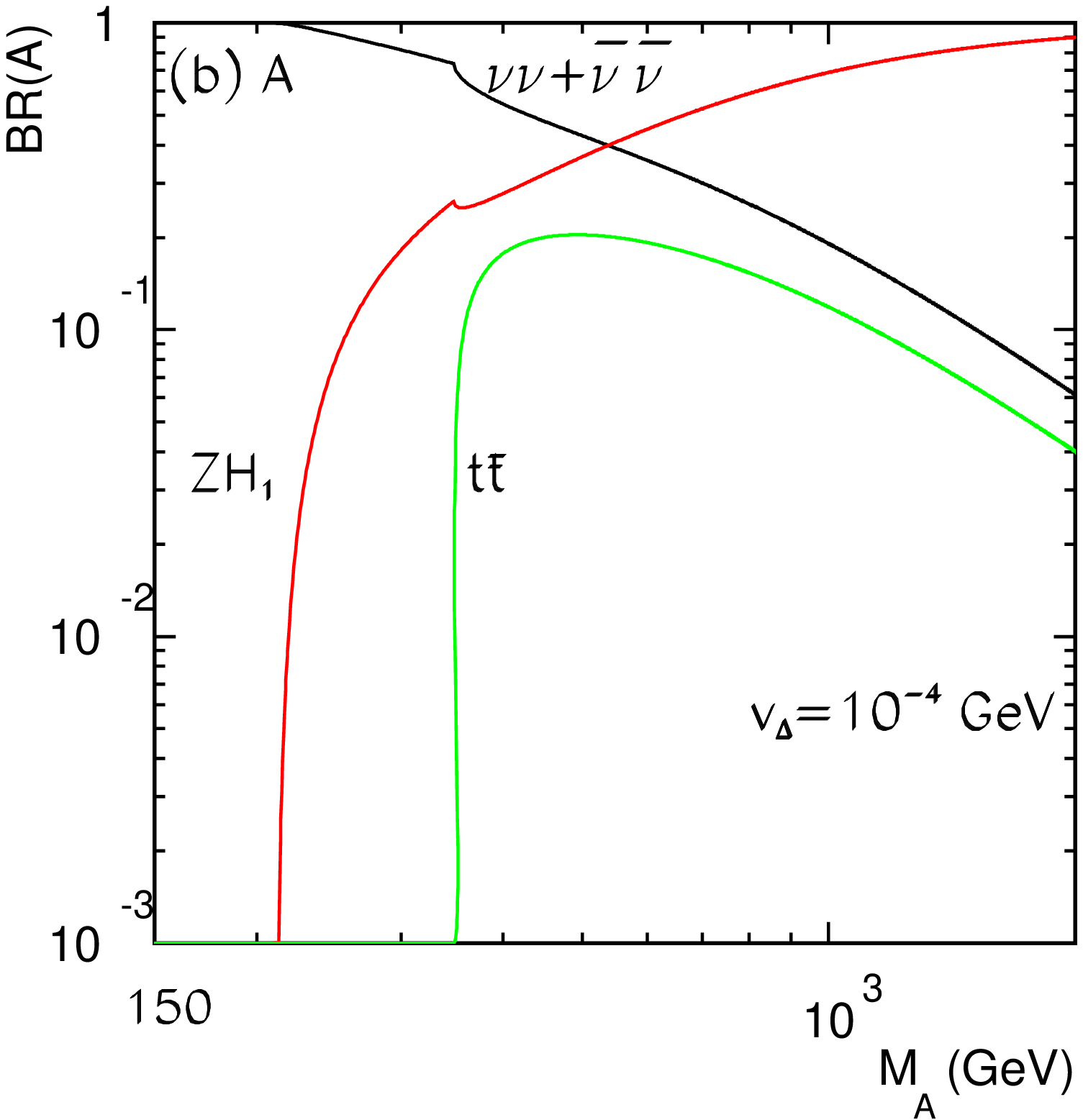}
\caption{Branching fractions of the heavy CP-odd Higgs boson decay
 versus its mass
for (a) $v_{\Delta}= 1$ GeV and (b) $v_{\Delta}= 10^{-4}$ GeV, respectively.}
\label{hs2}
\end{figure}

The relevant decays of the CP-odd scalar field $A$
are $A \to t \bar{t}, H_1 Z$ and the decays into
neutrinos and antineutrinos. The branching fractions
of $A$ for $M_{A}=300$ GeV and different values of $v_{\Delta}$
are shown in Fig.~\ref{hs1}. In Fig.~\ref{hs2} we
plot the different decays of $A$ for $v_{\Delta}= 1$ GeV and
$v_{\Delta}= 10^{-4}$ GeV, respectively. As in the previous cases the
decays into neutrinos and antineutrinos are the most relevant
for large Yukawa couplings and in the low mass region.
Notice that the decay $A \to Z H_1$ is the dominant
one for large values of $v_{\Delta}$.
From this discussion one can conclude that the lepton number violating
decays of the different physical Higgs bosons, $H_2, A, H^{\pm}$, and $H^{\pm \pm}$,
in the theory dominate for small values of the triplet vacuum expectation value.
\subsection{Mass Splitting And Heavy-to-Heavy Transition via Gauge Interactions}
In our discussions thus far, we have assumed the mass degeneracy for the triplet-like Higgs
bosons. According to Eq.~(\ref{Potential}), a tree level mass splitting can be generated and
the squared mass difference of the doubly and singly charged Higgs bosons
is given by $\lambda_4 v_0^2/4$. Even if there is no tree-level mass difference under
our assumption $\lambda_i=0$, the SM gauge bosons generate the
splitting of the masses via radiative corrections at one-loop~\cite{strumia},
leading to $\Delta M\equiv M_{H^{++}} - M_{H^{+}} \approx 540$ MeV.

\begin{figure}[tb]
\includegraphics[scale=1,width=8cm]{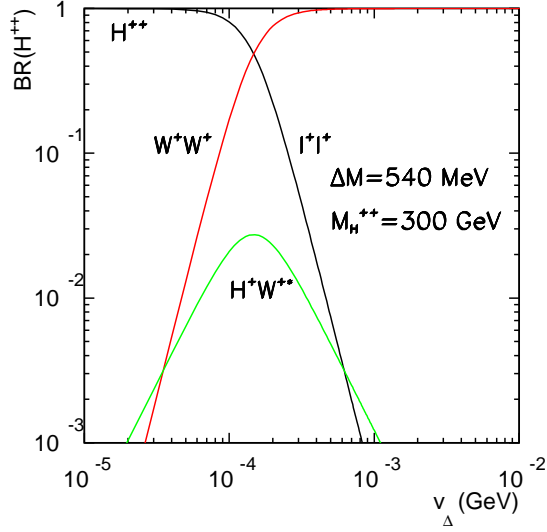}
\caption{Branching fractions of the doubly charged Higgs boson decay versus $v_{\Delta}$
for $\Delta M=540$ MeV.}
\label{hppsplit1}
\end{figure}

A small mass difference will make no appreciable effects for the Higgs production.
However, the transitions between two heavy triplet Higgs bosons via the SM gauge
interactions, such as
\be
H^{++} \to H^+ W^{+*},\quad H^{+} \to H^0 W^{+*}
\label{gaugemode}
\ee
may be sizable if kinematically accessible. Their partial decay widths are given
in Appendix B.
In Fig.~\ref{hppsplit2} we calculate the decay branching fractions
of the doubly charged Higgs versus the mass splitting for
$v_{\Delta}=10^{-4}$ GeV  and $v_{\Delta}=3 \times 10^{-4}$ GeV,
taking into account  $H^{++} \to H^+ M^+\ (M^+=\pi^+,K^+...), \
H^+ e_i^+ \nu\ (e_i=e, \mu, \tau)$ and $H^+ q\bar q'$.
We find that the heavy-to-heavy transition
can be dominant for $\Delta M > 1 $ GeV.
With our current assumption, $\Delta M=540$ MeV  \cite{strumia},
the decay branching fractions are shown in Fig.~\ref{hppsplit1} versus
 the triplet vev. We see that the decay mode $H^{++} \to H^+ (W^+)^*$
 is subleading and will be neglected in the rest of our discussions.

\begin{figure}[tb]
\includegraphics[scale=1,width=8cm]{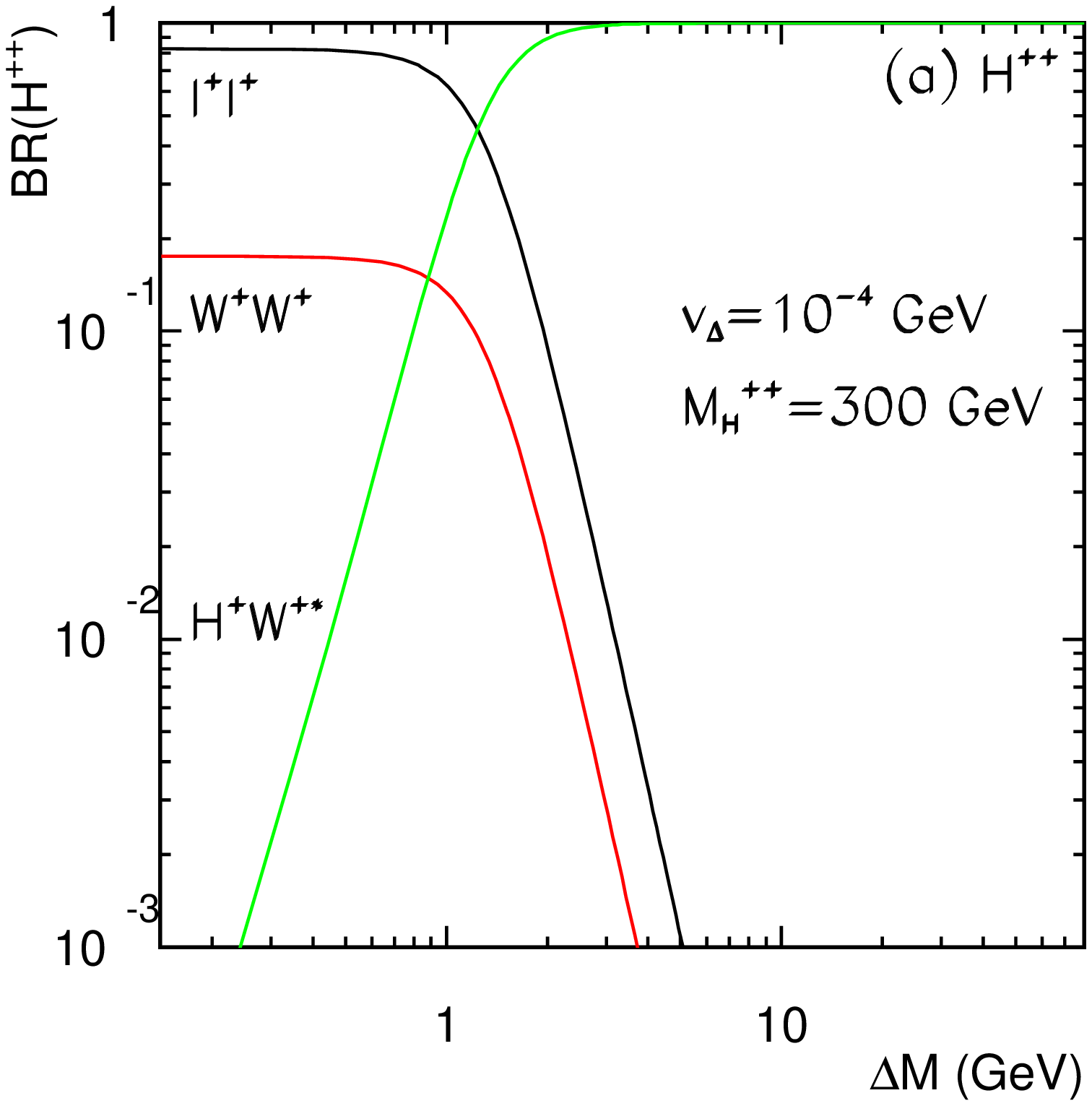}
\includegraphics[scale=1,width=8cm]{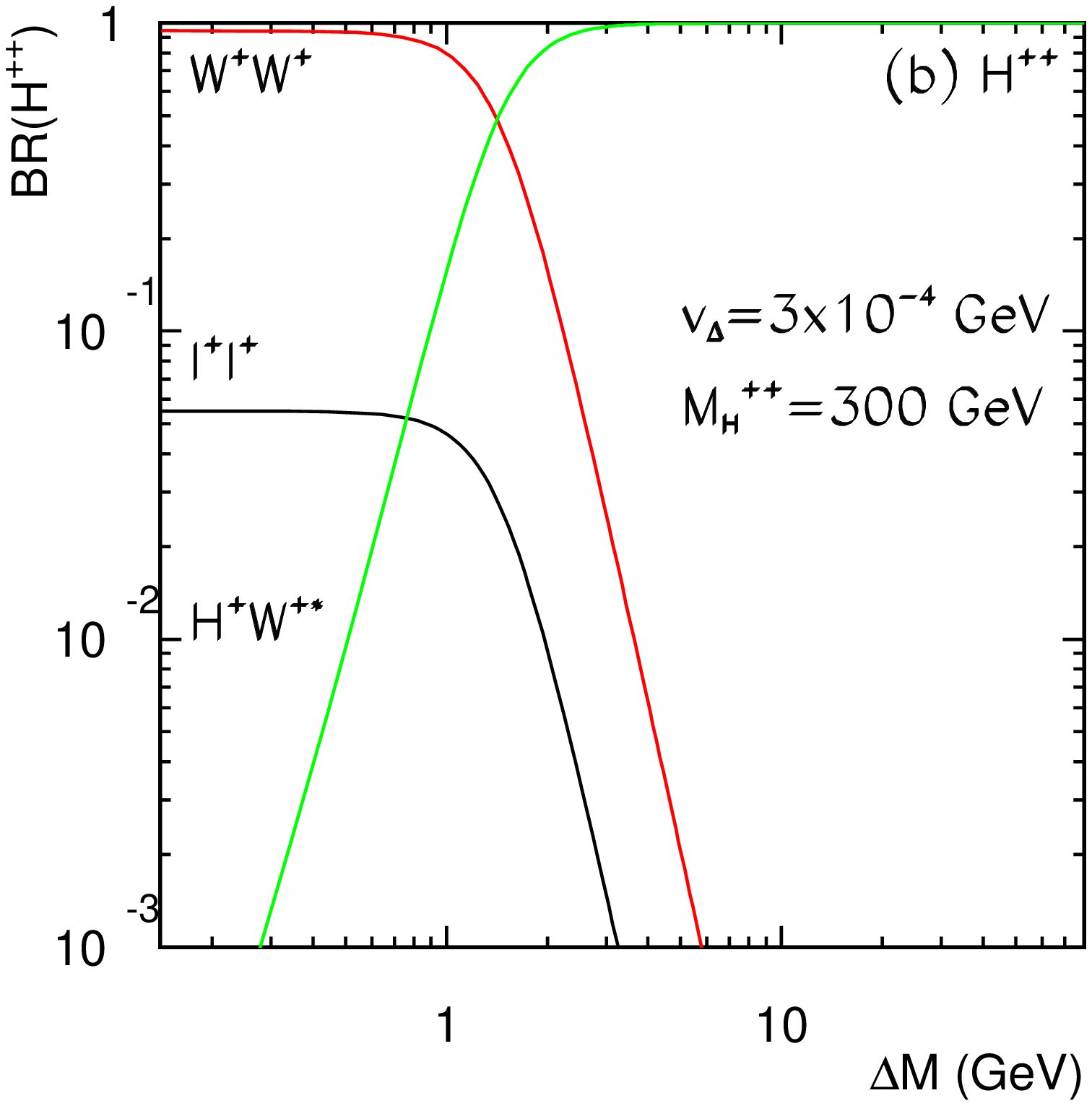}
\caption{Branching fractions of the doubly charged Higgs boson decay versus the mass splitting
$\Delta M\equiv M_{H^{++}} - M_{H^{+}}$
for (a) $v_{\Delta}=10^{-4}$ GeV and (b) $v_{\Delta}=3 \times 10^{-4}$ GeV,
respectively.}
\label{hppsplit2}
\end{figure}

\section{HIGGS Boson DECAYS IN CONNECTION TO  NEUTRINO PROPERTIES}

In this section we study the properties of the lepton number
violating Higgs decays taking into account the experimental
constraints on the neutrino masses and mixing.

\subsection{$H^{++} \ \to \ e_i^+ e_j^+$}

\begin{figure}[tb]
\includegraphics[scale=1,width=8.0cm]{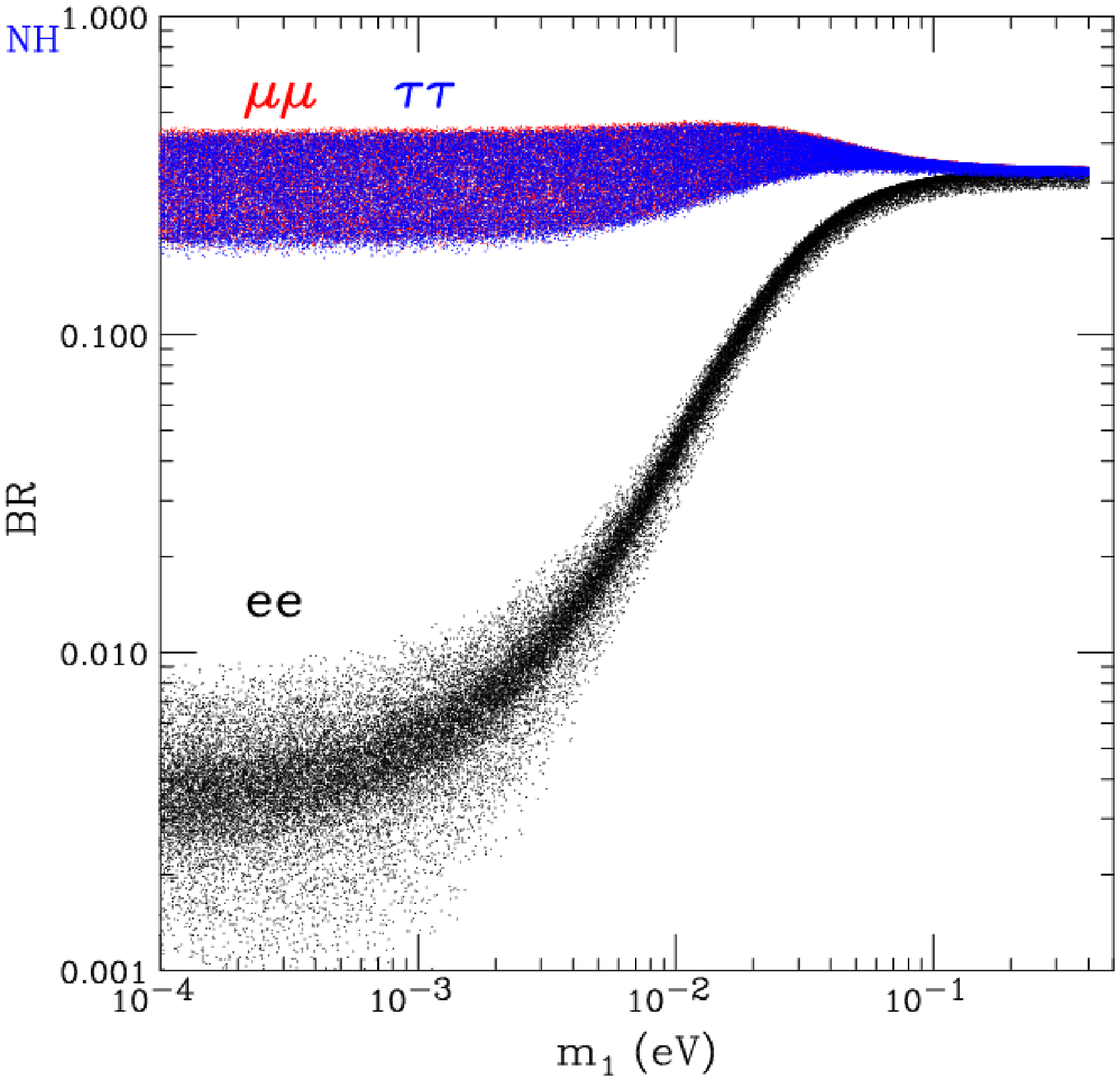}
\includegraphics[scale=1,width=8.0cm]{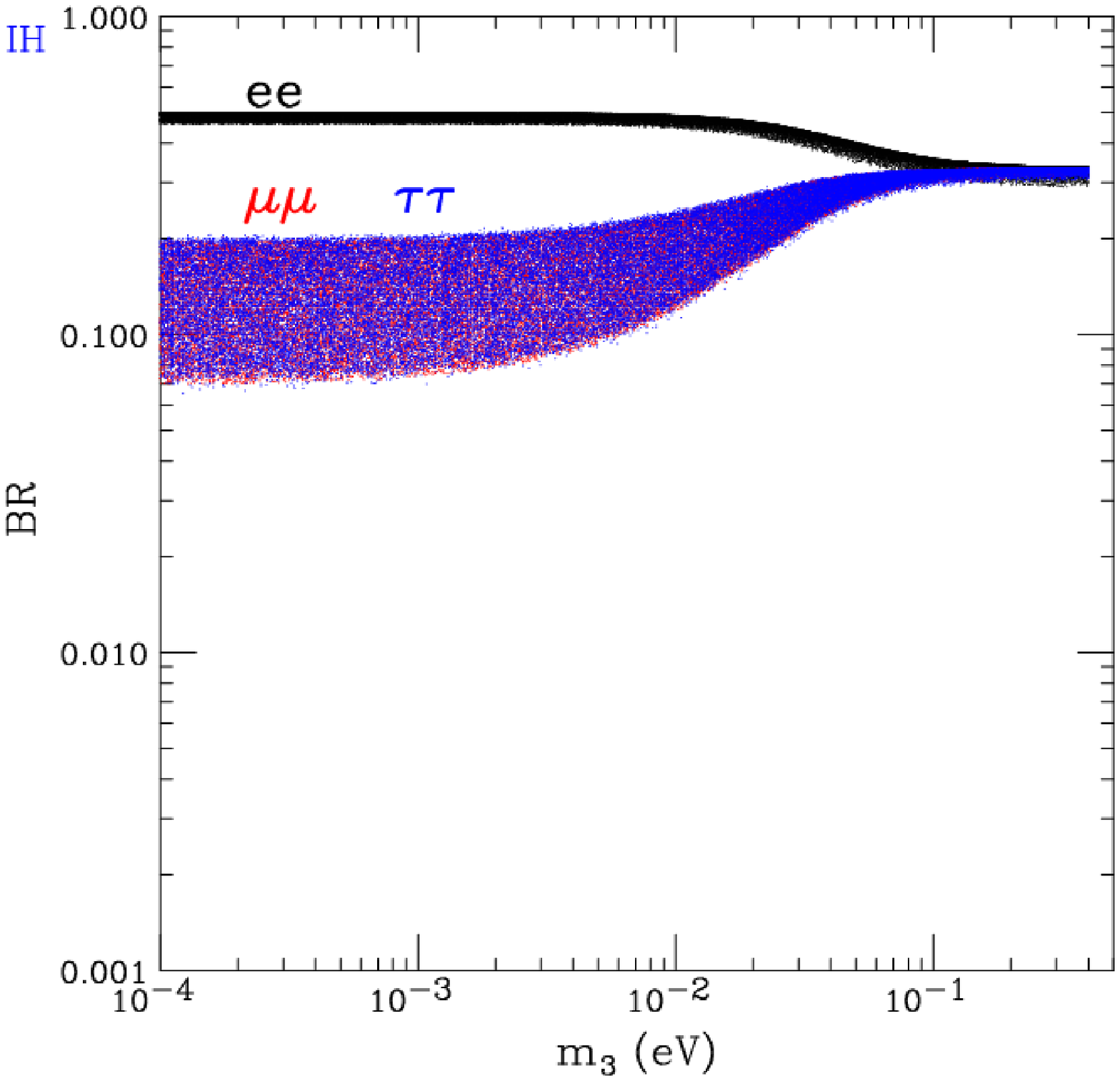}
\caption{Scatter plots for the $H^{++}$ decay branching fractions
 to the flavor-diagonal like-sign dileptons versus the lowest
neutrino mass for NH (left) and IH (right) with $\Phi_1 = \Phi_2 = 0$.}
\label{brii}
\end{figure}

\begin{figure}[tb]
\includegraphics[scale=1,width=8.0cm]{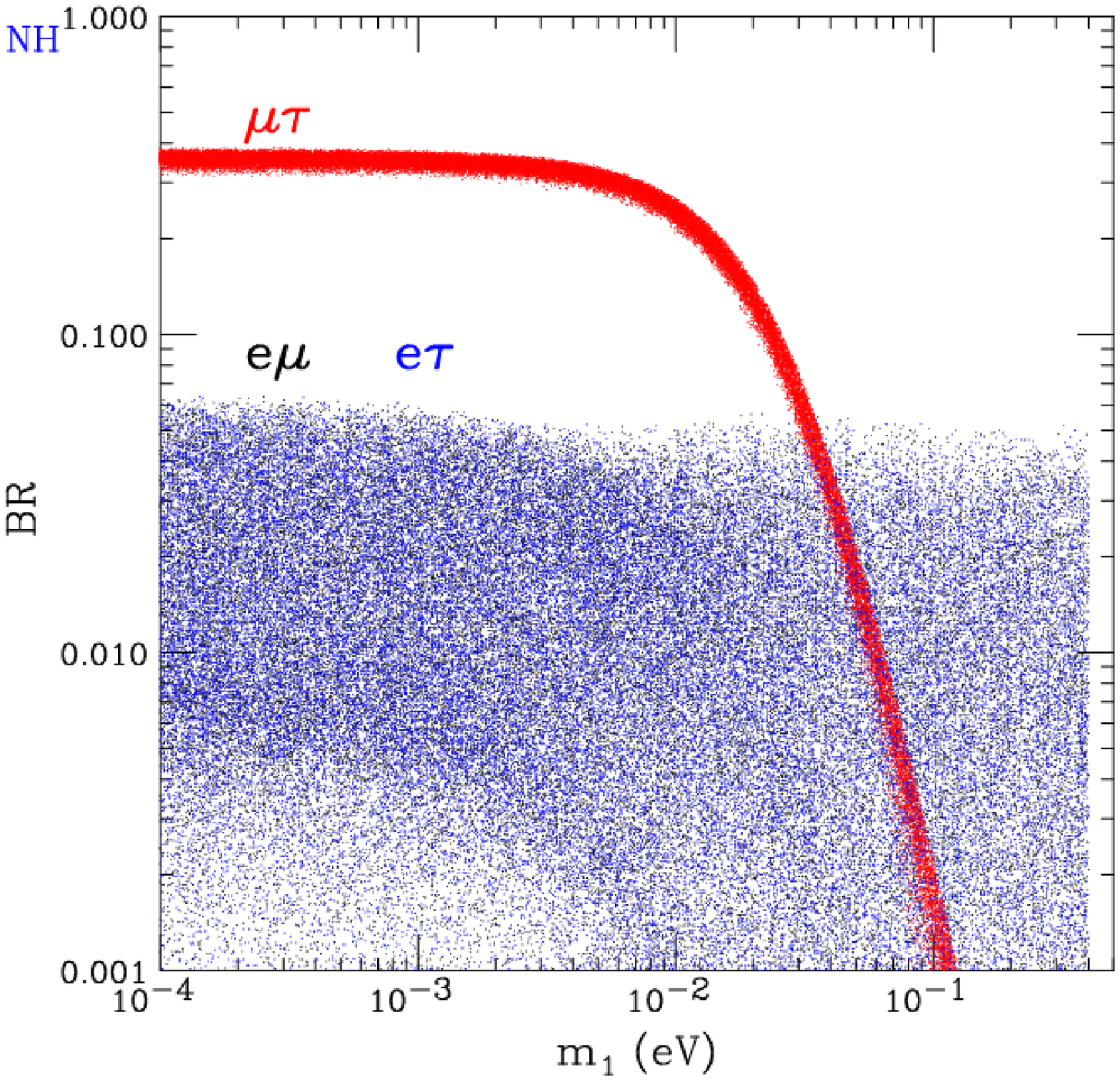}
\includegraphics[scale=1,width=8.0cm]{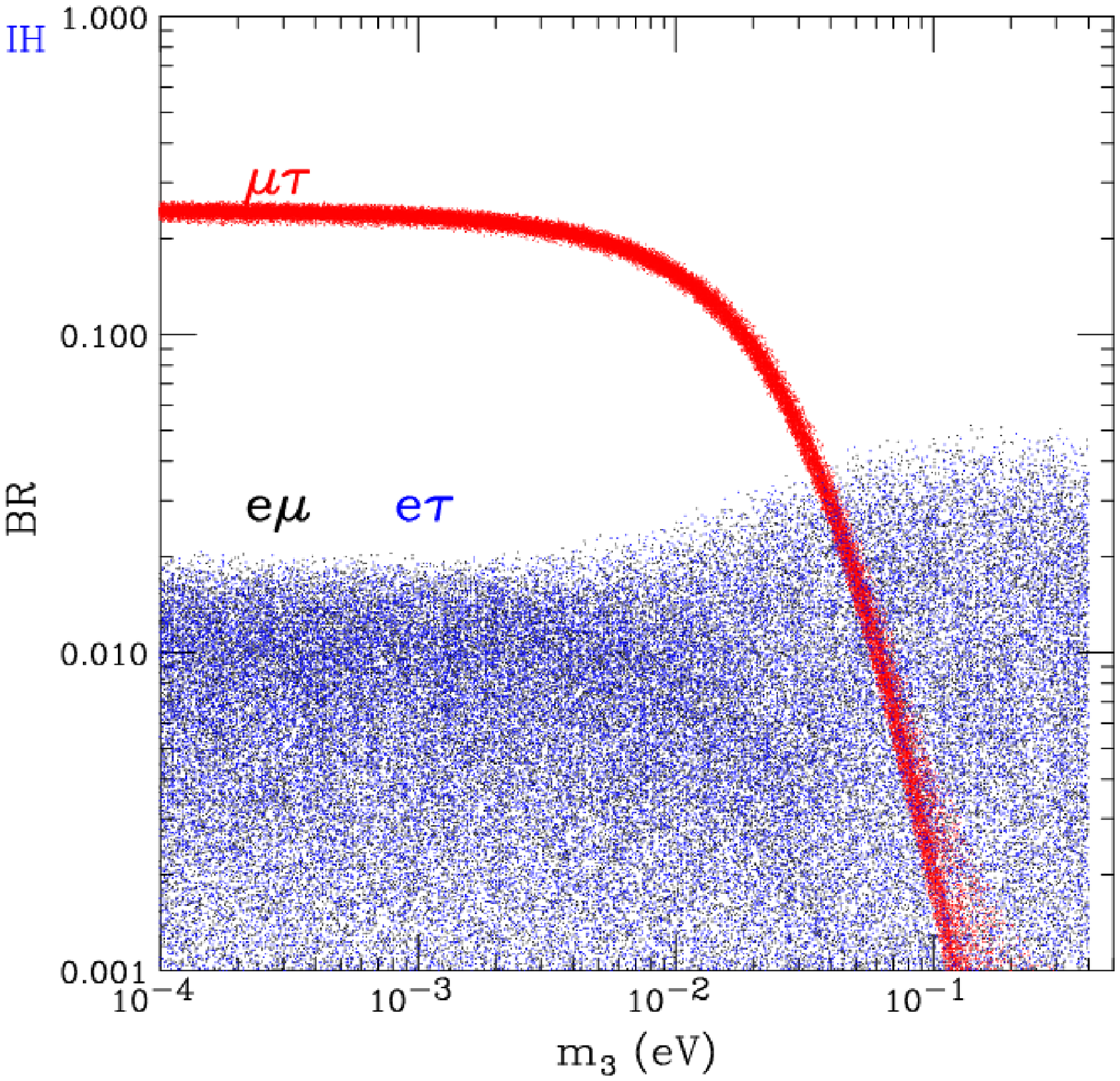}
\caption{Same as Fig.~\ref{brii}, but for  $H^{++}$ decay to
 the flavor-off-diagonal like-sign dileptons. }
\label{brij}
\end{figure}

In the previous section we have discussed the decays of the doubly charged
Higgs showing that below $v_{\Delta}\approx 10^{-4}$ GeV,
the decays of doubly charged Higgs $H^{++}$ are dominated by the leptonic channels.
For simplicity, we first ignore the effects of the Majorana phases $\Phi_1 = \Phi_2 = 0$.
In Figs.~\ref{brii} and~\ref{brij}, we show
the dramatic impact of the neutrino masses and mixing on the branching
ratios for the final states of the same and different flavors, respectively.
In the case of the decays with two identical (anti)leptons as in Fig.~\ref{brii},
the branching fraction can differ by two orders of magnitude in the
case of a normal hierarchy  with
BR$(H^{++} \to \tau^+ \tau^+)$, BR$(H^{++} \to \mu^+ \mu^+) \gg $ BR$(H^{++} \to e^+ e^+)$,
and about one order of magnitude in the inverted spectrum with
BR$(H^{++} \to e^+ e^+) >$BR$(H^{++} \to \mu^+ \mu^+)$, BR$(H^{++} \to \tau^+ \tau^+)$.
The impact is also dramatic for both spectra in the case of the decays with different
leptons in the final state with
BR$(H^{++} \to \mu^+ \tau^+)\gg$BR$(H^{++} \to e^+ \mu^+)$, BR$(H^{++} \to e^+ \tau^+)$,
as in Fig.~\ref{brij}.
These features directly reflect the neutrino mass and mixing patterns.
As one expects that all these channels are quite similar
when the neutrino spectrum is quasi-degenerate,
$m_1\approx m_2\approx m_3 \ge 0.1$ eV.
The rather large regions of the scatter plots reflect the imprecise
values for neutrino masses and leptonic mixings.
In the future~\cite{Experiments}, once those values will be known to a
better precision one can improve our
predictions for the lepton number violating Higgs decays.

\begin{figure}[tb]
\includegraphics[scale=1,width=8.0cm]{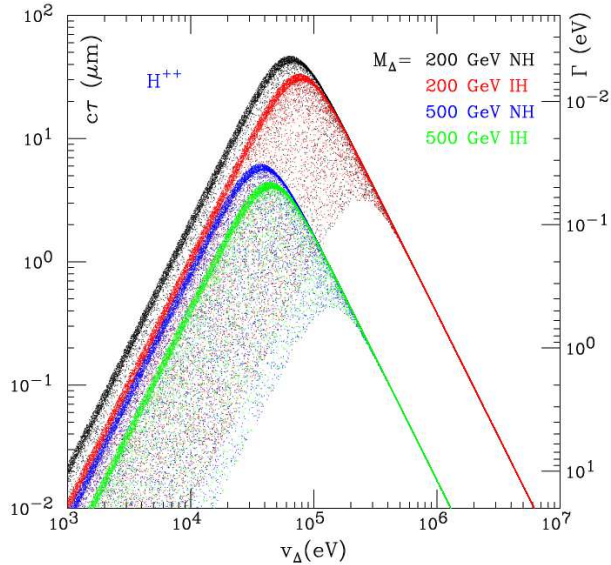}
\caption{Decay length and total width of the doubly charged Higgs boson
$H^{++}$ with $\Phi_1 = \Phi_2 = 0$.}
\label{decayl++}
\end{figure}

The total decay width of $H^{++}$ depends on the neutrino and Higgs
triplet parameters. In terms of $v_\Delta$, the minimal width
or the maximal decay length occur near the cross-over between
$WW$-dominant and $\ell\ell$-dominant regions  near $10^{-4}\ $ GeV.
As  seen in Fig.~\ref{decayl++}, the proper decay length can be as large
as $c\tau \gsim 10\ \mu m$. Although not considered as a long-lived charged
particle, the $H^{++}$ decay could lead to a visible displaced vertex in
the detector at the LHC.

\subsection{$H^{+} \ \to \ e_i^+ {\bar{\nu}}$}

\begin{figure}[tb]
\includegraphics[scale=1,width=8.0cm]{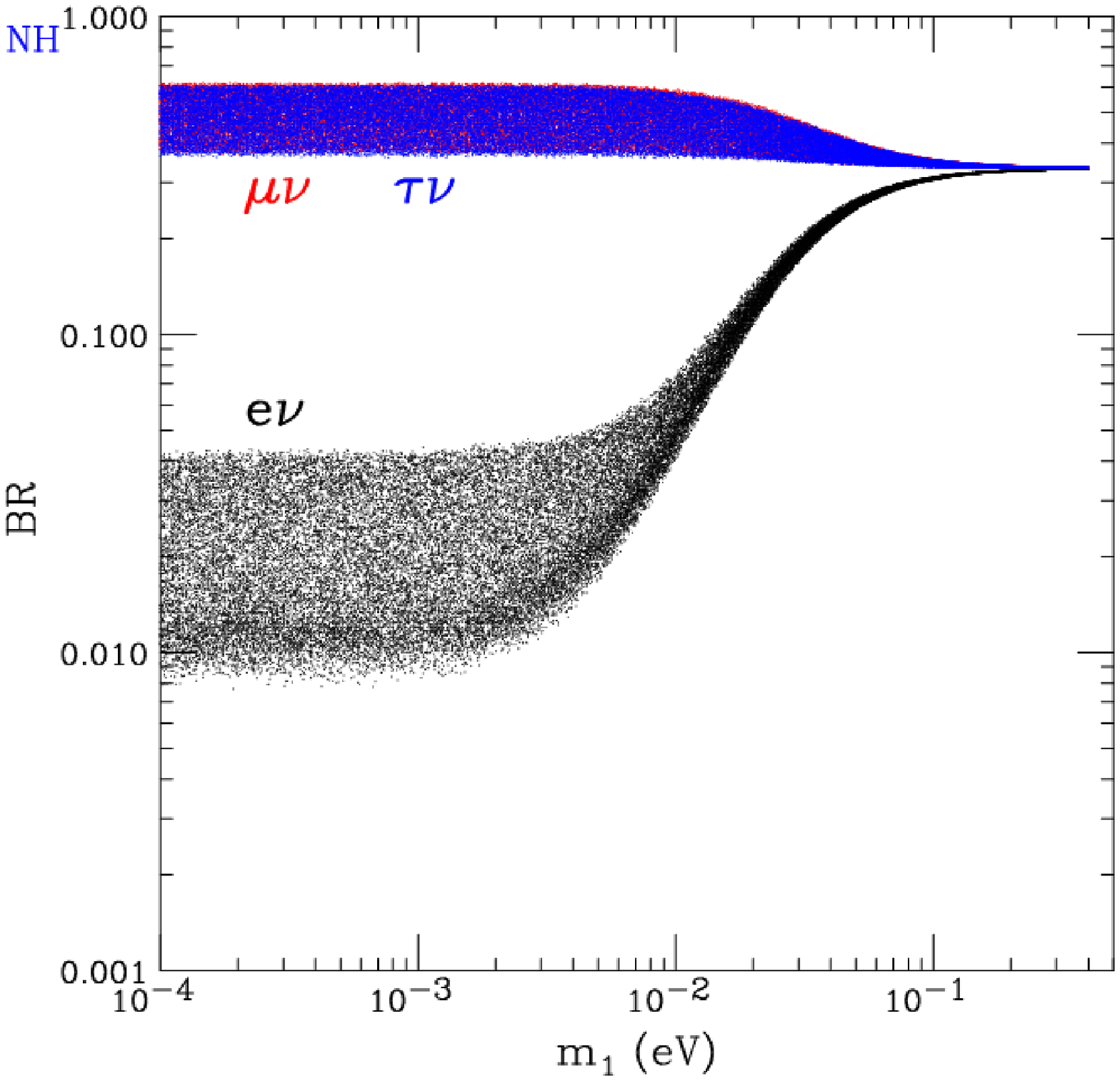}
\includegraphics[scale=1,width=8.0cm]{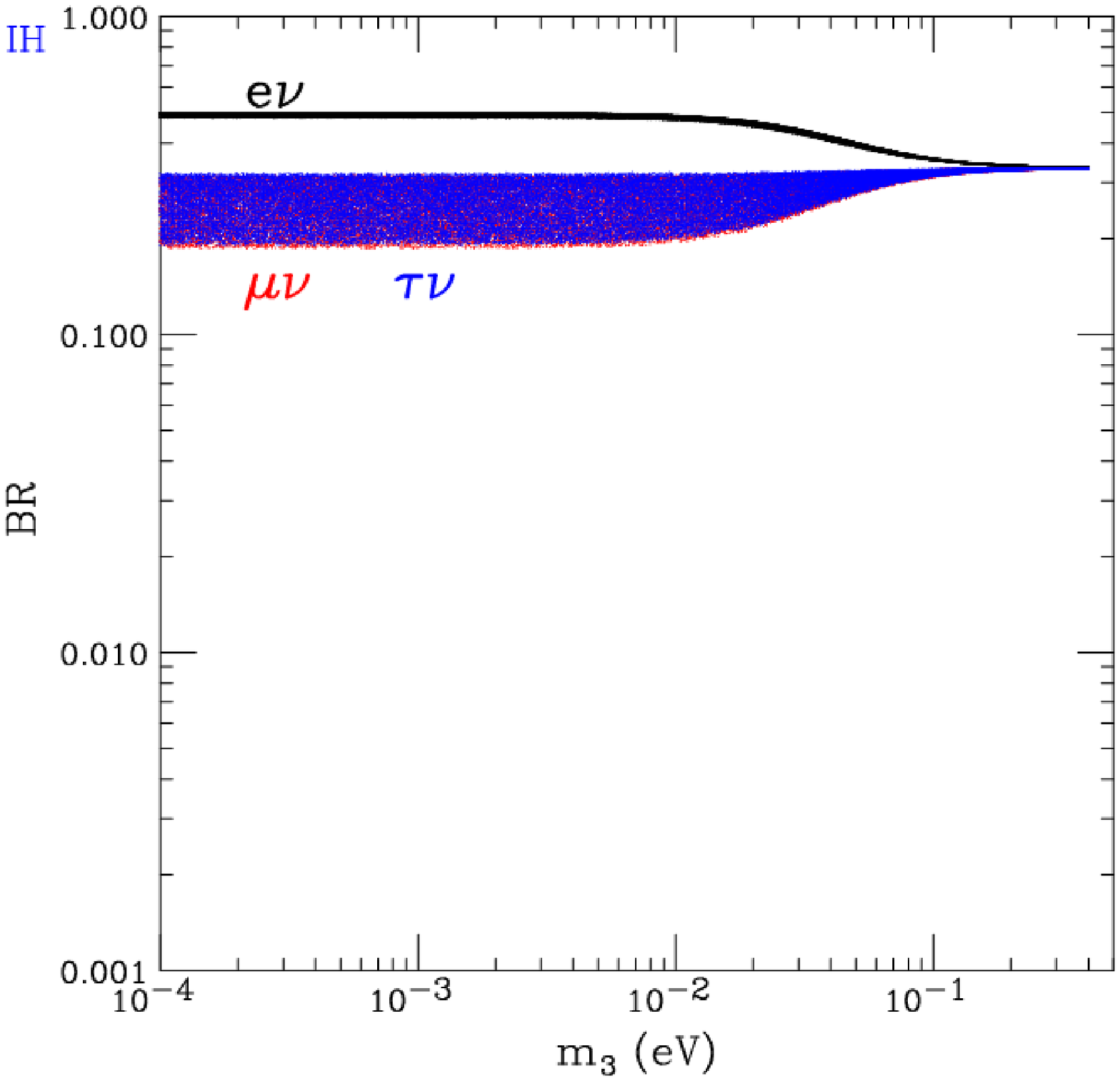}
\caption{Scatter plots for the $H^{+}$ decay branching fractions
 to leptons versus the lowest
neutrino mass for NH (left) and IH (right) with $\Phi_1 = \Phi_2 = 0$.}
\label{1vBi}
\end{figure}

\begin{figure}[tb]
\includegraphics[scale=1,width=8.0cm]{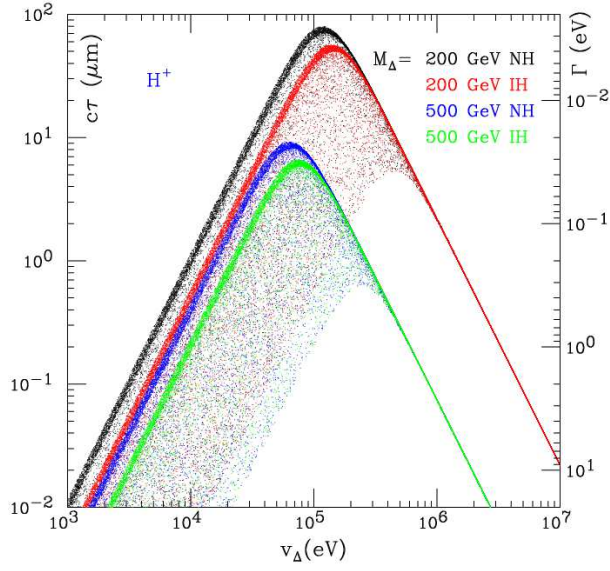}
\caption{Decay length and total width of the singly charged Higgs boson
$H^{+}$  with $\Phi_1 = \Phi_2 = 0$.}
\label{1xCt}
\end{figure}

The predictions for the decays of
singly charged Higgs boson taking into account the experimental
constraints on neutrino mass and mixing parameters are shown
in Fig.~\ref{1vBi}, again ignoring the effects of the Majorana phases $\Phi_1 = \Phi_2 = 0$.
The general features are similar to those of $H^{++}$ decays.
As one can see in the case of NH the 
BR$(H^+ \to \tau^+ \bar{\nu})$ and BR$(H^+ \to \mu^+ \bar{\nu})$ are dominant, 
while in the case of IH, the BR$(H^+ \to e^+ \bar{\nu})$ is the leading one.
The maximal decay lengths of the singly charged Higgs is about twice
that of the doubly charged Higgs, as shown in Fig.~\ref{1xCt}.
%


We now summarize the properties of the lepton-number violating Higgs decays,
that are intimately related to the patterns of the neutrino mass and mixing,
in Table I, where we have neglected the effects of the Majorana phases.

\begin{table}[tb]
\caption{\label{contributions} Relations among the branching fractions of the
lepton number violating Higgs decays
for the neutrino mass patterns of NH, IH, and QD,
with no Majorana phases $\Phi_1=\Phi_2=0$.}
\begin{ruledtabular}
\begin{tabular}{lcc}
 \text{Spectrum} & Relations
\\
\hline
Normal Hierarchy              & BR$( H^{++} \to \tau^+ \tau^+ )$, BR$( H^{++} \to \mu^+ \mu^+ ) \gg
$ BR$( H^{++} \to e^+ e^+ )$\\
$(\Delta m_{31}^2 > 0)$  & BR$( H^{++} \to \mu^+ \tau^+ ) \gg $ BR$( H^{++} \to e^+ \mu^+ )$,
BR$( H^{++} \to e^+ \tau^+ )$\\
                            & BR$( H^{+} \to \tau^+ \bar{\nu} )$, BR$( H^{+} \to \mu^+ \bar{\nu} ) \gg $ BR$(H^{+} \to e^+ \bar{\nu} )$
\\
\hline
Inverted Hierarchy            & BR$( H^{++} \to e^+ e^+ ) \ > \ $ BR$( H^{++} \to \mu^+ \mu^+ )$, BR$( H^{++} \to \tau^+ \tau^+ )$\\
$(\Delta m_{31}^2 < 0)$       & BR$( H^{++} \to \mu^+ \tau^+ ) \ \gg \ $ BR$( H^{++} \to e^+ \tau^+ )$,  BR$( H^{++} \to e^+ \mu^+ )$
\\                            & BR$( H^{+} \to e^+ \bar{\nu} ) \ > \ $ BR$( H^{+} \to \mu^+ \bar{\nu} )$, BR$( H^{+} \to \tau^+ \bar{\nu} )$
\\
\hline
Quasi-Degenerate         & BR$( H^{++} \to e^+ e^+ ) \sim $ BR$( H^{++} \to \mu^+ \mu^+ )\sim$ BR$( H^{++} \to \tau^+ \tau^+ )\approx 30\%$\\
$(m_1,m_2,m_3>|\Delta m_{31}|)$
& BR$( H^{+} \to e^+ \bar{\nu} ) \sim $BR$( H^{+} \to \mu^+ \bar{\nu} )\sim$BR$( H^{+} \to
\tau^+ \bar{\nu} )\approx 30\%$
\end{tabular}
\end{ruledtabular}
\end{table}

\subsection{Impact of Majorana Phases in Higgs Boson Decays}
Recently, the effects of  Majorana phases on the Higgs decays
have been investigated by several groups~\cite{Thomas,Raidal,Akeroyd}.
Wherever overlap exists, our results are in agreement with theirs. 
In fact, the effects can be made quite transparent under some
simple approximations.
\subsubsection{Normal Hierarchy with one quasi-massless neutrino: $m_1\approx 0$}
\begin{figure}[b]
\includegraphics[scale=1,width=8.0cm]{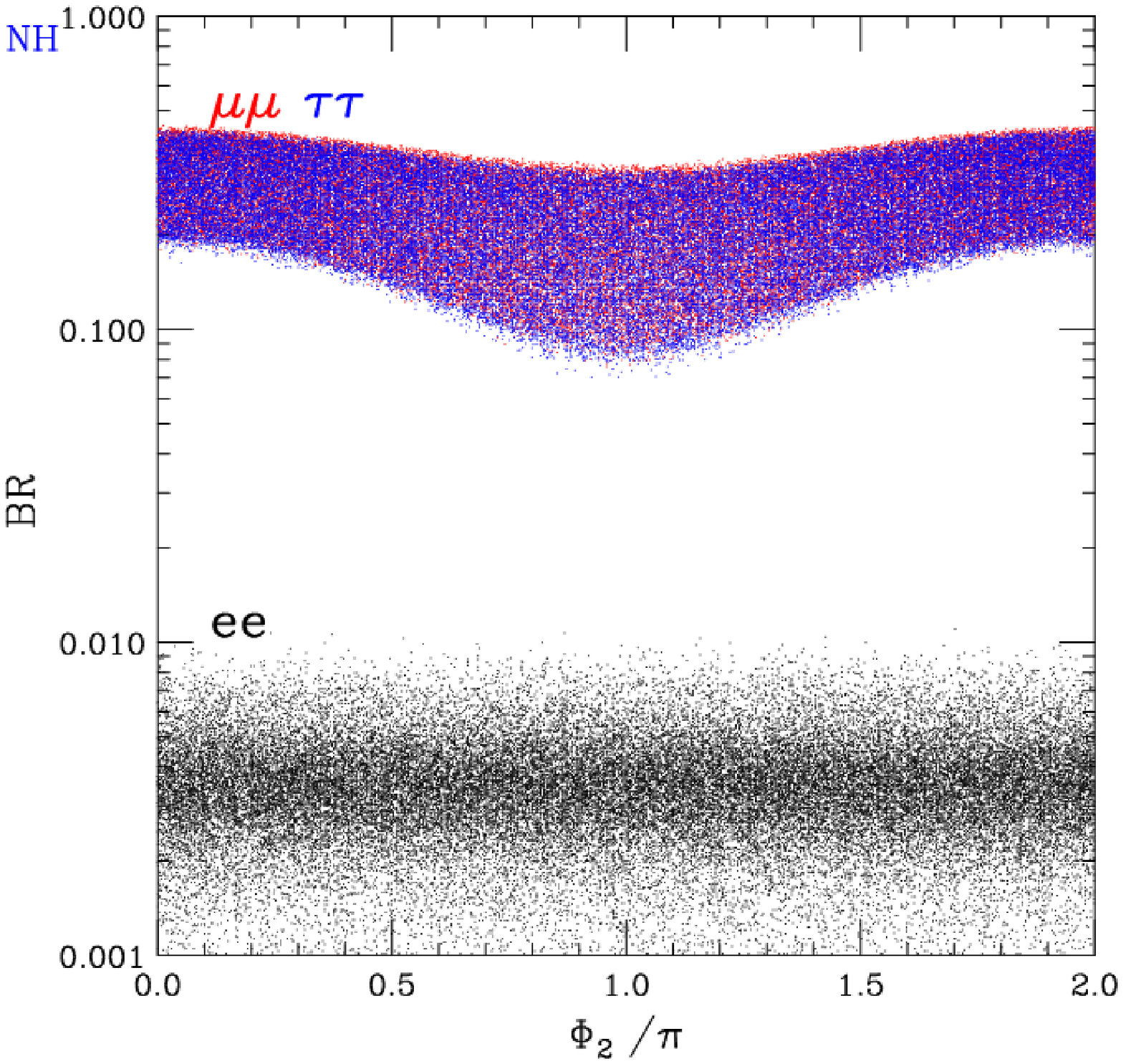}
\includegraphics[scale=1,width=8.0cm]{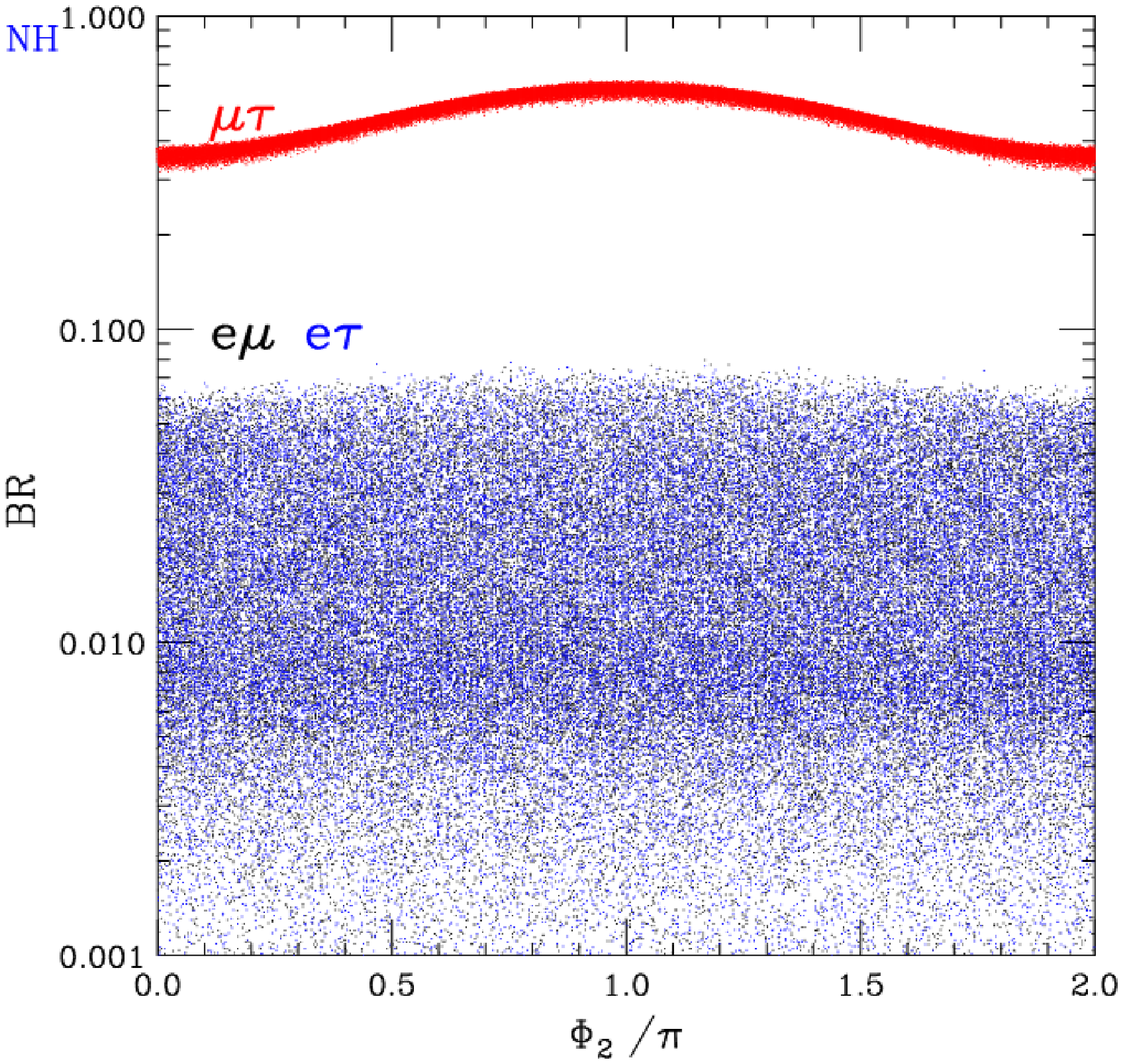}
\caption{Scatter plots of the same (left) and different (right) flavor leptonic branching
fractions for the $H^{++}$ decay versus the Majorana phase $\Phi_2$ for the NH
$m_1 = 0$  scenario. $\Phi_1 \in (0,2\pi)$.}
\label{Majorana1}
\end{figure}

As we have discussed in the previous section, the most important decay
channels of the doubly charged Higgs are $H^{++} \to \tau^+ \tau^+$,
$H^{++} \to \mu^+ \mu^+$ and $H^{++} \to \mu^+ \tau^+$. The leading couplings,
taking $s_{13}=0$ for simplicity, are
\begin{eqnarray}
\Gamma_{++}^{22} &=& \frac{1}{\sqrt{2} v_{\Delta}} \left( \sqrt{\Delta m_{21}^2} \ c_{12}^2 c_{23}^2 \ + \  \sqrt{\Delta m_{31}^2} \ e^{-i\Phi_2} s_{23}^2 \right) \\
\Gamma_{++}^{23} &=& \frac{s_{23} c_{23}}{\sqrt{2} v_{\Delta}} \left( - \sqrt{\Delta m_{21}^2} \ c_{12}^2 \ + \ \sqrt{\Delta m_{31}^2} \ e^{-i\Phi_2} \right) \\
\Gamma_{++}^{33} &=& \frac{1}{\sqrt{2} v_{\Delta}} \left( \sqrt{\Delta m_{21}^2} \ c_{12}^2 s_{23}^2 \ + \  \sqrt{\Delta m_{31}^2} \ e^{-i\Phi_2} c_{23}^2 \right)
\end{eqnarray}
The decay rates thus depend on only one Majorana phase $\Phi_2$.
The behavior of branching fractions for all channels is shown in Fig.~\ref{Majorana1}.
We see the rather weak dependence of the decay branching fractions on the phase,
which can be understood by realizing the large difference between the two interfering
terms $\Delta m_{21} \ll \Delta m_{31}$.
When the phase $\Phi_2=\pi$, one obtains the maximal suppression (enhancement)
for the channels
$H^{++} \to \tau^+ \tau^+$ and  $H^{++} \to \mu^+ \mu^+$ ($H^{++} \to \mu^+ \tau^+$)
by a factor of two at most.

\subsubsection{Inverted Hierarchy with one quasi-massless neutrino: $m_3\approx 0$}
\begin{figure}[tb]
\includegraphics[scale=1,width=8.0cm]{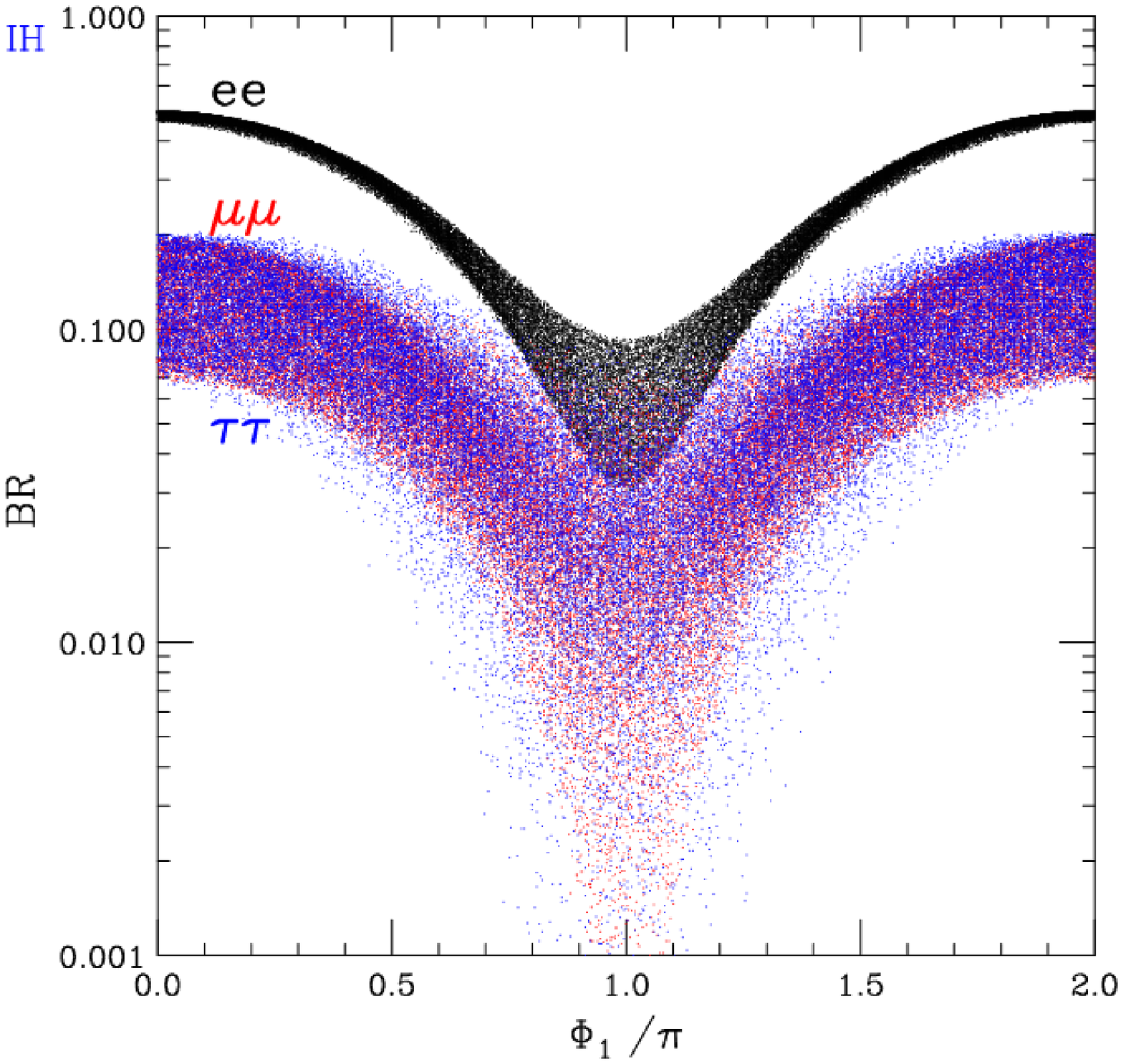}
\includegraphics[scale=1,width=8.0cm]{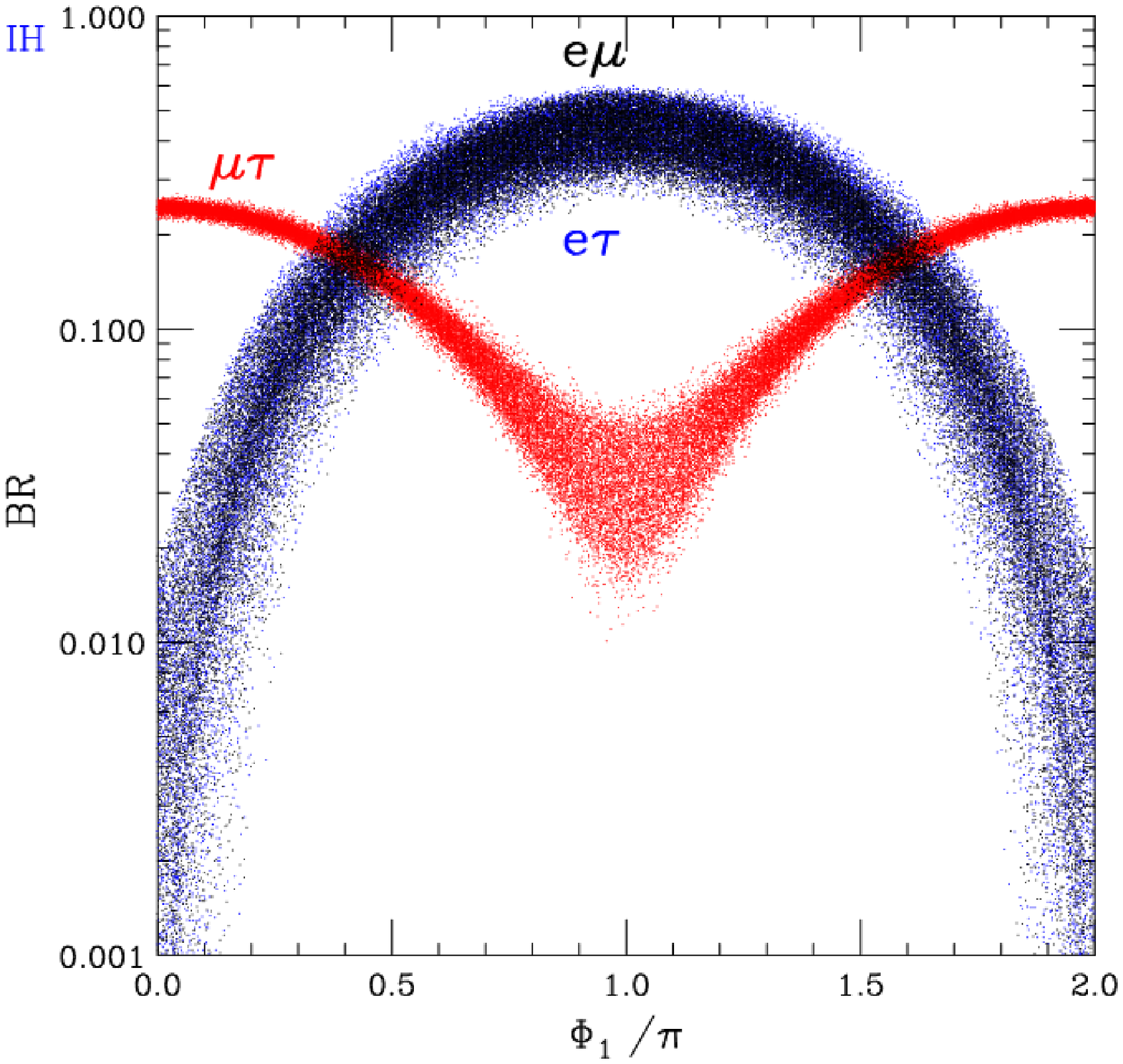}
\caption{Scatter plots of the same (left) and different (right) flavor leptonic branching
fractions for the $H^{++}$ decay versus the Majorana phase $\Phi_1$ for the IH
$m_3 = 0$  scenario. $\Phi_2 \in (0,2\pi)$.}
\label{Majorana2}
\end{figure}

In the case of Inverted Hierarchy the relevant channels are $H^{++} \to e^+ e^+,\  \mu^+ \tau^+$,
as well as $H^{++} \to e^+ \mu^+,\  e^+ \tau^+$. The couplings, taking $s_{13}=0$, read as
\begin{eqnarray}
\Gamma_{++}^{11} &=&
\frac{1}{\sqrt{2} v_{\Delta}} \left( \sqrt{ \Delta m_{21}^2 + |\Delta m_{31}^2|} \ s_{12}^2
+ \sqrt{|\Delta m_{31}^2|} \ e^{-i\Phi_1} c_{12}^2 \right) \approx
\sqrt{{ |\Delta m_{31}^2|} \over {2 v_{\Delta}^2 } }
\left( s_{12}^2 +  e^{-i\Phi_1} c_{12}^2 \right),~~ \\
\Gamma_{++}^{23} &=& - \frac{s_{23} c_{23}}{\sqrt{2} v_{\Delta}} \left( \sqrt{ \Delta m_{21}^2 + |\Delta m_{31}^2|} \ c_{12}^2
 + \sqrt{|\Delta m_{31}^2|} \ e^{-i\Phi_1} s_{12}^2 \right)
 \propto  c_{12}^2 +  e^{-i\Phi_1} s_{12}^2 , \\
\Gamma_{++}^{12} &=&
\frac{s_{12} c_{12} c_{23}}{\sqrt{2} v_{\Delta}} \left( - \sqrt{|\Delta m_{31}^2|} \ e^{-i \Phi_1}
\ + \ \sqrt{|\Delta m_{31}^2| + \Delta m_{21}^2} \right) \propto 1- e^{-i \Phi_1} , \\
\Gamma_{++}^{13} &=&
\frac{s_{12} c_{12} s_{23}}{\sqrt{2} v_{\Delta}} \left( \sqrt{|\Delta m_{31}^2|} \ e^{-i \Phi_1}
\ - \ \sqrt{|\Delta m_{31}^2| + \Delta m_{21}^2} \right)
 \propto -1+ e^{-i \Phi_1}.
\end{eqnarray}
All the relevant decays depend on only one phase $\Phi_1$, and the cancellations
due to the existence of the phase can be quite substantial as seen from the above equations.
In Fig.~\ref{Majorana2} we show the dependence of the branching fractions
on this Majorana phase. The maximal
suppression or enhancement takes places also when $\Phi_1=\pi$. However, in this scenario
the dominant channels swap from $H^{++} \to e^+e^+,\ \mu^{+} \tau^+$ when $\Phi_1 \approx 0$
to $H^{++} \to e^+ \mu^+,\ e^{+} \tau^+$  when $\Phi_1 \approx \pi$.
Therefore, this qualitative change can be made use of to
extract the value of the Majorana phase $\Phi_1$.

\begin{figure}[tb]
\includegraphics[scale=1,width=8.0cm]{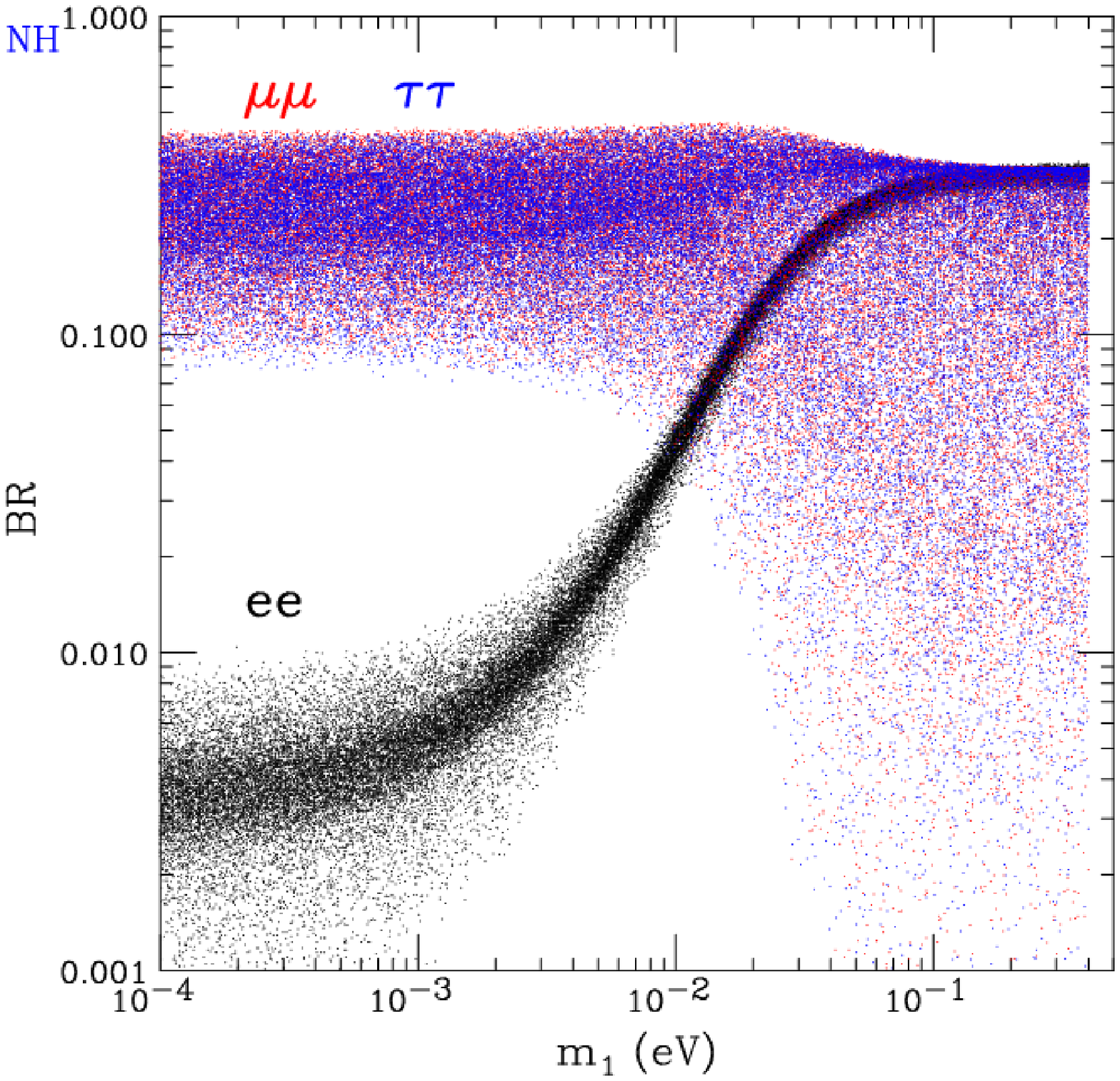}
\includegraphics[scale=1,width=8.0cm]{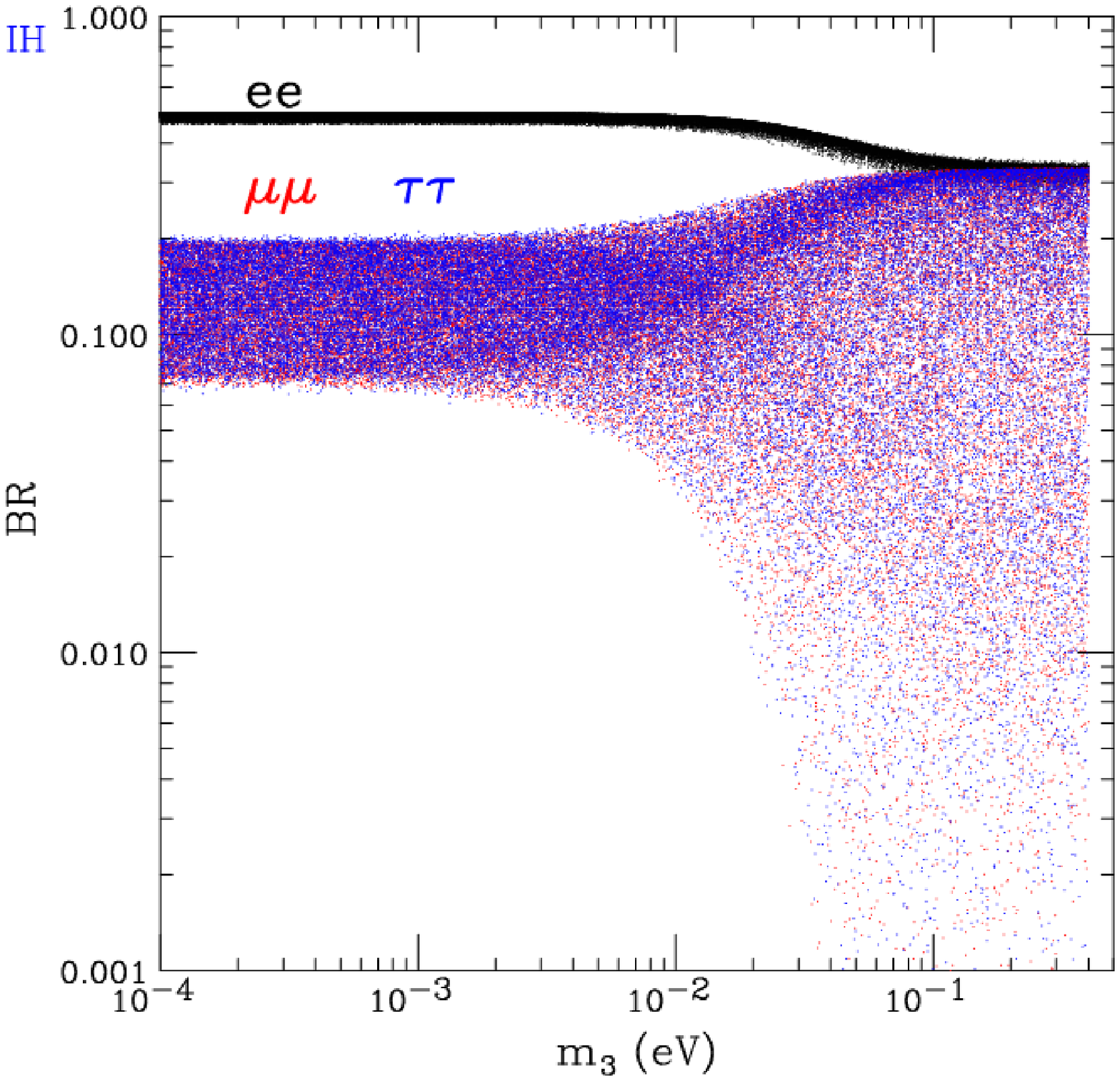}
\caption{Scatter plots for the $H^{++}$ decay branching fractions
 to the flavor-diagonal like-sign dileptons versus the lowest
neutrino mass for NH (left) and IH (right) with $\Phi_1 = 0$ and $\Phi_2 \in (0,2\pi)$.}
\label{Brii-Phi10}
\end{figure}
\begin{figure}[tb]
\includegraphics[scale=1,width=8.0cm]{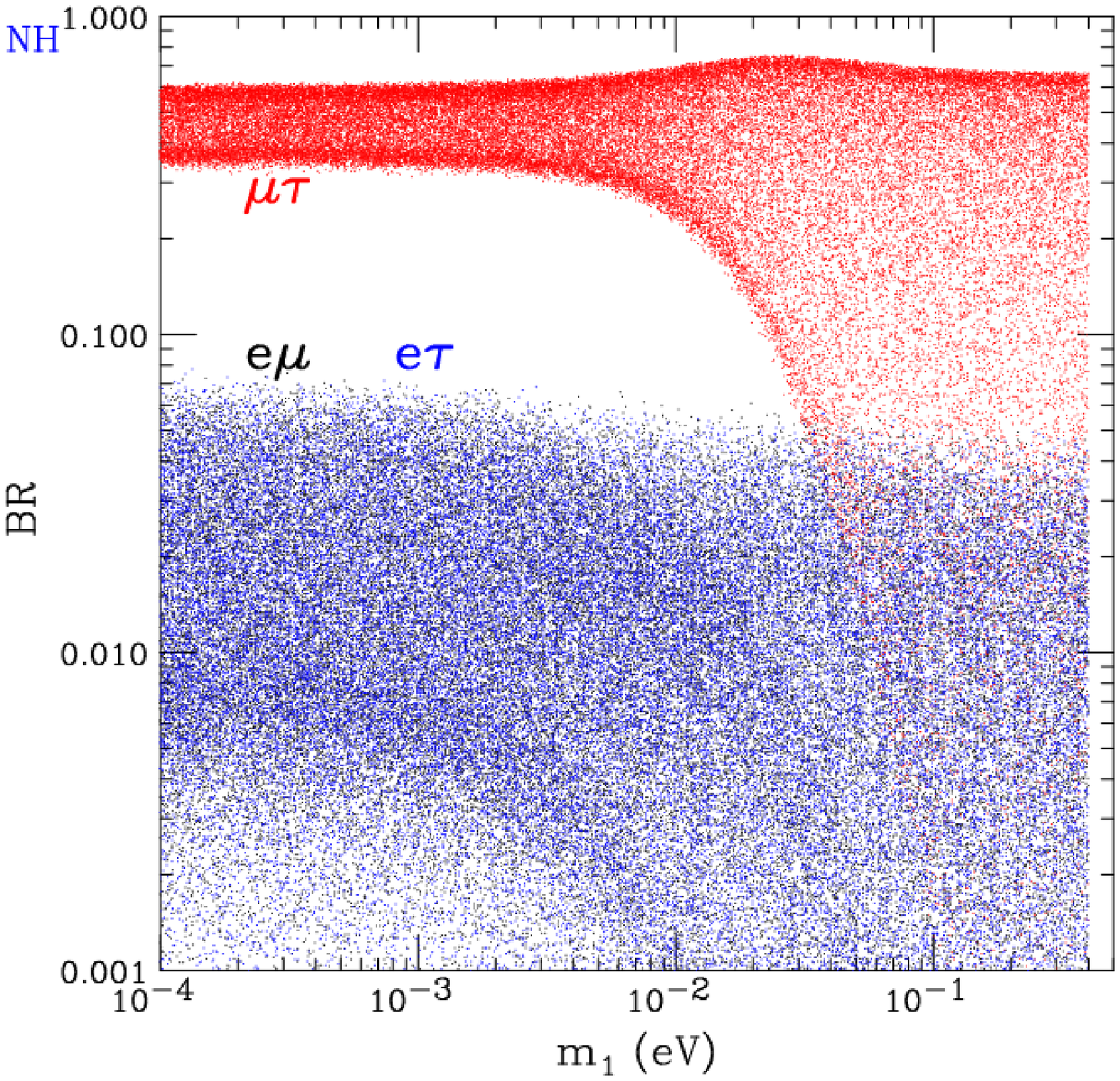}
\includegraphics[scale=1,width=8.0cm]{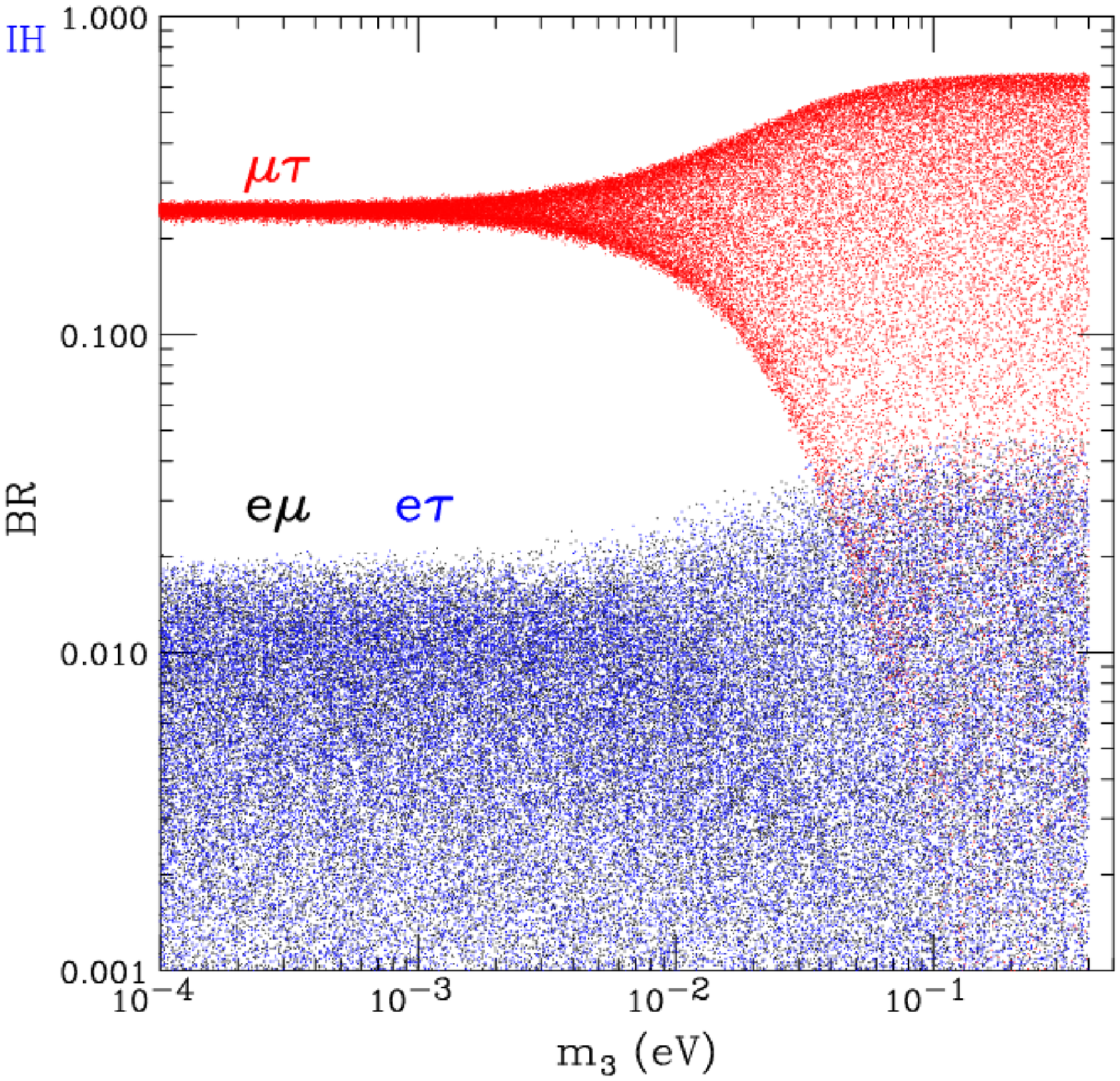}
\caption{Same as Fig.~\ref{Brii-Phi10}, but for  $H^{++}$ decay to
 the flavor-off-diagonal like-sign dileptons. }
\label{Brij-Phi10}
\end{figure}

In Figs.~\ref{Brii-Phi10} and \ref{Brij-Phi10}, we show the predictions of the leptonic
branching fractions of the doubly charged Higgs boson
for the same and different flavors versus the
lightest neutrino mass and $\Phi_1 =0$, and $\Phi_2 \in (0,2\pi)$.
These are to be compared with Figs.~\ref{brii} and \ref{brij} where $\Phi_1 = \Phi_2 = 0$.
Generically, the allowed ranges for the branching fractions are broadened
with nonzero phases, making the BR's less predictive and it is more difficult 
to determine the neutrino mass pattern. For small values of the lightest neutrino 
mass less than $10^{-2}$ eV, the BR's for the NH spectrum is more spread out 
than that for the IH with $\Phi_2 \ne 0$ as noticed earlier.
When the lightest neutrino mass is larger than $10^{-2}$ eV,
the BR's for both the NH and the IH spectra can be further spread out.

\begin{figure}[tb]
\includegraphics[scale=1,width=8.0cm]{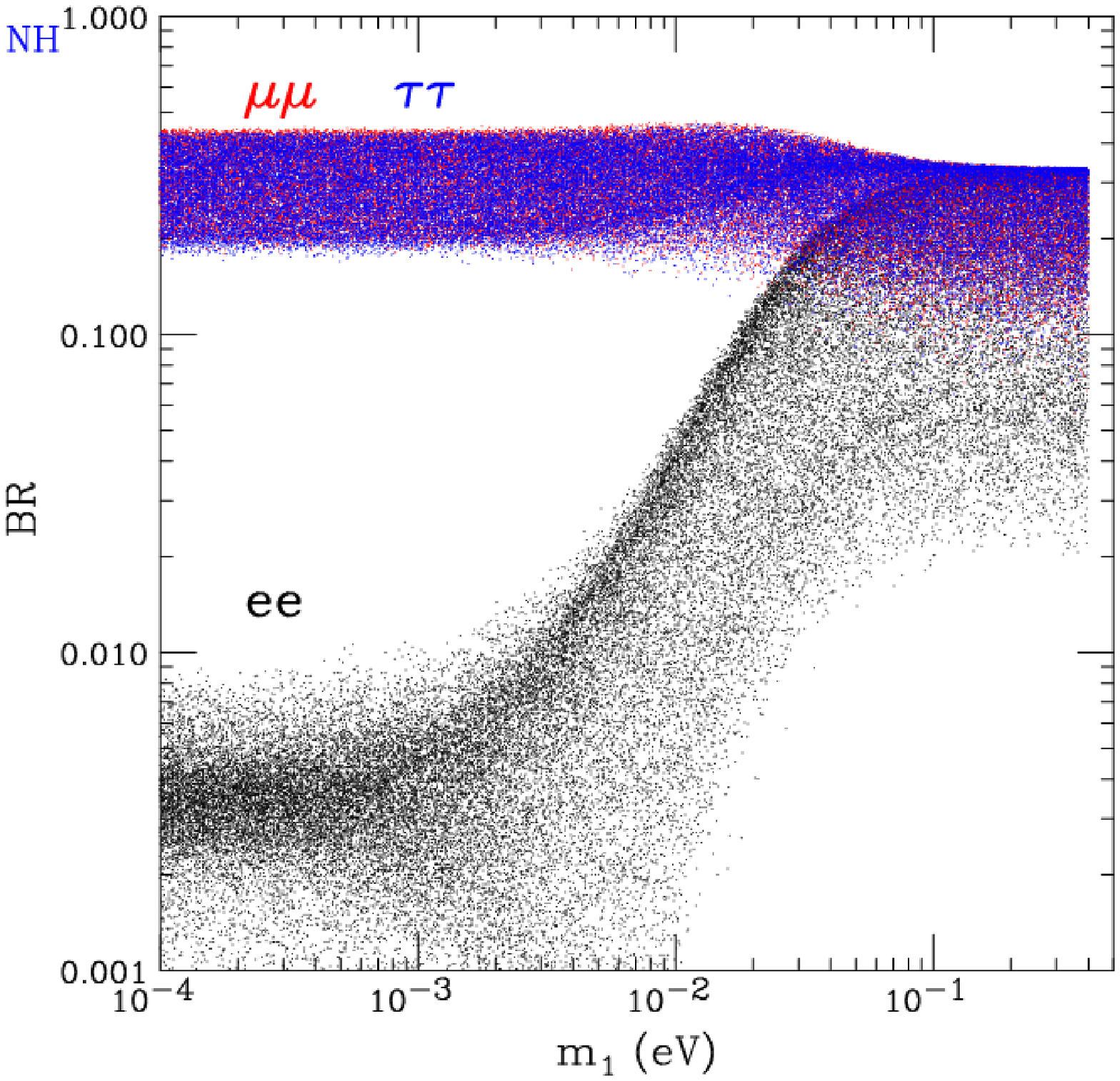}
\includegraphics[scale=1,width=8.0cm]{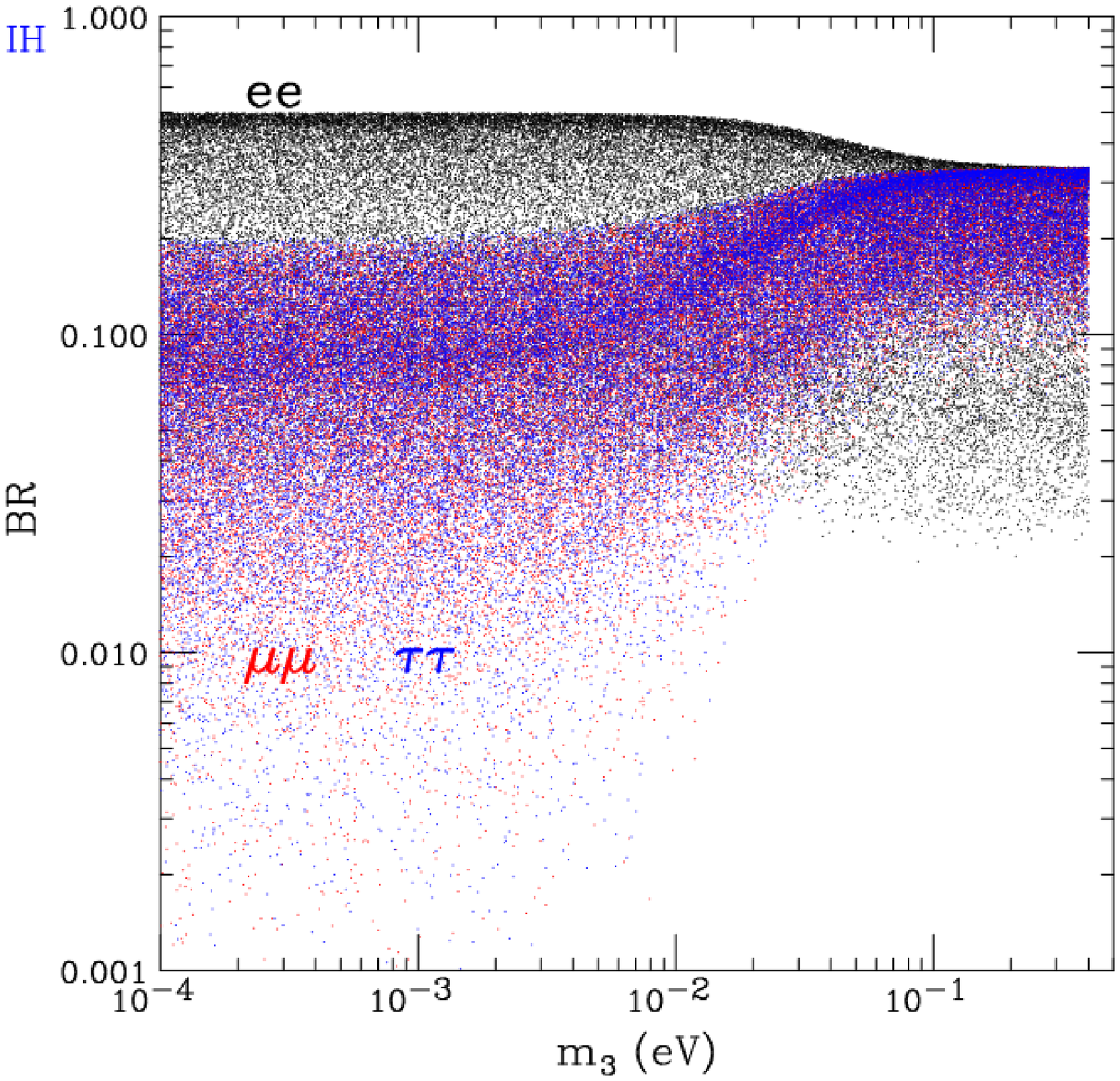}
\caption{Scatter plots for the $H^{++}$ decay branching fractions
 to the flavor-diagonal like-sign dileptons versus the lowest
neutrino mass for NH (left) and IH (right) with $\Phi_2 =0$ and $\Phi_1 \in (0,2\pi)$.}
\label{Brii-Phi20}
\end{figure}
\begin{figure}[tb]
\includegraphics[scale=1,width=8.0cm]{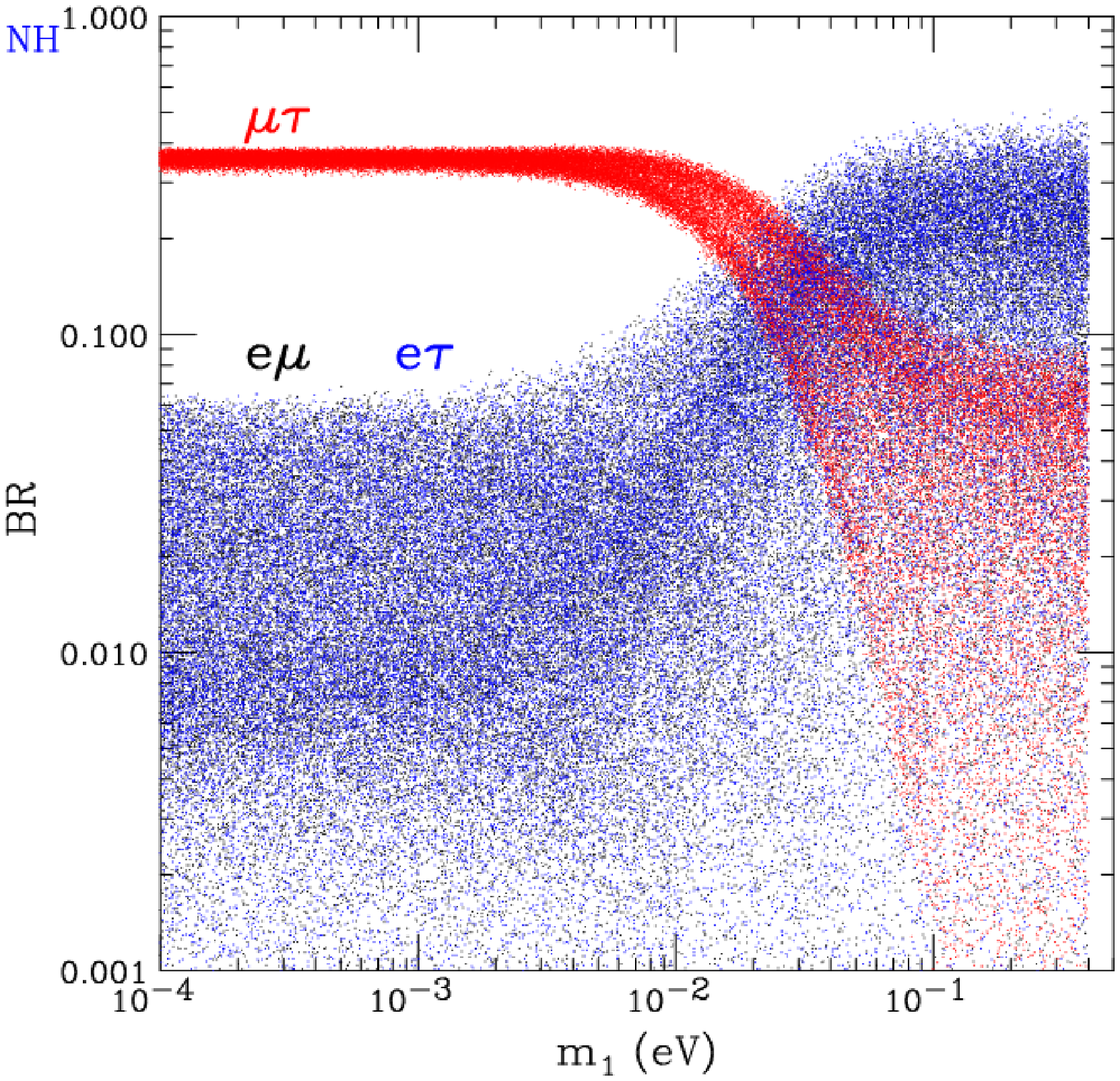}
\includegraphics[scale=1,width=8.0cm]{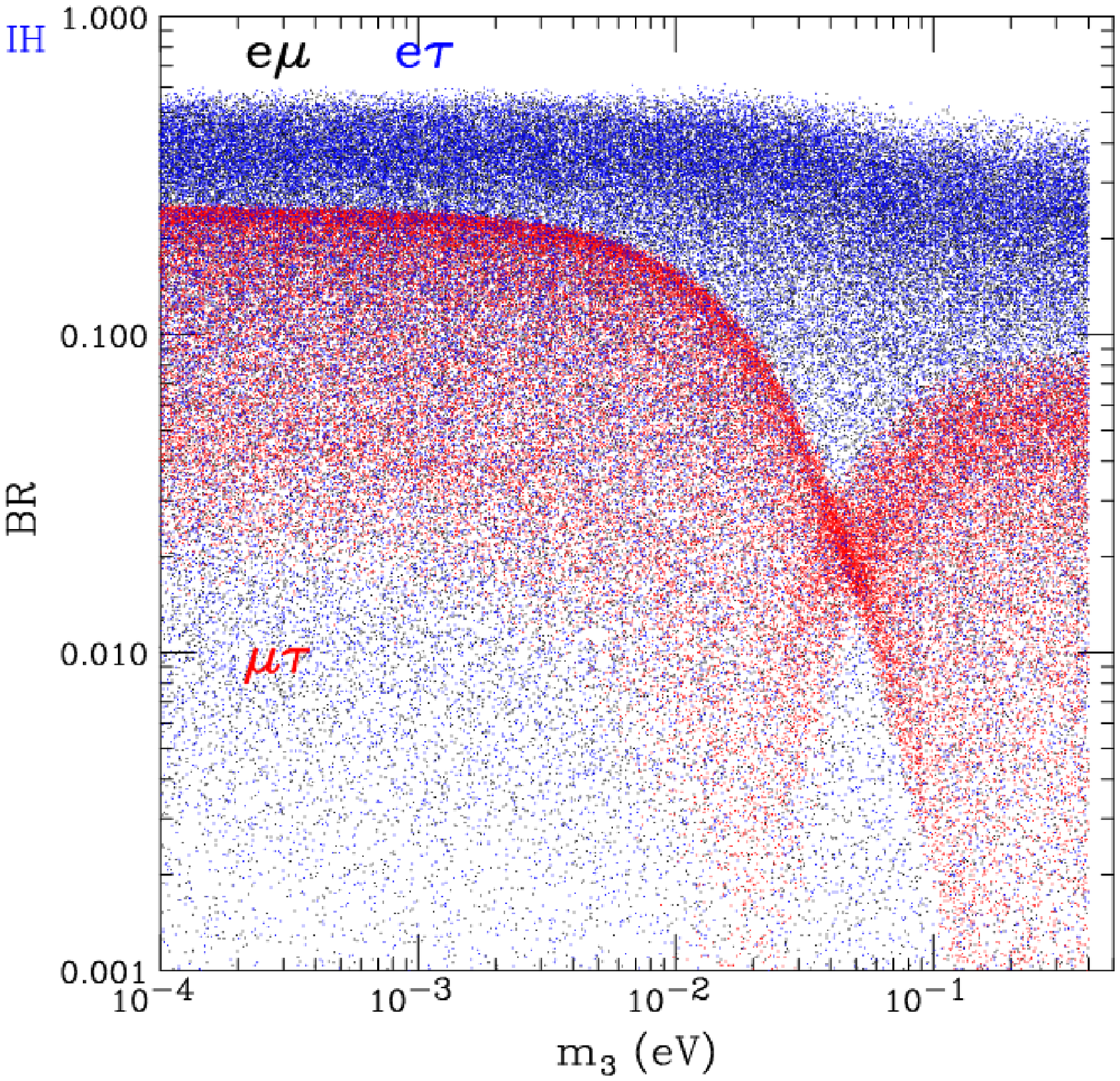}
\caption{Same as Fig.~\ref{Brii-Phi20}, but for  $H^{++}$ decay to
 the flavor-off-diagonal like-sign dileptons. }
\label{Brij-Phi20}
\end{figure}

Similar features can been seen in Figs.~\ref{Brii-Phi20} and \ref{Brij-Phi20}
where  $\Phi_1 \in (0,2\pi)$ and  $\Phi_2 =0$, again to be compared
with Figs.~\ref{brii} and \ref{brij} where $\Phi_1 = \Phi_2 = 0$.
The allowed ranges for the branching fractions are broadened
with nonzero phases, making the BR's less predictive.
For small values of the lightest neutrino mass less than $10^{-2}$ eV,
the BR's for the IH spectrum is more spread out than that for the NH with
$\Phi_1 \ne 0$ as noticed earlier.
When the lightest neutrino mass is larger than $10^{-2}$ eV,
the BR's for the NH can be completely spread out.

We thus conclude that the Majorana phases can change the branching
fractions of the doubly charged Higgs boson dramatically.
However, it is important to note that the decays of the singly charged Higgs
boson $H^+ \to e^+_i \bar{\nu}$ are independent of the Majorana phases.
Therefore, in order to distinguish the neutrino mass spectra non-ambiguously,
it is necessary to make use of the decays of the singly charged Higgs boson.
The combination of  the decays of both the singly and doubly
charged Higgs bosons may shed light on the Majorana phases, in particular
for the sensitive dependence on $\Phi_1$ in the case of IH.

\section{SEARCHING FOR SEESAW TRIPLET HIGGS AT THE LHC}
The leading production channels at hadron colliders for
these Higgs bosons are the following electroweak processes:
\begin{eqnarray}
q(p_1) \, + \, \bar{q}(p_2) \, &\rightarrow &
H^{++}(k_1)\, + \, H^{--}(k_2)\nonumber\\
q(p_1) \, + \, \bar{q}'(p_2) \, &\rightarrow &
H^{++}(k_1)\, + \, H^{-}(k_2)\nonumber\\
q(p_1) \, + \, \bar{q}'(p_2) \, &\rightarrow &
H^{+}(k_1)\, + \, H_2(k_2)\nonumber
\end{eqnarray}
In term of the polar angle variable
$y= \hat{p}_1\cdot \hat{k}_1$ in the parton c.m.~frame with energy $\sqrt{s}$, the parton level cross section for
these processes are
\begin{eqnarray}
{d\sigma\over dy}(q \bar{q}\rightarrow H^{++}H^{--}) &=& \frac{3\pi \alpha^2 \beta_i^3 (1-y^2)}{N_c {s}}
\Big\{ e_q^2 \,+ \,
\frac{ {s}}{({s}-M_Z^2)^2}
\, \frac{\cos 2\theta_W}{\sin^2 2\theta_W}\nonumber\\
&& \times\Big[4 e_q g_V^q  ({s}-M_Z^2)
\, + \,4 (g_V^{q2}+g_A^{q2}) {s}\  \frac{\cos 2\theta_W}{\sin^2 2\theta_W}
\Big]
\Big\} ,\\
{d\sigma\over dy}(q \bar{q}'\rightarrow H^{++}H^{-}) &=&
2 {d\sigma\over dy}(q \bar{q}'\rightarrow H^{+}H_2) = \frac{\pi \alpha^2 \beta_i^3(1-y^2)}{16 N_c \sin^4\theta_W}\frac{s}{(s-M^2_W)^2},
\end{eqnarray}
where $\beta_i =\sqrt{(1-(m_i+m_j)^2/s)(1-(m_i-m_j)^2/s)}$ is the speed factor of $H_i$
and $H_j$ in the c.m.~frame.

The production of $H^{\pm\pm}H^{\mp}$ \cite{single} and $H^\pm H_2$ can be crucial to test
its $SU(2)_L$ triplet nature at the collider. Doubly charged Higgs  and singly charged Higgs
can also be incorporated in other theories, for instance, the Zee-Babu model~\cite{Babu-Zee}
where $H^{\pm\pm}$ and $H^{\pm}$ are both $SU(2)_L$ singlets, and the
Majorana neutrino masses arise at two-loop level.
Both pair productions of $H^{++}H^-$ and $H^+H_2$ will vanish in
the Zee-Babu model due to the absence of the $SU(2)_L$ gauge couplings.
Drell-Yan production of $H^{++}H^{--}$ and $H^+H^-$ will be present
via the hypercharge interaction of $\gamma$ and $Z$.

\begin{figure}[tb]
\includegraphics[scale=1,width=8cm]{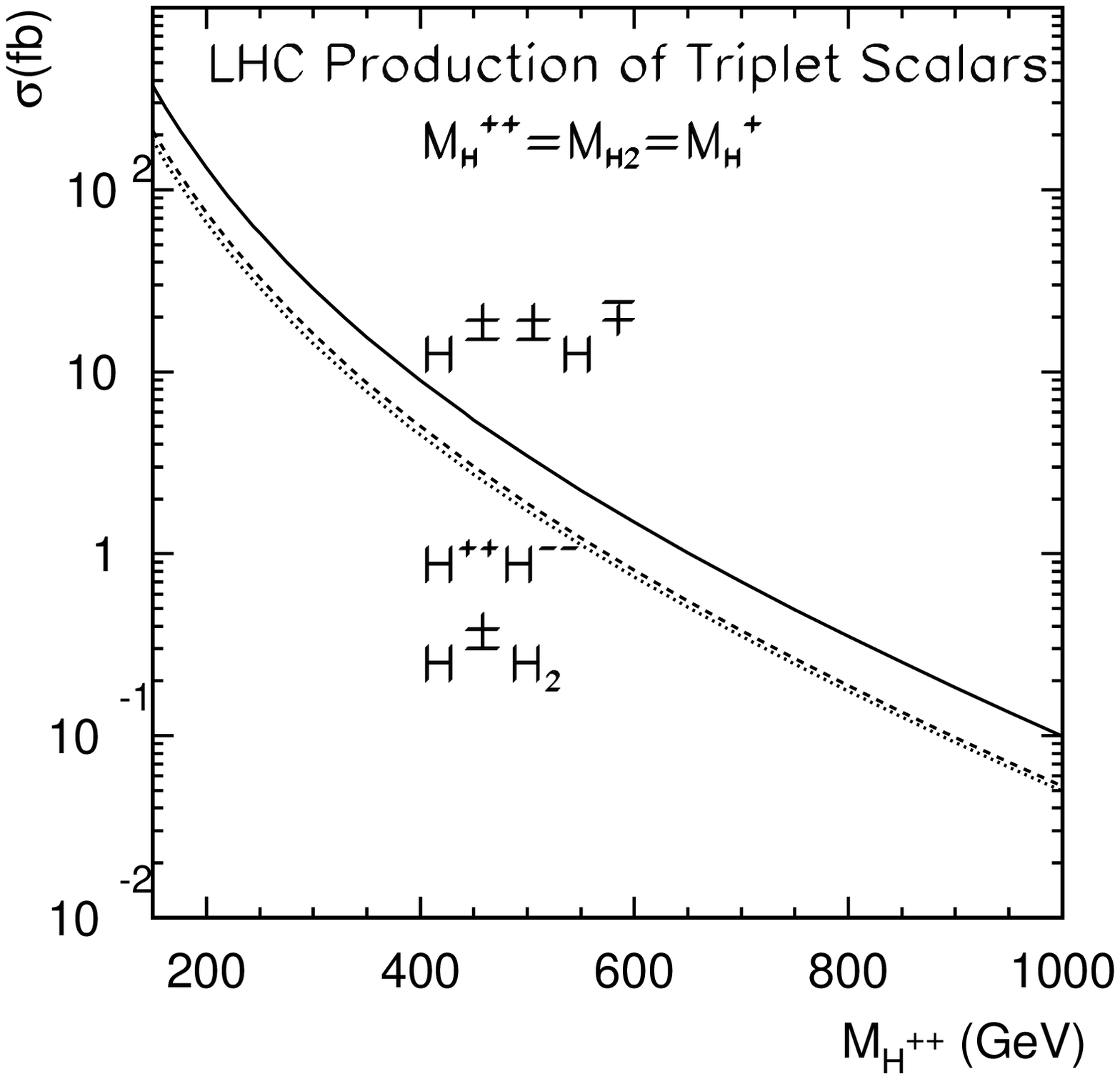}
\includegraphics[scale=1,width=8cm]{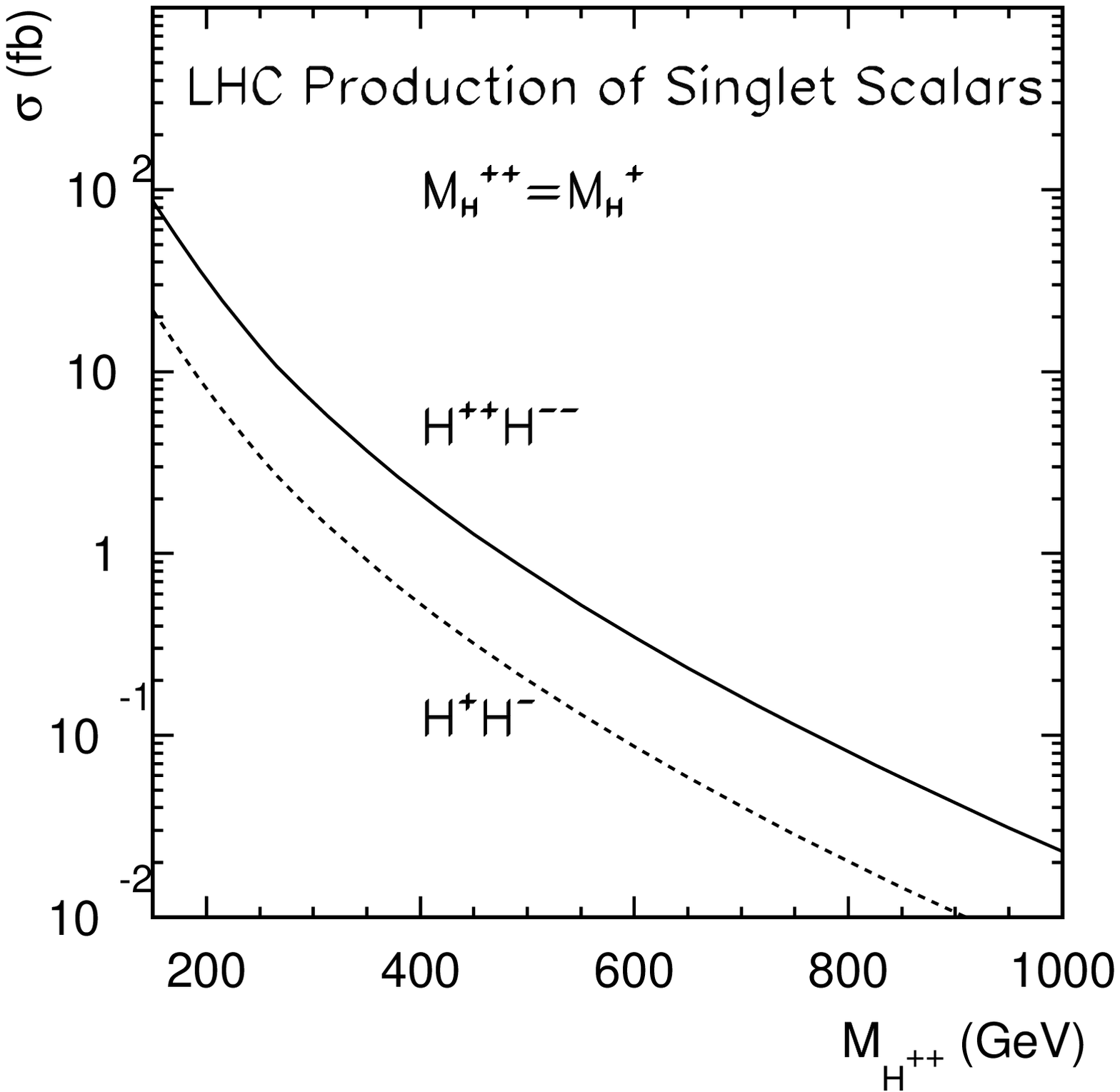}
\caption{Total production cross section at the LHC versus the heavy Higgs mass
for (a)  $H^{\pm}H_2$, $H^{\pm\pm}H^{\mp}$ and $H^{++}H^{--}$ processes in the
triplet model (left), and (b)  $H^{++}H^{--}$ and $H^+H^-$ processes in the
singlet model (right). }
\label{total}
\end{figure}

The production cross sections for all three channels are plotted in
Fig.~\ref{total}(a) ($H^+H^-$ is not presented since it is
phenomenologically less unique and we will not study it.)
For comparison, we also plot the production of $H^{++}H^{--}$ and
$H^+H^-$  in Zee-Babu model in Fig.~\ref{total}(b).
The production rate is lower by about
a factor of two comparing with the rates in the triplet model.
Only tree-level results are shown in these figures.
The QCD corrections to the process $H^{++}H^{--}$ have also been computed \cite{qcd}, and a
next-to-leading (NLO) $K$-factor of order 1.25 at the LHC for Higgs mass range from
150 GeV to 1 TeV is predicted.
QCD corrections to the production of $H^{\pm\pm} H^{\mp}$ and $H^{\pm}H_2$ are in principle very similar to $H^{++}H^{--}$ and we apply the same $K$-factor to these two processes in our numerical
analysis.
In the $H^{++}H^{--}$ production, contribution from real photon annihilation is
shown~\cite{last} to be an increase of  10\%
to the Drell-Yan production for the above mass range at the LHC. 
We will apply an overall $K$-factor of 1.35 for the $H^{++}H^{--}$ 
production, and 1.25 for the $H^{++}H^{-}$ production.

\subsection{Purely Leptonic Modes}
The light neutrino mass matrix and the leptonic decay branching fractions of triplet Higgs bosons
are related by the structure of triplet Yukawa matrix $\Gamma_{++}$ (or $Y_\nu$).
This direct correlation may enable
us to test the neutrino mass generation by collider observables of the decay branching fractions
for different flavor  combinations. Consider the case
of large Yukawa couplings ($v_\Delta < 10^{-4}$ GeV), the triplet Higgs
decays will be dominated by the leptonic modes
\begin{equation}
H^{++}  \rightarrow  e_i^+ e_j^+;~~H^{+}   \rightarrow  e_i^+ \bar{\nu} ;~~H_2\rightarrow  \nu\nu +\bar{\nu}\bar{\nu}~~~~(e_i=e,\mu,\tau).\nonumber
\end{equation}
The $H_2$ decays are experimentally invisible
and the reconstruction of $H_2$ becomes impossible.
Hence, we focus on the production of $H^{++} H^{-}$ and $H^{++} H^{--}$.
In the rest of this section, we establish the observability for the leading decay
channels at the LHC. We then discuss the measurement of their decay branching
fractions and connect the individual channels to the neutrino mass patterns.

\begin{figure}[tb]
\includegraphics[scale=1,width=9cm]{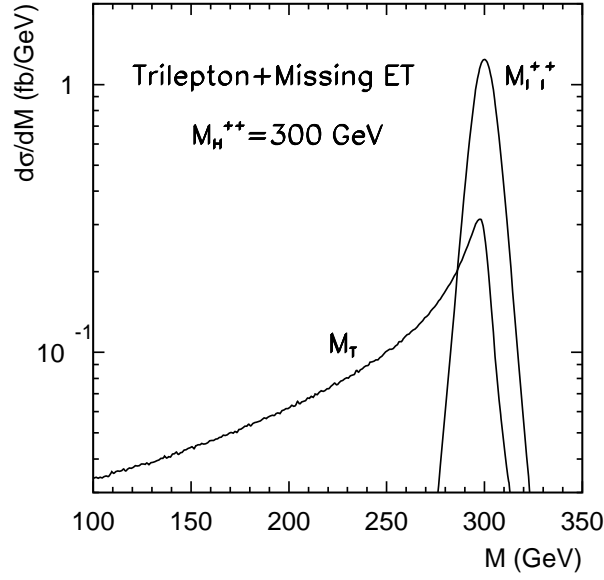}
\caption{ Reconstructed invariant mass of $M_{\ell^\pm\ell^\pm}$ and transverse
mass $M_{T}(\ell^\mp \nu)$ for the processes
$H^{\pm\pm}H^{\mp} \rightarrow \ell^\pm\ell^\pm\ \ell^\mp \nu$,
with a representative heavy Higgs mass 300 GeV. }
\label{fig:Rec}
\end{figure}

\subsubsection{$H^{\pm\pm}H^{\mp} \rightarrow \ell^\pm\ell^\pm\ \ell^\mp \nu \quad (\ell = e,\mu)$}
We start from the easy channels with $e,\mu$ in the final state of the Higgs decays.
The signal consists of one pair of same sign leptons and another opposite sign lepton
plus missing energy.
We employ the following basic acceptance cuts for the event selection \cite{CMS}
\bea
\nonumber
 && p_T(\ell_{\rm hard}) >  30\  {\rm GeV}, \ \ p_T(\ell) > 15\ {\rm GeV},\ \
 \cancel{E}_T> 40\ {\rm GeV},\\
 && |\eta(\ell)| < 2.5,\ \  \Delta R_{\ell\ell} > 0.4.
 \label{eq:basic}
\eea
To simulate the detector effects on the energy-momentum measurements,
we smear the electromagnetic energy
and the muon momentum by a Gaussian distribution whose width is parameterized as \cite{CMS}
\begin{eqnarray}
{ \Delta E\over E} &=& {a_{cal} \over \sqrt{E/{\rm GeV}} } \oplus b_{cal}, \quad
a_{cal}=5\%,\  b_{cal}=0.55\% ,
\label{ecal}\\
{\Delta p_T\over p_T} &=& {a_{track}\ p_T \over {\rm TeV}} \oplus {b_{track}\over \sqrt{\sin{\theta}} }, \quad
 a_{track}= 15\%,\ b_{track} =0.5\%.
\end{eqnarray}
For high $p_T$ leptons, the electromagnetic energy resolution is better than muon's
tracking resolution.

\begin{figure}[tb]
\includegraphics[scale=1,width=9cm]{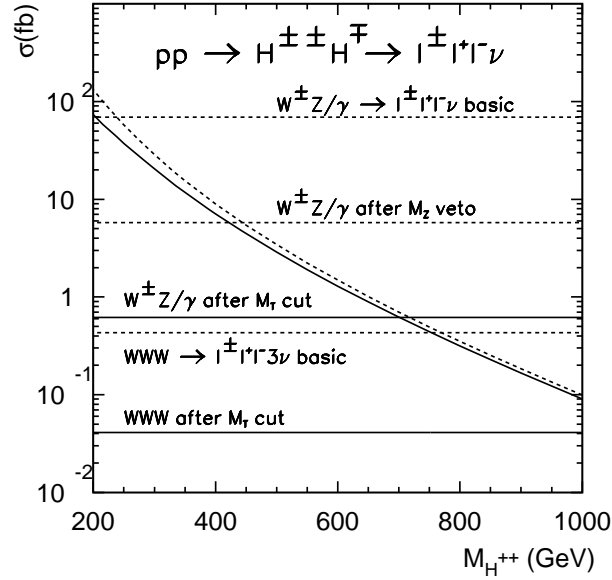}
\caption{Production cross section  of $H^{\pm\pm}H^{\mp} \to \ell^\pm\ell^\pm\ \ell^\mp \nu$
 at the LHC  versus the heavy Higgs mass
 with (solid curve) and without (dashed curve) the kinematical cuts.
Branching fractions for the Higgs decays are taken to be $100\%$ for illustration.
For comparison, the background processes
are also included with the sequential cuts as indicated. }
\label{fig:cuts}
\end{figure}

The irreducible SM backgrounds to this channel are
\beq
W^{\pm}Z/\gamma^*\rightarrow \ell^\pm \nu \ell^+\ell^-,\ \  W^\pm W^\pm W^\mp\rightarrow \ell^\pm \ell^+\ell^- +\cancel{E}_T.\nonumber
\eeq
Although the backgrounds are quite sizable with the basic leptonic cuts,
the order of 100 fb for $WZ$ and 1 fb for $WWW$, the kinematics is very different
between the signal and the backgrounds. We outline the characteristics and propose
some judicious cuts as follows.
\begin{itemize}
\item To remove the $WZ$ background, we veto the lepton pairs with the same flavor
but opposite charges in the $Z$-mass window $| M_{\ell^+ \ell^-} - M_Z |>15$ GeV.
\item

The mass reconstruction for $\ell^\pm\ell^\pm$ and $\ell^\mp \nu$ can be very indicative.
We first define a transverse mass $M_{T}$ by the opposite sign lepton
and missing transverse energy
\beq
M_T(\ell^\mp \nu)
=\sqrt{(E_T(\ell)+\cancel{E}_T)^2-(\vec{p}_T(\ell)+\vec{\cancel{p}}_T)^2}.
\nonumber
\eeq
This variable and the invariant mass of the like-sign dileptons are plotted in Fig.~\ref{fig:Rec}.
We then impose  a modest cut
\beq
M_T > 200\ {\rm GeV}.
\eeq
The cut can be further tightened up for heavier Higgs searches.
\item Finally, when we perform the signal significance analysis, we look for the
resonance in the mass distribution of $\ell^+\ell^+$.
For instance, if we look at a mass window of $M_\Delta \pm 25$ GeV in $M_{\ell^+\ell^+}$,
the backgrounds will be at a negligible level.
\end{itemize}

The production cross section  of $H^{\pm\pm}H^{\mp} \to \ell^\pm\ell^\pm\ \ell^\mp \nu$
with (solid curve) and without (dashed curve) the kinematical cuts are plotted in Fig.~\ref{fig:cuts}.
Branching fractions for the Higgs decays are taken to be $100\%$ for illustration.
For comparison, the background processes of $WZ$ and $WWW$
are also included with the sequential cuts as indicated. The backgrounds
are suppressed substantially.

As a remark, we would like to comment on the other potentially large, but reducible
backgrounds, the heavy quark production such as $t\bar t,\  W b\bar b$ etc.
The $t\bar t$ production rate is very high,
leading to the $\ell^+\ell^-\ X$ final state with about 40 pb.
Demanding another
isolated lepton presumably from the $b$ quarks and with the basic cuts, the
background rate will be reduced by about three to four orders of magnitude.
The stringent lepton isolation cut for multiple charged leptons can substantially
remove the $b$-quark cascade decays.
With the additional $M_T$ and  $M_{\ell^+\ell^+}$ cuts, the backgrounds should be
under control.

\subsubsection{$H^{\pm\pm}H^{\mp}\rightarrow \ell^\pm\ell^\pm\ \tau^\mp \nu\quad (\ell = e,\mu)$}

The $\tau$-lepton final state from $H^{\pm\pm}$ or $H^\pm$ decay
plays an important role in distinguishing different patterns of light neutrino masses.
Its identification and reconstruction are different from $e,\mu$ final states.
 There will always be a missing $\nu_\tau$ associated with the $\tau$ decay,
 and there is also a missing neutrino from $H^+$ decay as well. If the missing neutrinos
 are all from the same Higgs parent, one can still construct this Higgs boson
 via the transverse mass variable.
However, if the $\tau$ is from another Higgs decay like the $H^{\pm\pm}$,
the reconstruction will be difficult due to the multiple neutrinos from different parents.
Therefore in this section, we select the event involving a $\tau$ final state only  from
the decay  $H^\pm\rightarrow \tau^\pm \nu$.

Besides the two like-sign leptons that reconstruct the $H^{\pm\pm}$ and
are selected based on the basic cuts Eq.~(\ref{eq:basic}), we need to adjust
the threshold for the $\tau$ decay products that are significantly softer than
the direct decay from a heavy Higgs boson.
We accept isolated charged tracks as $\tau$ candidates (the ``1-prong" and ``3-prong" modes).
For the  muons and the other charged tracks, we take
\beq
p_T(\mu) > 5~{\rm GeV},\quad p_T({\rm track}) > 10~{\rm GeV}.
\nonumber
\eeq
With further kinematical selection similar to the last section, the irreducible
SM background is well under control. There may be additional backgrounds
with a jet to fake a $\tau$, such as $W^\pm W^\pm jj$.
According to ATLAS TDR~\cite{ATLAS}, for a hard $\tau$ in the range of $p_T\sim 70 -130$ GeV,
where $\tau$ identification efficiency is $60\%$,
the jet faking rate is $1\%$ into a hadronic decaying $\tau$.
Knowing the cross section for  $W^\pm W^\pm jj$ is the order of 15 fb after the
basic cuts, this leads to a faked background cross section to be way below 0.1 fb,
after vetoing the extra jet before the Higgs mass reconstruction.

There is one more complication for the event selection for the leptonic modes.
In order to identify the $\tau$ flavor, we must know if the $e$ or $\mu$ is from a $\tau$ decay or
from a heavy Higgs decay. Once again, we make use of the fact that the lepton
from a $\tau$ decay is softer.
We simulate the events and examine the fraction of wrong and correct $\tau$
identification with a given $p_T$ threshold and the results are presented in
Table \ref{tautable}. If an event contains a lepton
with $p_T$ less than the values shown in the table, it will be identified as $\tau$ leptonic
decay. Table \ref{tautable} gives the misidentification rate of $\tau$ from
$H^\pm \to e\nu, \mu\nu$ and the survival probability for $\tau \to e\nu\nu,\ \mu \nu\nu$.
To effectively keep the $\tau$ events, we choose in the rest of the analysis
the threshold $p_T < 100~{\rm GeV}$ for 
$M_{H^+}=300~{\rm GeV}$ and $p_T < 200~{\rm GeV}$ for $M_{H^+}=600~{\rm GeV}$.

\begin{table}
\begin{tabular}{| c|| c|c|c|c|c|c|}
\hline
& \multicolumn{3}{c|}{$M_{H^+}=300$ GeV}&\multicolumn{3}{c|}{$M_{H^+}=600$ GeV}\\
\hline
$p^\ell_T$ threshold (GeV)& 50  & 75 & 100 & 100 & 150 & 200 \\
\hline
$\ell$ misidentification rate & 2.9\% & 9.4\% & 17.6\%  & 4.6\% & 12.4\% & 22.2\% \\
$\tau$ survival probability & 57.0\% & 69.8\%  &78.8\% & 62.8\% & 75.7\% & 83.7\%\\
\hline
\end{tabular}
\caption{The misidentified rate of $\tau$ from
$H^\pm \to e\nu, \mu\nu$ and the survival probability for $\tau \to e\nu\nu,\ \mu \nu\nu$
 in the channels $H^{\pm\pm}H^\mp \to \ell^\pm\ell^\pm\ \tau^\mp \nu $.}
\label{tautable}
\end{table}

\subsubsection{$H^{++}H^{--}\rightarrow  \ell^+\ell^+\ \ell^-\tau^-, \quad
\ell^+\ell^+\ \tau^-\tau^-,\quad \ell^+\tau^+\ \ell^-\tau^-,\quad
 \ell^+\tau^+\ \tau^-\tau^-$}
The best channels for $H^{++}H^{--}\rightarrow \ell^+\ell^+\ell^-\ell^-\ (\ell=e,\mu)$
have been discussed extensively in the literature \cite{last}.
However, it has been strongly motivated in the early sections to look for
channels with $\tau$'s in the final state, such as
$H^{++}\rightarrow e^+\tau^+,\ \mu^+\tau^+,\ \tau^+\tau^+$.
Identifying decays of doubly charged Higgs bosons with $\tau$ final state is
crucial to distinguish different spectra of the neutrino mass.

\begin{figure}[tb]
\includegraphics[scale=1,width=9cm]{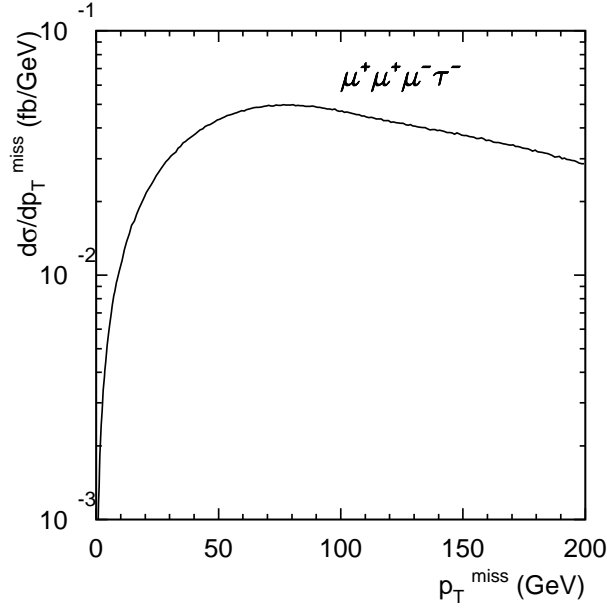}
\caption{ $\cancel{p}_T$ distribution in the channel $H^{++}H^{--}\rightarrow
\mu^+\mu^+\mu^-\tau^-$ with $\tau\to \ell\nu\bar \nu$ for $M_{H^{++}}=300$ GeV.}
\label{missing}
\end{figure}

 For signals with neutrinos only from $\tau$ decays, the $\cancel{p}_T$ spectrum will be softer.
 This is shown in Fig.~\ref{missing} for events of $\mu^+\mu^+\mu^-\tau^-$. Given the clean
 leptonic final state, we thus adjust the $\cancel{p}_T$ cut as
\begin{equation}
 \cancel{p}_T > 20 ~{\rm GeV}.
\end{equation}

\begin{figure}[tb]
\includegraphics[scale=1,width=9cm]{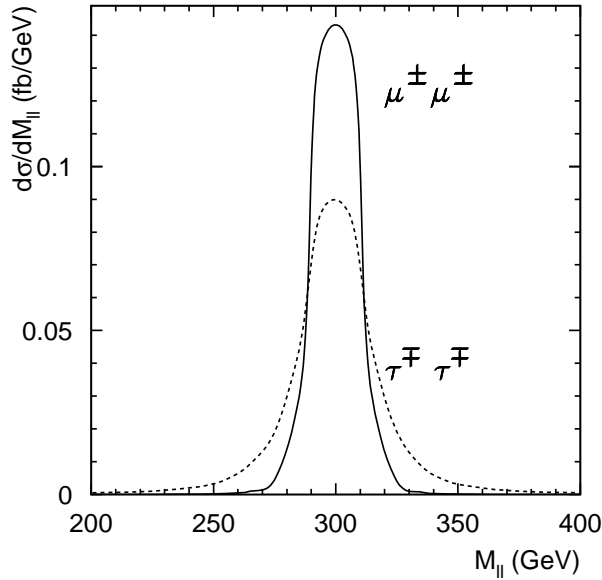}
\caption{Reconstructed invariant mass distributions for the like-sign $\mu\mu$ (solid)
and $\tau\tau$ (dotted) in $H^{++}H^{--}\rightarrow \mu^+\mu^+\ \tau^-\tau^-$
for $M_{H^{++}}=300$ GeV.}
\label{tautau}
\end{figure}

It is important to carefully consider the kinematical reconstruction of the events with $\tau$'s.
First of all, we note that all the $\tau$'s are very energetic, coming from the decay of a heavy Higgs
boson. For events with one $\tau$ and no other sources for missing particles,
the missing momentum will be
along the direction with the charged track. We thus have
\begin{equation}
{\vec{p}}\ ({\rm invisible}) =  k \vec{p}\ ({\rm track}),
\end{equation}
where the proportionality constant
$k$ is determined from the $\cancel{p}_T$ measurement by assigning
$\cancel{p}_T = k p_T({\rm track})$.
For events with two $\tau$'s, we generalize it to
\begin{equation}
{\vec{p}}\ ({\rm invisible}) =  k_1 \vec{p}\ ({\rm track}_1) + k_2 \vec{p}\ ({\rm track}_2).
\end{equation}
As long as the two $\tau$ tracks are not linearly dependent, $k_1$ and $k_2$
can be determined again from the $\cancel{p}_T$ measurement. The Higgs pair kinematics
is thus fully reconstructed.
In practice, we require that the invisible momenta pair with the two softer leptons
 to solve the combinatorics of the multiple charged leptons.
The Higgs masses reconstructed from the like-sign dileptons
are shown in Fig.~\ref{tautau} for the process
$H^{++}H^{--} \rightarrow \mu^+\mu^+\tau^-\tau^-$.
It is clear that the $\mu\mu$ mass reconstruction has a better resolution than the $\tau\tau$ pair.

One of the main features for the Higgs pair production is the equal heavy mass in the final
state, $M_{\ell^+\ell^+}=M_{\ell^-\ell^-}$ for the doubly charged Higgs production.
This serves as an important discriminator for the signal selection against the backgrounds.
This can also be used for momentum reconstruction with an additional $\tau$.
As long as we have less than 3 unknowns,
we will be able to determine the solutions. This extends the final states to contain
up to three $\tau$'s, such as $\ell^+\tau^+\ \tau^-\tau^-$~\cite{taurec}.

If the final state involves leptons plus one $\tau$ (e.g., $\ell^+\ell^+\ell^-\tau^-$)
with $\tau$ hadronic decay, the SM background will be $W^\pm Z+j$ and $W^\pm W^\pm W^\mp +j$.
As shown in last section, $W^\pm Z$ and $W^\pm W^\pm W^\mp$ is below 1 fb
after imposing $M_Z$ veto. With additional jet in final state and multiplied the
rate of jet fake hadronic $\tau$ which is $1\%$. It will be of the order
${\mathcal O}(10^{-3})$ fb and negligible. This remains true for events with two or more $\tau$s. For instance,
$\ell^+\ell^+\tau^-\tau^-$ may encounter $W^+W^+ jj$
background,  but the rate for both jets to fake hadronic $\tau$'s is $(1\%)^2$,
resulting in a background rate about $10^{-3}$ fb with basic cuts.
As for the other reducible background, the QCD  $t\bar{t}$ production, we expect that
the combination of the small fake rate of $b\to \ell, \tau$ and effective kinematical
cuts on $M_T, M_{\ell^\pm \ell^\pm}$ would be sufficient to bring the faked background
to a low level.

\begin{figure}[tb]
\includegraphics[scale=1,width=8cm]{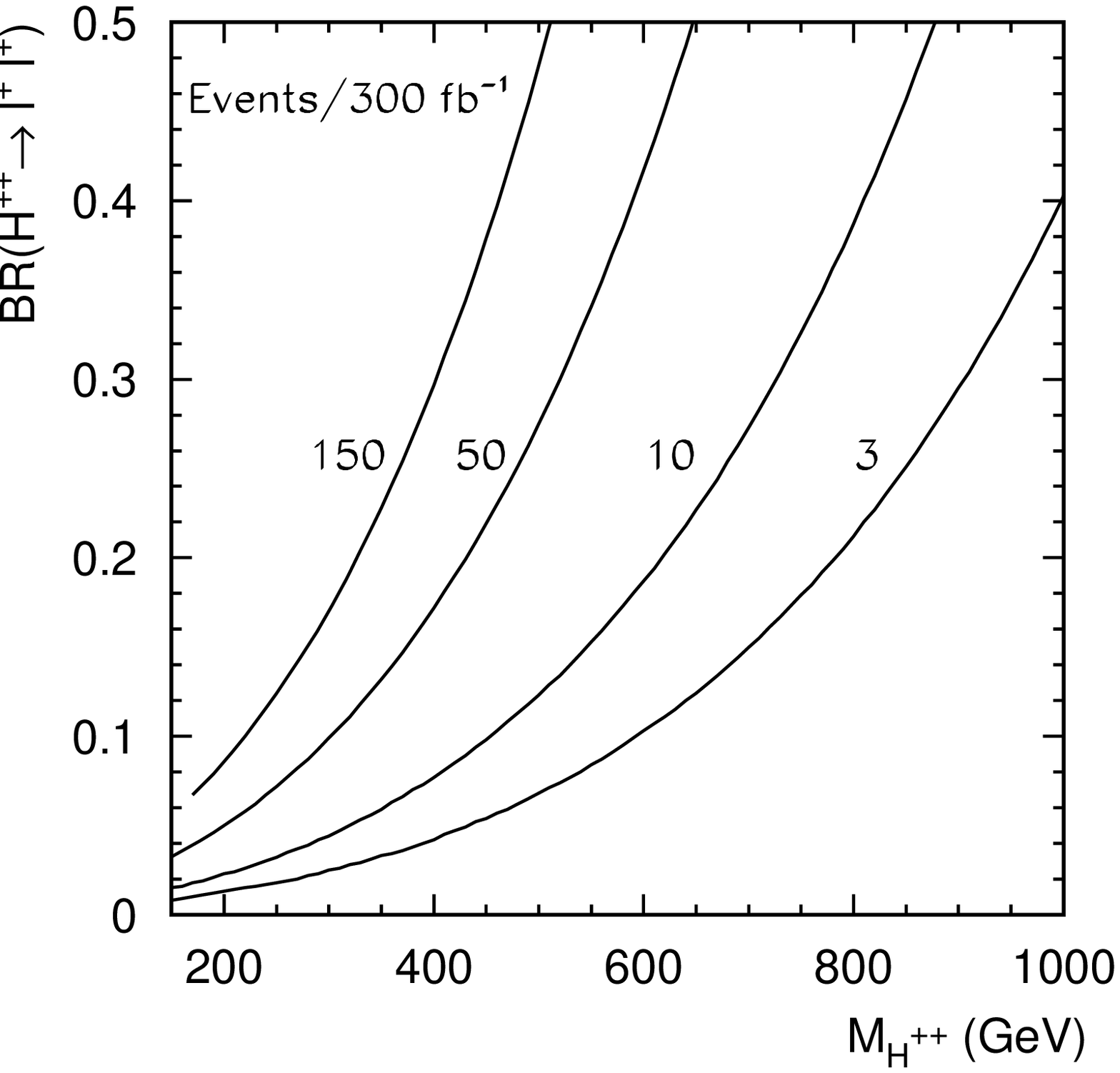}
\includegraphics[scale=1,width=8cm]{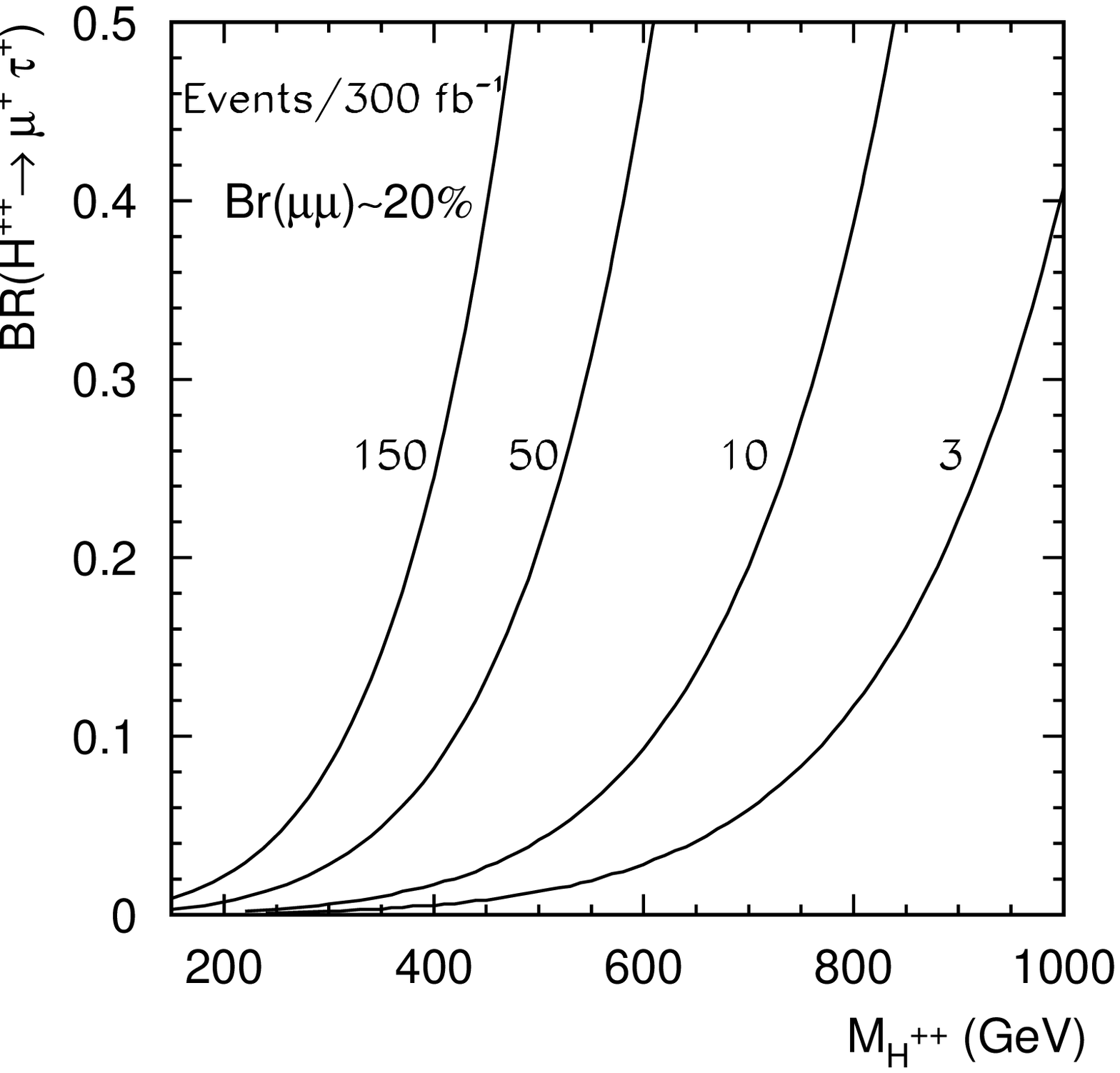}
\caption{Event contours in the BR$-M_{H^{++}}$ plane
for the doubly charged Higgs decay  at the LHC with an integrated luminosity
$300\ {\rm fb}^{-1}$  for $\mu^+\mu^+\mu^-\mu^-$ (left) and
 for $\mu^+\mu^+\mu^-\tau^-$ (right),
 assuming BR$(H^{++}\rightarrow \mu^+\mu^+)= 20\%$. }
\label{fig:BRll}
\end{figure}

\subsubsection{Measuring Branching Fractions and Probing the Neutrino Mass Pattern}
The direct correlation between leptonic branching fractions of triplet Higgs decay and realistic
light neutrino mass matrix is central for the Type II seesaw predictions.
Measuring the BR's of different flavor combinations becomes very crucial here.
For illustration, consider the cleanest channel with four muons first,
$H^{++}H^{--}\rightarrow \mu^+\mu^+\mu^-\mu^-$.
The event rate is written as
\begin{equation}
N_{4\mu} = { L} \times \sigma(pp\rightarrow H^{++}H^{--})\times {\rm BR}^2(H^{++}\rightarrow \mu^+\mu^+) ,
\label{br}
\end{equation}
where $ L$ is the integrated luminosity. Given a sufficient number of events
$N$, the mass of doubly charged Higgs boson is determined by the invariant mass
of the like-sign muons $M_{\mu^+\mu^+}$. We thus predict the
corresponding production rate $\sigma(pp\rightarrow H^{++}H^{--})$
for this given mass. The only unknown in the Eq.~(\ref{br}) is the decay branching fraction.

This procedure can be applicable for any channels that have been discussed for full
reconstruction earlier. In the Type II seesaw scheme,
we have ${\rm BR}(H^{++}\rightarrow \mu^+\mu^+) \sim 20-40\%$
for both NH and IH patterns as seen in Sec.~V.
Once we have measured this ${\rm BR}( \mu^+\mu^+) $, we can use it to
determine other channels, such
${\rm BR}(H^{++}\rightarrow \mu^+\tau^+,\  \tau^+\tau^+)$
and ${\rm BR}(H^{+}\rightarrow \tau^+\bar\nu)$.

With negligible SM backgrounds,
the only limitation would be the event rate that determines
the statistical error for the BR measurements, {\it i. e.}, a relative error
$1/\sqrt N$ if Gaussian statistics is applicable.
We present the event contours in the BR$-M_{H^{++}}$ plane in Fig.~\ref{fig:BRll}
for $300\  {\rm fb}^{-1}$.

To summarize our signal reconstruction in this section,
we list the leading reconstructable leptonic channels along with the branching
fractions in Tabel \ref{tab:llll}. We also
associate these channels  with predictions of the neutrino mass patterns.
These channels are not very sensitive to the Majorana phase $\Phi_2$,
and the maximal variation in the branching fractions can be up to a factor
of 2 in the case of NH. The sensitivity to $\Phi_1$ can be very significant
in the case of the IH. As for the case of quasi-degenerate spectrum, the Higgs
decay branching fractions for the three flavors of $e,\mu,\tau$ are equally
distributed as given in Table I, while the off-diagonal channels are negligibly small.

\begin{table}[tb]
\begin{tabular}{| c| c| c|}
\hline
Signal channels & Leading modes and BR range & Leading modes and BR range\\
 & Normal Hierarchy & Inverted Hierarchy \\
\hline
$H^{++} H^{--}$ & & \\
$\Phi_1=\Phi_2=0$
& $\mu^+\mu^+ \ \ \mu^- \mu^-\quad (20-40\%)^2$  & $e^+e^+\ \ e^-e^-\quad (50\%)^2$ \\
& $\mu^+\mu^+ \ \ \mu^- \tau^-\quad (20-40\%)\times 35\% $
&  $e^+e^+\ \ \mu^-\tau^-\quad 50\%\times 25\%$ \\
& $\mu^+\mu^+ \ \ \tau^- \tau^-\quad (20-40\%)^2$  &  $\mu^+\tau^+\ \ \mu^-\tau^-\quad (25\%)^2$ \\
& $\mu^+\tau^+ \ \ \mu^- \tau^-\quad (35\%)^2$  & \\
& $\mu^+\tau^+ \ \ \tau^- \tau^-\quad 35\%\times (20-40\%)$  & \\
$\Phi_1\approx \pi,\ \Phi_2=0$ & same as  above &
$ee, \mu \tau \to e\mu, e\tau~~~~ (30-60\%)^2$ \\
$\Phi_1= 0,\ \Phi_2\approx \pi$ &
$\mu\mu,\tau\tau: \times 1/2,\ \ \mu \tau: \times 2$ &  same as above  \\
\hline
$H^{\pm\pm} H^{\mp}$ &  &  \\
$\Phi_1=\Phi_2=0$
& $\mu^+\mu^+ \ \ \mu^- \nu\quad (20-40\%)\times(35- 60\%)$  &
$e^+e^+\ \ e^- \nu\quad (50\%)^2$ \\
& $\mu^+\mu^+ \ \ \tau^- \nu \quad (20-40\%)\times (35-60\%)$  &  \\
$\Phi_1\approx \pi,\ \Phi_2=0$ & same as  above &
$ee \to e\mu, e\tau~~~~ (30-60\%)\times 50\%$ \\
$\Phi_1= 0,\ \Phi_2\approx \pi$ &
$\mu\mu: \times 1/2$  &  same as above \\
\hline
\end{tabular}
\caption{
Leading fully reconstructable leptonic channels and the indicative ranges of their
branching fractions for $v_\Delta^{} \lsim 10^{-4}$ GeV.
The light neutrino mass patterns of the NH and IH, as well as vanishing
and large Majorana phases are compared. }
\label{tab:llll}
\end{table}

\subsection{Gauge Boson Decay Modes}

Although the triplet vev is constrained from above by the $\rho$-parameter
at the order of a GeV or so,
the pure gauge boson channel can still become dominant even for rather small
values of the triplet vev, {\it i.e.} $v_\Delta > {\cal O}(10^{-4}\ {\rm GeV})$,
especially for increasing the triplet mass. In this limit, the triplet Higgs bosons
will decay dominantly to the SM gauge boson pairs as discussed in the early sections.
Unfortunately, the absence of lepton number violation decays would prevent us
from extracting any information of neutrino mass patterns.
However, we would like to emphasize that the $\mu$-term in Eq.~(\ref{Potential})
has the identical gauge structure of the interactions as the Majorana mass generation
in Eq.~(\ref{Yukawa}). We therefore argue that confirmation of the existence of the
Higgs triplet  mixing with the SM doublets would strongly indicate the Majorana mass
generation to be at work.

\begin{figure}[tb]
\includegraphics[scale=1,width=9cm]{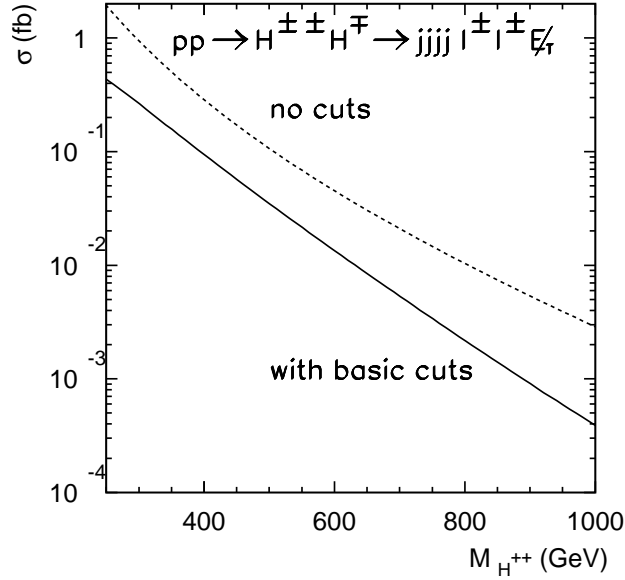}
\caption{ Total cross section for
$H^{\pm\pm} H^{\mp} \rightarrow jj b\bar{b} \ell^+\ell^+\cancel{E}_T$ at the LHC
versus the heavy Higgs mass before (dotted curve) and after the basic cuts (solid). }
\label{fig:ll}
\end{figure}

Collider searches for $pp\rightarrow H^{++}H^{--}\rightarrow W^+W^+W^-W^-$ has been
studied before \cite{last}. While the $W^\pm W^\pm$ channels are unique for the
signal identification, we would like to search for channels that confirm the mixing between
the Higgs triplet and the SM doublets. These include the decays
via the following channels directly proportional to $\mu$
\begin{equation}
H^{+}   \rightarrow  W^+ H_1,\ t\bar{b},\quad
H_2    \rightarrow H_1 H_1 ,\quad
A    \rightarrow H_1 Z,
\end{equation}
and those proportional to a combination of $\mu$ and $v^{}_\Delta$,
\begin{equation}
H^{+}   \rightarrow  W^+ Z ,\quad
H_2    \rightarrow W^+ W^-,\ Z Z.
\end{equation}

Both $H^{\pm} H_2$ and $H^{\pm\pm}H^{\mp}$ production channels are crucial to
test $SU(2)_L$ gauge coupling and confirm the triplet nature of the
Higgs fields.  However, it would be very challenging to study the channel
$H^{+}H_2\rightarrow W^+ H_1 H_1 H_1$,  which
consists  6 $b$-jets + $W^\pm$. The reconstruction of three light Higgs bosons
from the multiple $b$ jets would suffer from combinatorics, along with
the irreducible QCD backgrounds.
We will thus focus on $H^{\pm\pm}H^{\mp}$ for our study. We propose to reconstruct
the events by looking for two like-sign $W^\pm$'s from $H^{\pm\pm}$ decay through a pair of like-sign
dileptons; the $W^\mp$ in their hadronic decay modes and the SM-like
Higgs $H_1\rightarrow b\bar{b}$, both from $H^\mp$ decay,
\beq
pp \rightarrow H^{\pm\pm}H^\mp\rightarrow W^\pm W^\pm +W^\mp H_1/ W^\mp Z/ \bar{t}b(t\bar{b})\rightarrow
 jj\  b\bar{b}\  \ell^\pm\ell^\pm\cancel{E}_T.
\eeq
The decay branching fractions to final states are, respectively,
\beq
\br(W^\pm W^\pm, W^\mp H_1)\sim 2.2\%,\ \ \  \br(W^\pm W^\pm, W^\mp Z)\sim 2.3\%,\ \   \
\br(W^\pm W^\pm, \bar{t}b/t\bar{b})\sim 3.3\%.
\eeq
For a $M_{H_1}$ of $120 \ {\rm GeV}$,
the ${\rm BR}(H_1\rightarrow b\bar{b})$ is about $67.7\%$.
The decay branching fraction of the singly charged Higgs boson needs to
be included as given in Fig.~\ref{h22}(a).

We again start with some basic cuts. We demand
\bea
&& p_T(\ell) \geq  15\ {\rm GeV},\quad |\eta(\ell)| \leq 2.5,\quad \cancel{E}_T > 30 \ {\rm GeV},\\
&& p_T(j) \geq  25\ {\rm GeV},\quad |\eta(j)| \leq 3.0,\quad
\Delta R_{jj},\ \Delta R_{j\ell},\ \Delta R_{\ell\ell}>0.4.
\eea
The jet energies are also smeared using the same Gaussian formula as in Eq.~(\ref{ecal}),
but with  \cite{CMS}
\begin{equation}
a=100\%,\quad  b=5\%.
\end{equation}
We show the total cross section for the inclusive process
$H^{\pm\pm} H^{\mp} \rightarrow jj b\bar{b} \ell^\pm\ell^\pm\cancel{E}_T$
in Fig.~\ref{fig:ll} without any cuts (dotted curve) and after the basic cuts (solid curve).
We see that with the branching fractions included, the signal rate becomes rather low.

The leading irreducible background to our signal is
\be
pp \to t \bar{t} W^\pm \to jj\ b\bar b\ W^\pm W^\pm.
\ee
The QCD $jjjj+W^\pm W^\pm$ is much smaller. This is estimated based on the fact
that QCD $jjW^\pm W^\pm\rightarrow jj\ell^\pm\ell^\pm\cancel{E}_T$ is about 15 fb. With
an additional $\alpha^2_s$ and 6 body phase space suppression, it is much
smaller than $t\bar{t}W^\pm$.
To maximally retain the signal rate, we will not demand the $b$ tagging.
Instead, we tighten up the kinematical cuts
\be
p^{max}_T(\ell) > 50\ {\gev},\quad p^{max}_T(j)  > 100\ {\gev}.
\ee
Furthermore, for pair production of heavy particles like the two triplet Higgs bosons
of  several hundred GeV, the cluster mass of the system indicates the large
threshold. We define
\beq
M_{cluster} = \sqrt{M^2_{4j} + (\sum {\vec{p_T}^j})^2}  +
 \sqrt{M^2_{\ell\ell} + (\sum{\vec{p_T}^\ell})^2}  + {\cancel{E}}_T
\eeq
and will impose a high mass cut to select the signal events.
With $W^+ H_1$, $W^+ Z$, $t\bar{b}$ and  $W^+ W^+$ all decay hadronically,
we consider the mass reconstruction by the di-jets. We first impose a  cut
\begin{eqnarray}
|M^W_{jj}-M_W| <  15 \ {\rm GeV},
\end{eqnarray}
where $M^W_{jj}$ is the jet mass of six combinatorics that is closest to $M_W$.
The second reconstruction of $M_{jj}$ will give us the separation of $M_W,\ M_Z,$
or $M_{H_1}$.

\begin{figure}[tb]
\includegraphics[scale=1,width=9cm]{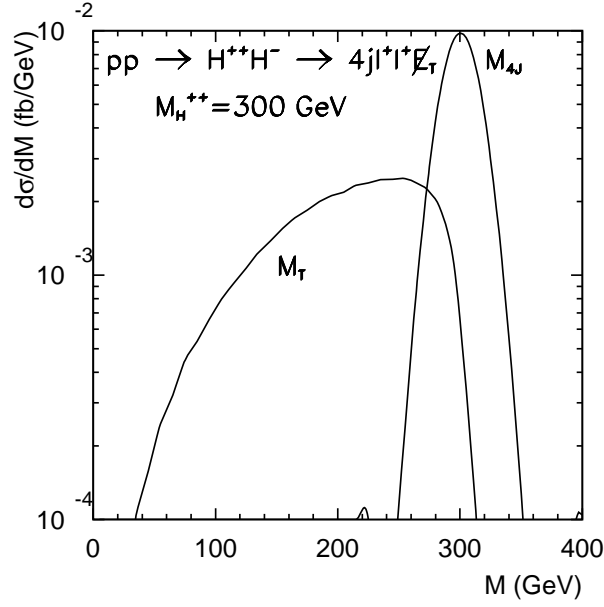}
\caption{ Reconstruction of triplet Higgs bosons via 4-jet invariant mass
$M_{jjjj}$ for $H^{\pm}$ and transverse mass $M_T$ for $H^{\pm\pm}$
with $M_{H^{++}}=300$ GeV. }
\label{fig:mass}
\end{figure}

The singly charged triplet $H^{\pm}$ decay has no missing
particles and we can fully reconstruct the $H^{\pm}$ by form a 4-jet
invariant mass $M_{jjjj}$. The doubly charged Higgs, on the other hand,
gives two like-sign dileptons plus large missing energy. We
define the leptonic transverse mass
\beq
M_T = \sqrt{ (\sqrt{M^2_{\ell\ell}  + (\sum {\vec{p_T}^\ell})^2} + \cancel{E}_T)^2  -
  (\sum{\vec{p_T}^\ell} + \vec{\cancel{E}}_T)^2}.
\eeq
These two variables are plotted in Fig.~\ref{fig:mass} for $M_{H^{++}}=300$ GeV.

In the leading background $t\bar{t}W$, there is another top quark that decays
leptonically. Taking the $b$-jet left over from the three jets of $m_t$ reconstruction,
we can construct two $M_{b\ell_1}$ and $M_{b\ell_2}$. If both $b$ and
$\ell$ come from the same top quark, there will be a strict constraint $M_{b\ell}<m_t$.
However, this cut will also reduce the signal by $70 \%$.
The wrong pair of $M_{b\ell}$ will be smaller in the $t \bar{t} W$
case since the $b$ and $\ell$ are both softer than the signal.
We impose a cut
\beq
M^{\rm max}_{b\ell} > 150~ {\rm GeV}.
\eeq

We show the effects of the cuts step by step in Table \ref{Tab:III} for both
the signal with $M_{H^{++}}=300$ GeV and the leading background  $t\bar{t}W$.
We combine the four decay channels in the table. We see that all the cuts
designed here are highly efficient in retaining the signal and suppressing
the background. One can reach a signal to background ratio of $2:1$ and
about 50 signal events/300 fb$^{-1}$.

\begin{table}[tb]
\begin{tabular}{| c || c | c |c | c| c| c | c| c | }
    \hline
     $\sigma$ (fb) & Basic & $p_T^\ell$ cut &  $p_T^j$ cut  & $M_{\rm Cluster}$  & $M_W$ rec.  & $M_X$ rec. & $M_T$ & $M_{jjjj}$  \\
cuts & Cuts & $>50~{\rm GeV}$ & $>100~{\rm GeV}$ & $>600~{\rm GeV}$  & $M_W\pm 15~{\rm GeV}$ & or $M_t$ veto & $<300~{\rm GeV}$ & $300\pm 50~{\rm GeV}$  \\
     \hline
      $t\bar{b}$ & 0.13  & 0.12 & 0.12 & 0.11 & 0.11 & $0.094^*$ & 0.094 & {\color{red}0.092} \\
\hline
      $WH$ & 0.074  & 0.069 & 0.065 & 0.061 & 0.06 & 0.046 & 0.045 & {\color{blue}0.045}  \\
\hline
      $WZ$ & 0.06  & 0.056 & 0.053 & 0.05 & 0.05 & 0.038 & 0.038 & {\color{green}0.038} \\
\hline
\hline
      $H^{\pm\pm}H^{\mp}$ sum  & 0.26 & 0.25 & 0.24 & 0.22 & 0.22 & 0.18 &  0.18&  0.17\\
\hline
\hline
      $H^{\pm\pm}H^{\mp\mp}$  & 0.24 & 0.23& 0.22 & 0.21 & 0.21 & 0.18 & 0.17 & 0.17 \\
\hline
\hline
      $t\bar t W$ & 3.1 & 2.5 & 1.8 & 1.4  & 1.4&  $0.88^*$ & 0.52 & {\color{red}0.095}  \\
&&&&& ($M_{H_1}$ rec.$\rightarrow$) &0.15& 0.097 & {\color{blue}0.045}\\
&&&&& ($M_Z$ rec.$\rightarrow$) &0.11& 0.071 & {\color{green}0.032}\\
&&&&& ($M_W$ rec.$\rightarrow$) &0.096&0.06 & 0.026\\
\hline
  \end{tabular}
  \caption{Production cross sections (in fb)  at the LHC for
$pp\rightarrow H^{\pm\pm}H^{\mp}\rightarrow W^{\pm} W^{\pm} W^{\mp} H_1 /W^{\pm} W^{\pm} W^{\mp} Z^0 \rightarrow jjjj+\ell^{\pm}\ell^{\pm} +\cancel{E}_T$ and $pp\rightarrow H^{++}H^{--}\rightarrow W^{+} W^{+} W^{-} W^{-}\rightarrow jjjj+\ell^{\pm}\ell^{\pm} +\cancel{E}_T$,
  and for the leading backgrounds.
  We take $M_{H^{\pm\pm}}=M_{H^\pm}=300$ GeV for illustration.
   The rates after imposing each selection criterion, as described in the text, are shown. }
\label{Tab:III}
\end{table}

For heavier Higgs bosons, the gauge boson decay modes of the singly charged Higgs
boson take over the $t\bar b$ mode. As an illustration, for $M_{H^+}=M_{H^{++}}=600~{\rm GeV}$,
the $H^{+}\rightarrow t\bar{b}$ is only $18\%$ so we don't include this channel.
Another important difference for a heavier Higgs boson is that
the $W$, $Z$, top and $H_1$ from $H^\pm$ decay become energetic
and their decay products will be highly collimated. The signal thus may look like
\be
 pp \rightarrow H^{\pm\pm}H^{\mp},\  H^{\pm\pm}H^{\mp\mp}
 \rightarrow W^{\pm} W^{\pm} JJ
 \rightarrow JJ+\ell^{\pm}\ell^{\pm} +\cancel{E}_T ,
\ee
where $J$ denotes a massive fat jet.

We note that the main source of the background is from $W^\pm W^\pm  +$QCD jets.
A light jet develops finite mass due to the QCD radiation and parton showering. Although
it is difficult to accurately quantify a jet mass, we parameterize
a jet mass as a function of its transverse energy $M_J\simeq 15\% E^J_T$, 
and require the jet mass to reconstruct $M_W$ and $M_X$ ($X=H_1,Z,W$).

The cross section for
 $jjW^+W^+$ is below ${\cal O}(10 ~{\rm fb})$ after some basic acceptance cuts.
The large jet mass cut will further reduce them. The results of the signal and
backgrounds are summarized in Table  \ref{Tab:IV} for
$M_{H^{\pm\pm}}=M_{H^\pm}=600\ {\rm GeV}$.
We see once again that the cuts are very efficient in retaining the signal and
the background can be suppressed to a negligible level. The difficulty is the
rather small signal rate to begin with, at the order of $5\times 10^{-2}$ fb.

\begin{table}[tb]
\begin{tabular}{| c || c | c |c | c| c| c |  }
    \hline
     $\sigma$ (fb) & Basic & $p_T^\ell$ cut &  $p_T^j$ cut  & $M_{J_1}$ rec.  & $M_{J_2}$ rec. & $M_{JJ}$  \\
cuts & Cuts & $>80~{\rm GeV}$ & $>200~{\rm GeV}$  & $M_W\pm 15~{\rm GeV}$ &  $M_X\pm 15~{\rm GeV}$
& $600\pm 75~{\rm GeV}$  \\
     \hline
\hline
      $WH$ & $1.1\times 10^{-2}$ & $9.5\times 10^{-3}$ & $9.5\times 10^{-3}$  & $9.4\times 10^{-3}$
 & $9.1\times 10^{-3}$  & $9.0\times 10^{-3}$  \\
\hline
      $WZ$ & $1.0\times 10^{-2}$  & $1.0\times 10^{-2}$ & $1.0\times 10^{-2}$  & $1.0\times 10^{-2}$
 & $9.9\times 10^{-3}$ & $9.8\times 10^{-3}$ \\
\hline
      $H^{\pm\pm}H^{\mp\mp}$  & $3.3\times 10^{-2}$ & $3.2\times 10^{-2}$& $3.1\times 10^{-2}$
 & $3.1\times 10^{-2}$ & $3.1\times 10^{-2}$ & $3.1\times 10^{-2}$\\
\hline
\hline
      $JJW^\pm W^\pm$ & 14.95  & 7. 65 & 4.69
 & 0.24  &  &   \\
&&&& ($M_{H_1}$ rec.$\rightarrow$) &$6\times 10^{-2}$ & $4.0\times 10^{-5}$\\
&&& &($M_Z$ rec.$\rightarrow$) & 0.13 & $1.4\times 10^{-4}$\\
&&& &($M_W$ rec.$\rightarrow$) & 0.1 & $1.6\times 10^{-4}$\\
\hline
  \end{tabular}
  \caption{Production cross sections (in fb)  at the LHC for
 $pp\rightarrow H^{\pm\pm}H^{\mp}\rightarrow W^{\pm} W^{\pm} W^{\mp} H_1 /W^{\pm} W^{\pm} W^{\mp} Z^0 \rightarrow JJ+\ell^{\pm}\ell^{\pm} +\cancel{E}_T$ and
 $pp\rightarrow H^{++}H^{--}\rightarrow W^{+} W^{+} W^{-} W^{-}
 \rightarrow JJ+\ell^{\pm}\ell^{\pm} +\cancel{E}_T$,
  and for the leading backgrounds.
  We take $M_{H^{\pm\pm}}=M_{H^\pm}=600$ GeV for illustration.
   The rates after imposing each selection criterion, as described in the text, are shown. }
\label{Tab:IV}
\end{table}

\section{Discussions and Conclusions}
\subsection{Discussion on Testing the Type II Seesaw Mechanism}
We have discussed the general properties of the Type II seesaw 
mechanism for neutrino masses where the Higgs sector of the 
Standard Model is extended by adding an $SU(2)_L$ Higgs 
triplet, $\Delta \sim (1,3,1)$. As is well-known, in this scenario 
the neutrino mass matrix is given by $M_\nu = \sqrt{2} \ Y_\nu  \ v_{\Delta}$, 
where  $v_{\Delta}$ is the vacuum expectation value (vev) of the 
neutral component of the triplet and $Y_\nu$ is the Yukawa coupling. 
Once the electroweak symmetry is broken 
$v_{\Delta}= \mu \ v_0^2/ \sqrt{2} \ M_{\Delta}^2$, where the 
dimension parameter $\mu$ defines the doublet-triplet mixing 
and $M_{\Delta}$ is the mass of the triplet. In the standard
``high-scale'' seesaw mechanism assuming $Y_\nu \approx 1$ 
and $\mu \sim M_{\Delta} \approx 10^{14-15}$ GeV one obtains 
the natural value for neutrino masses $m_\nu \approx 1$ eV. 
However, even if it is a natural scenario in this case one cannot 
hope to realize the direct test of the mechanism at future colliders. 
In this work we have focused on the possibility to observe at the 
LHC the fields responsible for the Type II seesaw mechanism. 
In this case assuming  $M_{\Delta} \lesssim 1$ TeV one finds that 
$Y_\nu  \times \mu \lesssim 1.7 \times 10^{-8}$ GeV. 
Therefore, if one assumes $Y_\nu \approx 1$, $\mu \approx 10^{-8}$ 
GeV and one can think about the $\mu$ term as a soft-breaking 
term of the global $U(1)_L$ (or $U(1)_{B-L}$) symmetry. 
Since this possibility is appealing and there is hope to test 
the mechanism at the LHC we have laid out the general properties 
of the Higgs bosons for both their leptonic decays and gauge boson
modes. We have also explored the sensitivity to search for those signals
at the LHC. We now outline our general proposal in order to
convincingly test the Type II seesaw mechanism.

We need the following necessary steps. First, the theory
must account for the experimentally measured values of
light neutrino masses and mixing angles, and then predict
the physical couplings of the doubly and singly charged
Higgs bosons. This was accomplished in Sec.~III.

\begin{figure}[tb]
\includegraphics[scale=1,width=8.0cm]{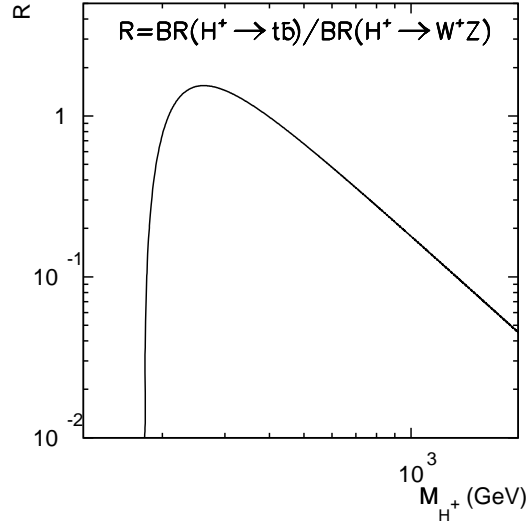}
\caption{Ratio between BR$(H^+ \to t \bar{b})$ and BR$(H^+ \to W^+ Z)$ versus $M^{H^+}$.}
\label{ratio}
\end{figure}

We need to establish the existence of the charged Higgs bosons
and further confirm the Higgs triplet nature.  This can be accomplished
by observing the associated production of the singly and doubly charged Higgs
bosons $H^{\pm\pm} H^\mp$. We wish to utilize the physics reach
at the LHC for this purpose, so we limit ourself  to the triplet
mass in the range
\be
110 \ \text{GeV} \lsim M_\Delta \lsim 1 \ \text{TeV},
\ee
where the lower limit comes from the direct experimental bound,
and the upper limit is roughly the LHC reach.
With our minimal model assumption, the only other crucial parameter,
the triplet vev $v_{\Delta}$, determines the Higgs phenomenology.
There are three typical regions which characterize the
different searching strategies.
\begin{itemize}
\item $\underline{{ 1 \ \text{eV} \lsim \ v_{\Delta} < \ 10^{-4} \ \text{GeV}}}$:
In this case the leading decays of the charged Higgs bosons are
$H^{++} \to e_i^+ e_j^+$ and $H^+ \to e_i^+ \bar{\nu}$.
There are in total
six lepton number violating channels for the doubly charged Higgs,
and three channels for the singly charged Higgs.
We thus expect  to test the theory once we discover the doubly and singly charged
Higgs bosons and determine their branching fractions of different flavor combinations,
in accordance with the model predictions in the Type II seesaw scheme as presented
in Table III.
\item $\underline{{ v_{\Delta} \approx 10^{-4} \ \text{GeV}}}$:
In this situation, $H^{++} \to e^+_i e^+_j$ and $H^{++} \to W^+ W^+$, 
as well as $H^+ \to e_i^+ \bar{\nu}$ and $H^+ \to W^+ H_1, \ W^+ Z, \ t \bar{b}$ 
are all comparable. One may thus wish to observe not only the clean dilepton 
signals of lepton number violation, but also the gauge boson pairs 
or $t\bar b$. The simultaneous observation of both channels will give 
a direct measurement for $\vd$.

\item  $\underline{ { 10^{-4} \ \text{GeV} \ < \ v_{\Delta} \lsim \ 1 \ \text{GeV}}}$:
In this case the lepton number violating Higgs decays are suppressed.
One then must confirm its mixing with SM doublets.
Through the decays of $H^+ \to t \bar{b}$ and $H^+ \to W^+ H_1$,
one can extract the $\mu$ parameter which
defines the key relation for seesaw scheme
$v_{\Delta}=\mu v_0^2 / \sqrt{2} M_{\Delta}^2$ since
\bea
&& \Gamma(H^+ \to W^+ H_1) \sim { \mu^2 \over M_{H^+}}, \quad
\Gamma(H^+ \to t \bar{b}) \sim  {\mu^2 m_t^2 \over M_{H^+}^3},
\eea
and
\bea
&& \Gamma(H^+ \to W^+ Z) \sim
\left( g_1^2 \ \frac{\mu \ v_0^2}{M_{\Delta}^2} - \sqrt{2} ( 2 g_1^2 \ + \ g_2^2 ) \ v_{\Delta} \right)^2 \
{M_{H^+}^3\over v_0^4}.
\eea
In Fig.~\ref{ratio}  the ratio between BR$(H^+ \to t \bar{b})$ and BR$(H^+ \to W^+ Z)$
is shown which can be predicted once one uses the seesaw relation.
The decay $H^+ \to t \bar{b}$ is dominant at low mass, and
 $H^+ \to W^+ Z$  takes over for a heavier mass. Both channels should be
 searched for and they are complementary.
\end{itemize}

\subsection{Conclusions}
The possibility to test one of the most appealing mechanisms for neutrino mass generation,
the so-called Type II seesaw mechanism, at the Large Hadron Collider has been investigated.
We first emphasize the importance to observe the associated production $H^{\pm\pm} H^{\mp}$
to establish the gauge triplet nature of the Higgs field.
We have  found very encouraging results for further testing the theory.

In the optimistic scenarios,
$1 \ \text{eV} \lsim v_{\Delta} < 10^{-4}$ GeV, one can test  this theory
to a great detail by looking for the clear signals of lepton number violation in
the decays of  doubly and singly charged Higgs bosons, at the LHC up to
a mass about 1 TeV.
\begin{itemize}
\item Observing the difference in rate by comparing the decay channels for
 $H^{++} \to  \mu^+ \mu^+,\ \mu^+ \tau^+,\  \tau^+ \tau^+$
 and $H^{++} \to e^+ e^+,\ \mu^+ \tau^+$, one could distinguish
between the Normal Hierarchy and Inverted Hierarchy for the light neutrino
mass spectrum,  when the effect of the Majorana phases is not appreciable.
\item If the Majorana phases play an important role, then the decay channels
of $H^{++}$ are less predicable. However, it is still possible to distinguish the neutrino spectrum by using
the singly charged Higgs decay $H^+ \to e^+_i \bar{\nu}\ (e_i=e,\mu,\tau)$,
which are independent of the Majorana phases. For a special case in IH, the
significant changes in decay rate for the doubly charged Higgs
$e^+ e^+, \mu^+\tau^+ \leftrightarrow e^+ \mu^+, e^+\tau^+$ will probe the
phase $\Phi_1$.
\end{itemize}

In the least favorable region of the parameter space,
$ v_{\Delta} > 10^{-4}$ GeV, where the lepton number violating
processes are suppressed, we need to study the decays to SM gauge boson
pairs or heavy quarks. Using the decays $H^+ \to t \bar{b}$ and $H^+ \to W^+ H_1$
one could extract the $\mu$ parameter which defines the mixing between the SM
Higgs doublet and the triplet, which in turn implies the existence of the same
gauge interaction between the lepton doublet and the Higgs triplet.
Therefore, we can check the seesaw relation
$v_{\Delta}=\mu v_0^2 / \sqrt{2} M_{\Delta}^2$
and the prediction for $H^+ \to W^+ Z$.

In the most optimistic situation, $v_{\Delta} \sim10^{-4}$ GeV, both channels of the
lepton pairs and gauge boson pairs may be available simultaneously. The determination
of their relative branching fractions would give a measurement for the value of $v_{\Delta}$.

\subsection*{Acknowledgment}
We would like to thank Goran Senjanovi\'c and Lian-Tao Wang for discussions.
The work of P. F. P. was supported in part by the U.S. Department of Energy
contract No. DE-FG02-08ER41531 and in part by the Wisconsin Alumni
Research Foundation. The work of T. H., G. H. and K. W. is supported in part by the U.S.
Department of Energy under grant No. DE-FG02-95ER40896, and by the Wisconsin
Alumni Research Foundation.
The research at the KITP was supported in part by the National Science Foundation 
under Grant No. PHY05-51164. T.L.~would like to thank the Ministry 
of Education of China for support and would also like to acknowledge 
the hospitality of the Phenomenology Institute, University of Wisconsin-Madison while the
work was carried out.

\section*{APPENDIX A: TYPE II SEESAW AND FEYNMANN RULES}
As we have discussed in the previous sections the Type II seesaw
mechanism~\cite{TypeII} is one of the most appealing scenarios
for the generation of neutrino masses. In this appendix  we discuss
in detail this mechanism. In this extension of the Standard Model
the Higgs sector is composed of the SM Higgs, $H \sim (1,2,1/2)$,
and a complex triplet, $\Delta \sim (1,3,1)$:
\begin{equation}
H = \left( \begin{array} {c}
  \phi^+ \\
 \phi^0
\end{array} \right) ,~~ \text{and} ~~
\Delta = \left( \begin{array} {cc}
 \delta^+/\sqrt{2}  &  \delta^{++} \\
 \delta^0 & - \delta^+/\sqrt{2}
\end{array} \right)
\end{equation}
The kinetic terms and relevant interactions in this theory are given in Eq.~(\ref{Lagrangian})
and the new interactions for the leptons read as
\begin{eqnarray}
{\cal L}_{Y} & = & - Y_\nu \ l_L^T \ C \ i \sigma_2 \ \Delta \ l_L \ + \ h.c. \nonumber \\
& = & - Y_\nu \ \nu^T_L \ C \ \delta^0 \ \nu_L \ + \ \sqrt{2} \ Y_\nu \ \nu^T_L \ C \ \delta^+ \ e_L \ + \
Y_\nu \ e^T_L \ C \ \delta^{++} \ e_L \ + \ \text{h.c.}
\label{Yukawa1}
\end{eqnarray}
The scalar potential for $H$ and $\Delta$ is given in Eq. (\ref{Potential}).
The simultaneous presence of the Yukawa coupling in Eq.~(\ref{Yukawa1})
and the trilinear term proportional to the $\mu$ parameter in Eq.~(\ref{Potential}) tell us
that the lepton number or $U(1)_L$ is explicitly broken.

Imposing the conditions of global minimum one finds that
\begin{eqnarray}
&& - m_H^2 \ + \ \frac{\lambda}{4} v_0^2 \ - \ \sqrt{2} \ {\mu} \ v_{\Delta}= 0,\quad
v_{\Delta} = \frac{\mu \ v_0^2}{ \sqrt{2} \ M_{\Delta}^2},\quad {\rm and}\quad
{\lambda} M_\Delta^2 -4\mu^2 > 0,
\end{eqnarray}
where $v_0$ and $v_\Delta$ are the vacuum expectation values (vev)
of the Higgs doublet and triplet, respectively, with
$v_0^2 + v_\Delta^2 \approx (246\ {\rm GeV})^2$.
Once the neutral component in $\Delta$ gets
the vev, $\langle \delta^0 \rangle =v_{\Delta}/\sqrt{2}$, the neutrino mass matrix is given
by $M_{\nu}= \sqrt{2} \ Y_\nu \ v_{\Delta}$.

\subsection*{Higgs boson spectrum and gauge interactions}

Let us compute the spectrum of the different Higgses present in the theory. Using
\beq \phi^0 = ( v_0 \ + \ h^0 \ + i \xi^0 )/ \sqrt{2}, ~~ \text{and}~~
\delta^0 = ( v_{\Delta} \ + \ \Delta^0 \ + i \eta^0)/\sqrt{2}
\eeq
one finds that the mass matrix and the mixing angle for the CP-even states read as
\begin{eqnarray}
&& {\cal M}_{\text{even}}^2 = \left( \begin{array} {cc}
 \lambda \ v_0^2 /2     &  - \sqrt{2} \ \mu \ v_0  \\
 - \sqrt{2} \ \mu \ v_0  & M_{\Delta}^2
\end{array} \right) \quad \text{and} \quad \tan 2 \theta_0 = - \frac{4 \ M_{\Delta}^2 \ v_{\Delta}}{v_0 \left( M_{H_1}^2 \ + \ M_{H_2}^2 \ - \ 2 M_{\Delta}^2 \right)},\\
&&
H_1 = \cos \theta_0 \ h^0 \ + \ \sin \theta_0 \ \Delta^0,\quad
H_2 = - \sin \theta_0 \ h^0 \ + \ \cos \theta_0 \ \Delta^0.
\end{eqnarray}
where $\sqrt{v_0^2+v_\Delta^2} \approx 246$ GeV.
The mass matrix and the mixing angle for the CP-odd states are given by:
\begin{eqnarray}
&& {\cal M}_{\text{odd}}^2=\left( \begin{array} {cc}
 2 \sqrt{2} \ \mu \ v_\Delta      &  - \sqrt{2} \ \mu \ v_0  \\
 - \sqrt{2} \ \mu \ v_0  & M_{\Delta}^2
\end{array} \right) \quad \text{and} \quad \tan 2 \alpha = - \frac{4 \ M_{\Delta}^2 \ v_{\Delta}}
{v_0 \left( M_{A}^2 - 2 M_{\Delta}^2\right)},\\
&&
G= \cos \alpha \ \xi^0 \ + \sin \alpha \ \eta^0, \quad
A = - \sin \alpha \ \xi^0 \ + \ \cos \alpha \ \eta^0.
\end{eqnarray}
In the singly charged Higgs sector
$(\phi^+, \delta^+)$, the mass matrix and the mixing angle for the physical states read as
\begin{eqnarray}
&& {\cal M}_{\pm}^2 = \left( \begin{array} {cc}
 \sqrt{2} \mu \ v_\Delta      &  - \mu \ v_0  \\
 - \mu \ v_0  & M_{\Delta}^2
\end{array} \right) \qquad \text{and} \qquad \tan 2 \theta_+= - \frac{ 2 \sqrt{2} \ v_{\Delta} \ M_{\Delta}^2}
{v_0 \left( M_{H^+}^2 - 2 M_{\Delta}^2 \right)},\\
&&
G^{\pm}= \cos \theta_{\pm} \ \phi^{\pm} \
+ \sin \theta^{\pm} \ \delta^{\pm}, \quad
H^{\pm} = - \sin \theta_{\pm} \ \phi^{\pm} \ + \ \cos \theta_{\pm} \ \delta^{\pm}.
\end{eqnarray}

There are thus seven  physical mass eigenstates:
$H_1$  (SM-Like), $H_2$ ($\Delta$-Like), $A$, $H^{\pm}$, and $H^{\pm \pm}=\delta^{\pm \pm}$.
In this minimal setting,
\beq
M_{H_2}\simeq M_{A} \simeq M_{H^{\pm}} \simeq M_{H^{\pm\pm}}=M_{\Delta}.
\eeq
The mixing angles in all sectors are very small since $v_{\Delta} \ll v_0$.
It is thus useful to write down the approximations
\bea
\tan 2 \theta_0 \approx {-2\sqrt{2}\mu v_0\over M_{H_1}^2-M_\Delta^2}\approx
4{v_\Delta\over v_0},\quad
\tan 2 \alpha \approx {2\sqrt{2}\mu v_0\over M_\Delta^2}=4{v_\Delta\over v_0},\quad
\tan 2 \theta_+ \approx {2\mu v_0\over M_\Delta^2}= 2\sqrt{2}{v_\Delta\over v_0}.
\eea

The Feynmann rules for the Higgs boson gauge interactions are listed
in Table \ref{rules}.
\begin{center}
\begin{table}[tb]
\begin{tabular}[t]{|c|c|c|}
\hline
 Vertices & Gauge Couplings  & Approximation
\\
\hline
 $H^{++}H^-W^-_\mu$ & $- ig_2\cos\theta_+(p_1-p_2)_\mu$ &
$- ig_2(p_1-p_2)_\mu$\\
      $H^{++}W^-_\mu W^-_\nu$ &  $i\sqrt{2}g^2_2 v_{\Delta} g_{\mu\nu}$   &
$i\sqrt{2}g^2_2 v_{\Delta} g_{\mu\nu}$\\
\hline
 $H^+H_2W^-_\mu$ & $-i{g_2\over
  2}(\sqrt{2}\cos\theta_+\cos\theta_0+\sin\theta_0\sin\theta_+)(p_1-p_2)_\mu$ &
$- i{g_2\over \sqrt{2}}(p_1-p_2)_\mu$ \\
           $H^+AW^-_\mu$ & ${g_2\over
2}(\sqrt{2}\cos\theta_+\cos\alpha+\sin\alpha\sin\theta_+)(p_1-p_2)_\mu$ &
          ${g_2\over \sqrt{2}}(p_1-p_2)_\mu$
\\
          $H^+H_1W^-_\mu$ & $- i{g_2\over
2}(\sqrt{2}\cos\theta_+\sin\theta_0-\cos\theta_0\sin\theta_+)(p_1-p_2)_\mu$
& $-i{g_2\over 2} { \mu v_0  \over M_\Delta^2} (p_1-p_2)_\mu$
\\
     $H^+Z_\mu W^-_\nu$ & $i {\cos\theta_W\over 2}(g^2_1\sin\theta_+
v_0-\sqrt{2}\cos\theta_+(2g^2_1+g^2_2)v_{\Delta}) g_{\mu\nu}$
& $i {g_2^2\sin^2\theta_W\over 2\cos\theta_W}({\mu v_0^2\over
M_\Delta^2}-\sqrt{2}(2+\cot^2\theta_W)v_{\Delta}) g_{\mu\nu}$  \\
\hline
 $H_2H_1H_1$ & $i{1\over 4}\cos\theta_0(3\sin2\theta_0\lambda v_0
+4\sqrt{2}\cos2\theta_0 \mu - 4\sqrt{2}\sin^2\theta_0\mu)$
& $i(\sqrt{2}\mu+{3\over 2}\lambda {\sqrt{2}\mu v_0^2\over M_\Delta^2})$
\\
     $H_2W^+_\mu W^-_\nu$ & $-i{1\over 2} g^2_2 (\sin\theta_0 v_0
-2\cos\theta_0 v_{\Delta}) g_{\mu\nu}$ & $-i{1\over 2} g^2_2({\sqrt{2}\mu
v_0^2\over M_\Delta^2}-2v_\Delta) g_{\mu\nu}$
\\
   $H_2Z_\mu Z_\nu$ & $-i{1\over 2}{g^2_2\over \cos^2\theta_W}(\sin\theta_0
v_0 -4\cos\theta_0 v_{\Delta})g_{\mu\nu} $
& $-i{1\over 2}{g^2_2\over \cos^2\theta_W}
({\sqrt{2}\mu v_0^2\over M_\Delta^2}-4v_{\Delta})g_{\mu\nu}$
\\
\hline
 $AH_1Z_\mu$ & ${g_2\over
2\cos\theta_W}(\cos\theta_0\sin\alpha-2\cos\alpha\sin\theta_0)(p_1-p_2)_\mu$
&  $-{g_2\over \sqrt{2}\cos\theta_W} {\mu v_0 \over M_\Delta^2}(p_1-p_2)_\mu$
\\
\hline
\end{tabular}
\caption{Feynman rules for the heavy Higgs boson gauge Interactions. The momenta
are all assumed to be incoming, and $p_1\ (p_2)$ refers to the first (second) scalar field
listed in the vertices.
The approximation is based on $v_0\gg v_\Delta^{},\ M_\Delta > M_{H_1}$.}
\label{rules}
\end{table}
\end{center}

\subsection*{Heavy Higgs boson Yukawa interactions via mixing}
The triplet fields mix with the Higgs doublet via the dimensional parameter $\mu$ .
Thus the Standard Model Yukawa
interactions will yield the heavy Higgs couplings to the SM fermions. The Feynman
rules are listed in Table \ref{rules1}.

\begin{center}
\begin{table}[tb]
\begin{tabular}[t]{|c|c|c|}
\hline
 Vertices & Yukawa Couplings  & Approximation
\\
\hline
$H^+\bar{t}b$ & $-i\sqrt{2}{m_tP_L+m_bP_R\over v_0}\sin\theta_+$ &
$-i\sqrt{2}{m_t \mu \over M_\Delta^2} P_L $
\\
\hline
 $H_2 f\bar{f}$ & $-i {m_f\over v_0}\sin\theta_0$ &
$-i\sqrt{2}{m_f \mu \over M_\Delta^2}$
\\
\hline
 $A f\bar{f}$ & $\gamma_5{m_f\over v_0}\sin\alpha$ &
$\sqrt{2}\gamma_5{m_f \mu \over M_\Delta^2}$
\\
\hline
\end{tabular}
\caption{Feynman rules for the heavy Higgs boson Yukawa Interactions
via mixing $\mu$. }
\label{rules1}
\end{table}
\end{center}

\subsection*{Heavy Higgs boson $\Delta L=2$ Yukawa interactions}

The physical interactions
in the Yukawa sector are given in Eq.~(\ref{Physical-Yukawa}) and Eq.~(\ref{gamma}).
We present the Yukawa couplings for lepton number violating vertices in Table~\ref{rules2}.
\begin{center}
\begin{table}[tb]
\begin{tabular}[t]{|c|c|c|c|}
\hline
Fields & Vertices & Yukawa Couplings  & Approximation
\\
\hline
$H^{++}$ & $H^{++}e^{-T}_i e^-_j$ & $C2\Gamma^{ij}_{++} P_L$ &
$C2\Gamma^{ij}_{++} P_L$
\\
\hline
$H^+$ & $H^+ \nu_i^T e^-_j $ & $C\Gamma_+^{ij} P_L$ & $C\Gamma_+^{ij} P_L$\\
\hline
$H_2$ & $H_2 \nu_i^T \nu_i\ (\bar{\nu}_i\bar{\nu}_i)$ & $C\cos\theta_0\ (m_{\nu_i}/v_\Delta)  \ P_L$ &
$C (m_{\nu_i}/v_\Delta) \ P_L$
\\
\hline
$A$ & $A \nu_i^T \nu_i \ (\bar{\nu}_i\bar{\nu}_i)$ & $C\cos\alpha\ (m_{\nu_i}/v_\Delta) \ P_L$ &
$C (m_{\nu_i}/v_\Delta) \ P_L$
\\
\hline
\end{tabular}
\caption{Yukawa Interactions for lepton number violating vertices.}
\label{rules2}
\end{table}
\end{center}

The explicit couplings in terms of the neutrino mass and mixing parameters
are as follows:
\begin{eqnarray}
\Gamma_+  =  \cos \theta_+ \ \frac{ m_{\nu}^{diag} \ V_{PMNS}^\dagger}{v_{\Delta}}, \quad \text{and} \quad
\Gamma_{++} =  \frac{V_{PMNS}^* \ m_{\nu}^{diag} \ V_{PMNS}^\dagger}{\sqrt{2} \ v_{\Delta}}
\equiv Y_\nu,
\end{eqnarray}
and in the text we have defined squared sum relevant for the singly charged Higgs
processes
\begin{eqnarray}
Y_+^j  =  \sum_{i=1}^3 |\Gamma_+^{ij}|^2 \times v_{\Delta}^2,
\end{eqnarray}
where
\begin{eqnarray}
   \non
Y_+^{1}&=&m_1^2 c_{12}^2 c_{13}^2 +m_2^2 c_{13}^2  s_{12}^2 +m_3^2 s_{13}^2
\\ \non
Y_+^{2}&=&\left(c_{23}^2 m_2^2+m_1^2 s_{13}^2 s_{23}^2\right) c_{12}^2+2 \cos (\delta )
   c_{23} \left(m_1^2-m_2^2\right) s_{12} s_{13} s_{23} c_{12}
\\ \non
      &+& c_{23}^2 m_1^2 s_{12}^2+\left(c_{13}^2 m_3^2+m_2^2 s_{12}^2 s_{13}^2\right) s_{23}^2
\\ \non
Y_+^{3}&=&\left(c_{23}^2 m_1^2 s_{13}^2+m_2^2 s_{23}^2\right) c_{12}^2-2 \cos (\delta )
   c_{23} \left(m_1^2-m_2^2\right) s_{12} s_{13} s_{23} c_{12}
\\ \non
      &+& c_{13}^2 c_{23}^2 m_3^2+s_{12}^2 \left(c_{23}^2 m_2^2 s_{13}^2+m_1^2 s_{23}^2\right)
\end{eqnarray}
\begin{eqnarray}
   \non
\sqrt{2} v_{\Delta} \Gamma_{++}^{11} &=&
 m_1 e^{-i \Phi_1} c_{12}^2 c_{13}^2
+m_2 s_{12}^2 c_{13}^2
+m_3 e^{i(2 \delta -\Phi_2)} s_{13}^2
\\ \non
\sqrt{2} v_{\Delta} \Gamma_{++}^{22} &=&
 m_1 e^{-i \Phi_1} \left(-s_{12} c_{23} -e^{-i \delta } c_{12} s_{13} s_{23}\right){}^2
+m_2 \left(c_{12} c_{23}-e^{-i \delta } s_{12} s_{13} s_{23}\right){}^2
+m_3e^{-i \Phi_2} c_{13}^2 s_{23}^2
\\ \non
\sqrt{2} v_{\Delta} \Gamma_{++}^{33} &=&
 m_1 e^{-i \Phi_1} \left(s_{12} s_{23}-e^{-i\delta } c_{12} s_{13} c_{23}\right){}^2
+m_2 \left(-c_{12} s_{23}-e^{-i \delta } s_{12} s_{13} c_{23}\right){}^2
+m_3 e^{-i \Phi_2} c_{13}^2 c_{23}^2
\\ \non
\sqrt{2} v_{\Delta} \Gamma_{++}^{12} &=&
 m_1 e^{-i \Phi_1} c_{12} c_{13} \left(-s_{12} c_{23}-e^{-i \delta } c_{12} s_{13} s_{23}\right)
+m_2 s_{12} c_{13} \left(c_{12} c_{23}-e^{-i \delta } s_{12} s_{13} s_{23}\right)
\\ \non
&+&m_3 e^{i (\delta -\Phi_2)} s_{13} c_{13} s_{23}
\\ \non
\sqrt{2} v_{\Delta} \Gamma_{++}^{13} &=&
 m_1 e^{-i \Phi_1} c_{12} c_{13} \left(s_{12} s_{23}-e^{-i \delta } c_{12} c_{23} s_{13}\right)
+m_2 c_{13} s_{12} \left(-c_{12} s_{23}-e^{-i\delta } s_{12} s_{13} c_{23}\right)
\\ \non
&+&m_3 e^{i (\delta -\Phi_2)} s_{13} c_{13} c_{23}
\\ \non
\sqrt{2} v_{\Delta} \Gamma_{++}^{23} &=&
 m_1 e^{-i \Phi_1} \left(s_{12} s_{23}-e^{-i \delta } c_{12} s_{13} c_{23}\right)
 \left(-s_{12} c_{23}-e^{-i \delta } c_{12} s_{13} s_{23}\right)
\\ \non
&+&m_2 \left(-c_{12} s_{23}-e^{-i \delta } s_{12} s_{13} c_{23}\right)
 \left(c_{12} c_{23}-e^{-i \delta } s_{12} s_{13} s_{23}\right)
+m_3 e^{-i \Phi_2} c_{13}^2 s_{23} c_{23}
\end{eqnarray}

\section*{APPENDIX B: DECAYS OF $H^{++}$, $H^+$, $H_2$ AND $A$}
The expressions for the relevant partial decay widths are the following:
\begin{flushleft}
\underline{Doubly Charged Higgs}:
\end{flushleft}
\begin{eqnarray}
\Gamma (H^{++}\rightarrow e^+_i e^+_j) &=&
\frac{1}{4\pi (1+\delta_{ij})}|\Gamma^{ij}_{++}|^2 M_{H^{++}}
\\
\Gamma (H^{++}\rightarrow W^+_T W^+_T) &=&
{g_2^4 v^2_{\Delta} \over 8 \pi M_{H^{++}}} \lambda^{\frac{1}{2}}(1,r^2_W,r^2_W)
\approx {g_2^2 M_W^2 v_{\Delta}^2\over 2 \pi M_{H^{++}} v_0^2}
\\
\Gamma (H^{++}\rightarrow W^+_L W^+_L) &=&
{g_2^4 v^2_{\Delta} \over 16 \pi M_{H^{++}}} \lambda^{\frac{1}{2}}(1,r^2_W,r^2_W)\frac{(1 - 2r_W^2)^2}{4r_W^4}
\approx {M_{H^{++}}^3 v_{\Delta}^2\over 4\pi v_0^4}
\\
\Gamma(H^{++}\to H^+ \pi^+)&=&{g^4_2 V_{ud}^2 \Delta M^3 f^2_\pi\over 16\pi
M_W^4}
\\
\Gamma(H^{++}\to H^+ e^+(\mu^+) \nu_e(\nu_\mu))&=&{g^4_2 \Delta M^5\over
240\pi^3 M_W^4}
\\
\Gamma(H^{++}\to H^+ q \bar{q}')&=&3\Gamma(H^{++}\to H^+ e^+(\mu^+) \nu_e(\nu_\mu))
\end{eqnarray}
where $\Delta M=M_{H++} - M_{H^+}$ and $r_i=M_i/M_\Delta$.
\begin{flushleft}
\underline{Singly Charged Higgs}:
\end{flushleft}
\begin{eqnarray}
\Gamma(H^+\rightarrow \ell^+_i \bar{\nu_j}) &=&
{1\over 16\pi}|\Gamma_+^{ij}|^2 M_{H^+}
\\
\Gamma(H^+\rightarrow W^+_T Z_T) &=&
{\text{cos}^2\theta_W\over 32\pi M_{H^+}} \left( g_1^2\text{sin}\theta_+ \ v_0 \ - \ \sqrt{2}\text{cos}\theta_+ \ (2g_1^2+g_2^2)
\ v_{\Delta} \right)^2 \lambda^{{1\over 2}}(1,r^2_W,r^2_Z) \nonumber \\
&\approx& {g_2^2\sin^4\theta_W M_Z^2\over 8\pi M_{H^+}v_0^2}({\mu v_0^2\over M_{H^+}^2}-\sqrt{2}(2+\cot^2\theta_W)v_\Delta)^2={g_2^2 M_Z^2 v_{\Delta}^2\over 4\pi M_{H^+}v_0^2}
\\
\Gamma(H^+\rightarrow W^+_L Z_L) &=&
{\text{cos}^2\theta_W\over 64\pi M_{H^+}} \left( g_1^2\text{sin}\theta_+ \ v_0 \ - \ \sqrt{2}\text{cos}\theta_+ \ (2g_1^2+g_2^2)
\ v_{\Delta} \right)^2 \lambda^{{1\over 2}}(1,r^2_W,r^2_Z){(1-r^2_W-r^2_Z)^2\over 4r^2_Wr^2_Z}\nonumber \\
&\approx& {M_{H^+}^3\sin^4\theta_W\over 16\pi v_0^4}({\mu v_0^2\over M_{H^+}^2}-\sqrt{2}(2+\cot^2\theta_W)v_\Delta)^2={M_{H^+}^3 v_{\Delta}^2\over 8\pi v_0^4}
\\
\Gamma(H^+\rightarrow W^+_L H_1) &=&
{M_{H^+}g_2^2\over 64\pi r_W^2} \left( \sqrt{2}\text{cos}\theta_+ \ \text{sin}\theta_0 \ - \ \text{sin}\theta_+ \ \text{cos}\theta_0 \right)^2 \lambda^{{3\over 2}}(1,r^2_W,r^2_{H_1})\nonumber \\
&\approx& {\mu^2\over 16\pi M_{H^+}}={M_{H^+}^3 v_{\Delta}^2\over 8\pi v_0^4}
\\
\Gamma(H^+\rightarrow t\bar{b}) &=&
{N_cM_{H^+}m_t^2\text{sin}^2\theta_+\over 8 \pi v_0^2}\lambda^{{1\over 2}}(1,r^2_t,r^2_b)(1-r^2_t-r^2_b)
\approx {N_c\mu^2m_t^2\over 8\pi M_{H^+}^3}={N_cM_{H^+}m_t^2 v_{\Delta}^2\over 4\pi v_0^4}
\end{eqnarray}
\begin{flushleft}
\underline{Heavy CP-even Higgs}:
\end{flushleft}
\begin{eqnarray}
\Gamma(H_2\rightarrow \nu_i\nu_i+\bar{\nu_i}\bar{\nu_i}) &=&
{1\over 16\pi}\text{cos}^2\theta_0{m_{\nu_i}^2\over v_\Delta^2}M_{H_2}\approx {m_{\nu_i}^2\over 16\pi v_\Delta^2}M_{H_2}
\\
\Gamma(H_2\rightarrow Z_T Z_T)&=&
{g_2^4\over 64\pi M_{H_2}\text{cos}^4\theta_W} \left( \text{sin}\theta_0  \ v_0 \ - \ 4 \text{cos}\theta_0 \ v_{\Delta} \right)^2 \lambda^{{1\over 2}}(1,r^2_Z,r^2_Z)\nonumber \\
&\approx& {g_2^2m_Z^2\over 16\pi M_{H_2}\cos^2\theta_Wv_0^2}({\sqrt{2}\mu v_0^2\over M_{H_2}^2}-4v_\Delta)^2={g_2^2 m_Z^2 v_{\Delta}^2\over 4\pi M_{H_2} \text{cos}^2\theta_W v_0^2}
\\
\Gamma(H_2\rightarrow Z_L Z_L)&=&
{g_2^4\over 128\pi M_{H_2}\text{cos}^4\theta_W} \left( \text{sin}\theta_0 \ v_0 \ - \ 4 \text{cos}\theta_0 \ v_{\Delta} \right)^2\lambda^{{1\over 2}}(1,r^2_Z,r^2_Z){(1-2r^2_Z )^2 \over 4r^4_Z}\nonumber\\
&\approx& {M_{H_2}^3\over 32\pi v_0^4}({\sqrt{2}\mu v_0^2\over M_{H_2}^2}-4v_\Delta)^2={M_{H_2}^3 v_{\Delta}^2\over 8\pi v_0^4}
\end{eqnarray}
\begin{eqnarray}
\Gamma(H_2\rightarrow H_1H_1)&=&
{1\over 512\pi
M_{H_2}}\text{cos}\theta_0^2
\left( 3\text{sin}2\theta_0\lambda v_0 \ + \ 4\sqrt{2}\text{cos}2\theta_0 \ \mu \ - \ 4\sqrt{2}\text{sin}^2\theta_0 \ \mu \right)^2 \lambda^{{1\over
2}}(1,r^2_{H_1},r^2_{H_1})
                \nonumber \\
        &\approx&{1\over 32\pi M_{H_2}}({6M_{H_1}^2 v_{\Delta} \over
v_0^2}+\sqrt{2}\mu)^2\approx{\mu^2\over 16\pi M_{H_2}}={M_{H_2}^3 v_{\Delta}^2 \over
8\pi v_0^4}
\\
\Gamma(H_2\rightarrow t\bar{t})&=&
{N_cM_{H_2}m_t^2 \text{sin}^2\theta_0\over 8 \pi v_0^2}\lambda^{{1\over 2}}(1,r^2_t,r^2_t)(1-4r^2_t)
\approx {N_cm_t^2\mu^2\over 4\pi M_{H_2}^3}={N_cM_{H_2}m_t^2 v_{\Delta}^2\over 2\pi v_0^4}
\\
\Gamma(H_2\rightarrow b\bar{b})&=&
{N_cM_{H_2}m_b^2 \text{sin}^2\theta_0\over 8 \pi v_0^2}\lambda^{{1\over 2}}(1,r^2_b,r^2_b)(1-4r^2_b)
\approx {N_cm_b^2\mu^2\over 4\pi M_{H_2}^3}={N_cM_{H_2}m_b^2 v_{\Delta}^2\over 2\pi v_0^4}
\end{eqnarray}
\begin{flushleft}
\underline{CP-odd Higgs}:
\end{flushleft}
\begin{eqnarray}
\Gamma(A\rightarrow \nu_i\nu_i+\bar{\nu_i}\bar{\nu_i}) &=&
{1\over 16\pi}\text{cos}^2\alpha{m_{\nu_i}^2\over v_\Delta^2}M_{A}\approx {m_{\nu_i}^2\over 16\pi v_\Delta^2}M_{A}
\\
\Gamma(A\rightarrow Z_L H_1)&=&
{M_A g_2^2\over 64\pi r_Z^2 \text{cos}^2\theta_W} \left( \text{cos}\theta_0 \text{sin}\alpha \ - \ 2\text{sin}\theta_0 \text{cos}\alpha \right)^2
\ \lambda^{{3\over 2}}(1,r^2_{H_1},r^2_Z)\nonumber \\
&\approx& {\mu^2\over 8\pi M_A}={M_A^3 v_{\Delta}^2\over 4\pi v_0^4}
\\
\Gamma(A\rightarrow t\bar{t})&=&
{N_cM_{A}m_t^2 \text{sin}^2 \alpha \over 8\pi v_0^2}\lambda^{{1\over 2}}(1,r^2_t,r^2_t)
\approx {N_c\mu^2m_t^2\over 4\pi M_A^3}={N_cM_{A}m_t^2 v_{\Delta}^2\over 2\pi v_0^4}
\\
\Gamma(A\rightarrow b\bar{b})&=&{N_cM_{A}m_b^2 \text{sin}^2 \alpha
\over 8 \pi v_0^2}\lambda^{{1\over 2}}(1,r^2_b,r^2_b)
\approx {N_c\mu^2m_b^2\over 4\pi M_A^3}={N_cM_{A}m_b^2 v_{\Delta}^2\over 2\pi v_0^4}
\end{eqnarray}

%
%

\end{document}